\documentclass[3p,final]{elsarticle}

\pdfoutput=1

\usepackage{graphicx}
\usepackage[version=3]{mhchem}
\usepackage[T1]{fontenc}

\biboptions{sort&compress}

\usepackage{xcolor}
\definecolor{B}{HTML}{006699}
\usepackage[implicit=false,colorlinks=true,urlcolor=blue,bookmarks=false,pdfpagelabels=false]{hyperref}
\usepackage{libertine}
\usepackage[libertine]{newtxmath}
\usepackage[scaled=0.95]{inconsolata}
\usepackage{microtype}

\usepackage{multicol}
 
\begin{document}

\title{Quantitative wave function analysis for excited states of transition metal complexes}

\author[UoV]{Sebastian Mai}
\ead{sebastian.mai@univie.ac.at}
\author[UoV]{Felix Plasser}
\ead{felix.plasser@univie.ac.at}
\author[UoV]{Johann Dorn}
\author[US]{Maria Fumanal}
\ead{fumanal@unistra.fr}
\author[US]{Chantal Daniel}
\ead{c.daniel@unistra.fr}
\author[UoV]{Leticia Gonz\'alez\corref{cor1}}
\ead{leticia.gonzalez@univie.ac.at}

\cortext[cor1]{Corresponding author}

\address[UoV]{Institute of Theoretical Chemistry, Faculty of Chemistry, University of Vienna, W\"ahringer Str. 17, 1090 Vienna, Austria.}
\address[US]{Laboratoire de Chimie Quantique, Institut de Chimie Strasbourg, UMR-7177 CNRS/Universit\'e de Strasbourg, 1 Rue Blaise Pascal BP 296/R8, Strasbourg, 67008 France.}

\date{\today}

\hypersetup{urlcolor=B}

%


\begin{abstract}

The character of an electronically excited state is one of the most important descriptors employed to discuss the photophysics and photochemistry of transition metal complexes.
In transition metal complexes, the interaction between the metal and the different ligands gives rise to a rich variety of excited states, including metal-centered, intra-ligand, metal-to-ligand charge transfer, ligand-to-metal charge transfer, and ligand-to-ligand charge transfer states. 
Most often, these excited states are identified by considering the most important wave function excitation coefficients and inspecting visually the involved orbitals.
This procedure is tedious, subjective, and imprecise. 
Instead, automatic and quantitative techniques for excited-state characterization are desirable.
In this contribution we review the concept of charge transfer numbers---as implemented in the TheoDORE package---and show its wide applicability to characterize the excited states of transition metal complexes.
Charge transfer numbers are a formal way to analyze an excited state in terms of electron transitions between groups of atoms based only on the well-defined transition density matrix.
Its advantages are many: it can be fully automatized for many excited states, is objective and reproducible, and provides quantitative data useful for the discussion of trends or patterns.
We also introduce a formalism for spin-orbit-mixed states and a method for statistical analysis of charge transfer numbers.
The potential of this technique is demonstrated for a number of prototypical transition metal complexes containing Ir, Ru, and Re.
Topics discussed include orbital delocalization between metal and carbonyl ligands, nonradiative decay through metal-centered states, effect of spin-orbit couplings on state character, and comparison among results obtained from different electronic structure methods.


\end{abstract}


\maketitle
\clearpage
{
\sloppy
\tableofcontents
\noindent\rule{\textwidth}{0.5pt}
}


\section{Introduction}
\label{sec:intro}

Transition metal complexes (TMCs) are a very rich class of molecules with a vast array of possible applications.
One very interesting aspect of TMCs is their collection of exceptional excited-state properties---i.e., their spectroscopic, photophysical, and photochemical properties.
Due to these properties, various classes of TMCs are actively employed, or show high potential, for a large number of innovative technological applications~\cite{Garino2013PTRSA}.
They are used in devices which convert solar energy, either directly to electrical current in dye-sensitized solar cells~\cite{Gratzel2003JPPC,Polo2004CCR,Fantacci2011CCR,Jager2015CCR}, to chemical energy of fuels in artificial photosynthesis~\cite{Gust2009ACR,Teets2011CC}, or to complex chemicals in TMC-based photocatalysis~\cite{Hennig2008,Bitterwolf2011,Twilton2017NRC}.
Furthermore, TMCs are useful in photodynamic therapy~\cite{Farrer2009DT,Dolmans2003NRC,Ruggi2011CCR}, where light activates therapeutically beneficial chemical reactions that initiate apoptosis.
TMCs can also be used to produce light, either for illumination or display purposes in organic light-emitting diodes (OLEDs)~\cite{Choy2014AM,Fantacci2011CCR,Yersin2011CCR,Powell2015CCR}, or for various kinds of luminescent probes in biochemical research~\cite{Lakowicz2006,Gray2003QRB,Bitterwolf2011,Vlcek2015ACR}, biological imaging~\cite{Fernandez-Moreira2010CC,Ruggi2011CCR}, or sensoring~\cite{Ruggi2011CCR}.
Moreover, TMCs can also be used to modify light, by means of their non-linear optical properties~\cite{Zhang2007O,Fantacci2011CCR} or photoswitching abilities~\cite{Coppens2002CR,Polo2006CCR}.

From a chemical point of view, TMCs provide a highly flexible toolbox: by varying the central metal atom or the ligand sphere, the complexes can be functionalized in order to fine-tune properties and integrate them into different chemical or biological environments.
The specific interaction between the metal and the different ligands leads to several distinct ``textbook'' classes of excited states in TMCs---metal-centered states (MC), intra-ligand states (IL), metal-to-ligand charge transfer states (MLCT), ligand-to-metal charge transfer states (LMCT), and ligand-to-ligand charge transfer states (LLCT)---whose properties and interplay are of central importance ~\cite{Turki2001JACS, Daniel2003CCR, Daniel2015CCR, Daniel2016TCC, Daniel2017CCR}.
Actually, most of the above-mentioned applications are based on achieving some favorable properties of particular excited states \cite{Vlcek2009TOC,Kumar2010TOC,Powell2015CCR}.
For example, redox photocatalysis exploits the fact that MLCT states can be both better electron acceptors and electron donors than the ground state~\cite{Bitterwolf2011}.
TMC-based OLEDs rely on MLCT states which provide ultrafast excited-state dynamics---internal conversion (IC) and intersystem crossing (ISC)---and a long life time of the lowest excited state~\cite{Yersin2011CCR,Powell2015CCR}.
On the other hand, the population of MC states often leads to highly efficient nonradiative decay to the ground state or to the dissociation of metal-ligand bonds~\cite{Yersin2011CCR, Wagenknecht2011CCR, Baryshnikov2017CR}.
Which type of electronic states are involved upon excitation is also relevant for sensoring/imaging/photolabeling applications because states of MLCT and LLCT character often show a high sensitivity to the environment, leading, e.g., to solvato-~\cite{Sone1990RIC} or rigidochromism~\cite{Lees1995CIC}.
Moreover, in certain TMCs, such as metal carbonyls, the population of MLCT or MC states affects the vibrational spectra such that these states can be conveniently probed with IR spectroscopy~\cite{Butler2007CCR}.

Due to the very large interest in the excited states of TMCs, there is an effort to describe these states accurately, mostly in terms of their transition energies, optical intensity, and state character.
Theoretically, two general approaches can be distinguished to this aim~\cite{Powell2015CCR}.
Phenomenological or semi-empirical methods---like the ligand-field theory~\cite{Griffith1957QRCS,Ballhausen1962,Figgis2000}---excel at providing simple, unified, and qualitative models with significant predictive power~\cite{Powell2015CCR}.
In contrast, quantum chemistry, based either on wave function approaches or density functional theory (DFT), allows for quantitative predictions related to individual compounds.
Nowadays, quantum chemical computations can be considered the main work horse for the theoretical prediction of excited-state properties of TMCs.
However, to boil down the large amount of computed data into simple and general statements is challenging because comprehensive analysis tools are not as developed as the quantum chemical methods themselves.

That the accurate computational description of excited states of TMCs is a formidable challenge is evidenced by the large amount of literature devoted to this topic, see, e.g., Refs~\cite{Gisbergen1999JPCA, Dreuw2005CR, Neese2007CCR, Vlcek2007CCR, Cramer2009PCCP, Vancoillie2011JCTC, Labat2012ACR, Gonzalez2012C, Almeida2014, Daniel2015CCR, Garino2013PTRSA,Daniel2016TCC,Daniel2003CCR}.
The different classes of excited states depend to a different extent on the description of electronic correlation and exchange effects, necessitating a well-balanced method.
Among the many electronic structure methods available, two approaches are most often applied to TMCs.
One are multi-configurational/multi-reference methods~\cite{Neese2007CCR, Vancoillie2011JCTC, Gonzalez2012C}, including complete active space self-consistent field (CASSCF)~\cite{Roos1987ACP}, CAS second-order perturbation theory (CASPT2)~\cite{Pulay2011IJQC}, as well as the related NEVPT2~\cite{Angeli2007TCA} or MRPT2~\cite{Chattopadhyay2016WCMS}, and multi-reference configuration interaction (MRCI)~\cite{Szalay2012CR}.
The other avenue is to employ DFT with its excited-state extension TD-DFT (time-dependent DFT)~\cite{Casida1995, Marques2004ARPC, Casida2012ARPC, Ferre2016}, which is the standard practice due to its favorable balance between accuracy, computational cost, and usability~\cite{Gonzalez2012C,Daniel2003CCR,Neese2007CCR}.
A detailed description of these methods is beyond the scope of this work.
However, regardless of the method chosen, several computational aspects are important to describe accurately TMCs~\cite{Daniel2015CCR}.
Foremost, metal atoms induce strong relativistic effects due to their large nuclear charge, especially scalar relativistic effects and spin-orbit coupling (SOC), and those should be included in the calculations~\cite{Bakova2011CCR, Heydova2012JPCA}.
Additionally, environment effects can be critical in obtaining results with predictive power regarding experimental observables~\cite{Very2014CEJ}.

Once an accurate calculation of the excited states has been performed, detailed insight into the excited-state characters is indispensable in order to tune TMCs for technological applications.
In TMCs, such an analysis of the excited-state wave functions is very demanding, as often the excited states are represented as linear combinations of different configurations composed of delocalized orbitals whose character is not well defined.
In this context, it should be remembered that these orbitals are simply a way to represent the many-electron wave function, whereas it is ``fallacious to conflate the molecular orbitals with anything in the real world'' \cite{Powell2015CCR}.
Moreover, a given computed state might be a non-trivial linear combination of the ``textbook'' classes of excited states~\cite{Vlcek2009TOC}, making assignment difficult.
Also, the comparison between excited-state properties computed using wave function methods and DFT is not an easy task and encounters limitations, as, e.g., illustrated for third-row TMCs~\cite{Heydova2012JPCA, Fumanal2016JCC}.

The factors above challenge the analysis of excited-state computations, so that in many cases quantitative or even qualitative insight from the computational results is difficult to obtain.
In this direction, recent developments have been proposed---mostly focused on the DFT framework---for rationalizing and interpreting charge transfer excited-states properties~\cite{LeBahers2011JCTC, Ehara2013JCC, Savarese2013JPCB, LeBahers2014PCCP, Savarese2017JPCA, Etienne2014JCTC, Etienne2014JCTCa, Peach2008JCP, Ronca2014JCP}.
These methods are very suitable for analyzing the strength of charge transfer of low lying states in molecular systems with a clear donor-acceptor structure.
However, in TMCs different parts of the system could act as donors or acceptors and thus it not enough to quantify the charge transfer character but also to specify its direction, i.e., to clearly discriminate between MLCT, LMCT and LLCT states.

With the goal to ameliorate this situation, an automatic procedure to analyze wave functions as well as electron densities has been recently designed, providing a variety of visual and quantitative exploration methods for excited-state computations \cite{Plasser2012JCTC, Plasser2014JCP1, Bappler2014PRA, Plasser2015JCC, Plasser2016JCP}.
This comprehensive toolbox has shown to be useful whenever a large number of excited states is to be analyzed, e.g., ensemble calculations of spectra, trajectory simulations, or higher excited states~\cite{Plasser2012JCP, Plasser2012JPCA, Plasser2014TCC, Nogueira2017CS}.
Therefore, it is expected to be particularly helpful in the case of TMCs where a high density of electronic excited states of various characters is present.
One additional advantage of these quantitative tools is that they eliminate the role of subjective reasoning in the discussions.
State characters are defined based on well-defined equations rather than on the individual subjective interpretation of a molecular orbital (MO).
Quantitative tools can also reveal subtle trends, such as ``substitution at some position by a fluorine atom increases the MLCT character of a state by 5\%.''
Last but not least, the application of sophisticated analysis tools has allowed identifying physical phenomena such as excitonic correlation \cite{Mewes2016PCCP, Mewes2017JPCL} and secondary orbital relaxation \cite{Plasser2015JPCA}, which could otherwise not be understood even qualitatively in the canonical MO picture.

These wave function analysis tools are implemented in the TheoDORE package \cite{Plasser2017TheoDORE}, which provides interfaces to a wide range of quantum chemistry programs and electronic structure methods.
As part of this work, we report a new interface between the TheoDORE and ADF \cite{ADF2017} program packages that allows for a quantitative analysis of large-scale TD-DFT computations and also provides the possibility of analyzing spin-orbit coupled states.
In addition, we use integrated analysis utilities in the Q-Chem \cite{QChem4, Plasser2014JCP1} and Molcas \cite{Aquilante2015JCC, Plasser2017JCTC} program packages in connection with wave function-based single- and multi-reference methods, where appropriate.

The goal of this review is to present a variety of powerful ways in which wave function analysis can be applied to TMC excited-state computations.
First, we shall introduce several orbital transformation techniques which can be used to represent excitations (Section~\ref{sec:orbs}).
By means of descriptors like single-excitation contribution, charge transfer numbers, or entanglement measures we move away from orbital visualization to a purely numeric representation of the excitations and we show how these descriptors can be used to compactly depict large number of excited states (Section~\ref{ssec:CTnumbers} and \ref{ssec:CTnumbersSOC}).
We also show that a statistical analysis of the numeric descriptors can lead to further insight, which is otherwise not attainable by simple inspection of the wave functions (Section~\ref{ssec:correlation}).

The applicability of these tools will be demonstrated in a series of calculations on prototypical TMCs containing Ir(III), Ru(II), and Re(I) centers (Section~\ref{sec:cases}).
Five case studies will illustrate the potential of quantitative analysis techniques at deciphering the excited state characters in different situations, such as:
(i) How are the excited-state characters affected by nuclear relaxation?
(ii) How does SOC change the mixing ratios between the different classes of excited states?
(iii) How are the excited-state characters influence by the ligand substitution?
(iv) How can the interaction of the metal center and the ligands be quantified?
(v) And finally, how does the electronic structure method affect the electronic excited states, and how can one easily compare a multitude of electronic states computed at differing levels of theory?


\section{Visual inspection of orbitals}
\label{sec:orbs}


For pedagogical purposes, we begin reviewing qualitative procedures to analyze electronic wave functions.
Without any doubt, the most common way to do so is by inspecting visually the involved orbitals.
Despite this being an intuitive procedure, it should be remembered that the orbitals are only mathematical objects without a direct physical meaning~\cite{Powell2015CCR}.
As a consequence, there is not one ``true'' set of orbitals but a number of different visualization techniques that can be applied, leading to different sets of orbitals.
Depending on the visualization technique chosen, the orbital shapes and, thus, the apparent state characters can be quite different.
Hence, we commence this section by introducing the different techniques available to carry out a qualitative wave function analysis, based on the visual inspection of MOs.
The goal could be formulated as identifying the ``textbook'' state character of a given electronic excited state.
For this purpose,  we shall focus on linear-response (LR) TD-DFT, but most of the techniques described in this and the following sections can be applied to states computed with other electronic structure methods as well.

In TD-DFT, one considers the response of the electronic density of some molecule to a small time-dependent perturbation~\cite{Casida1995,Marques2004ARPC,Casida2012ARPC}.
The excitation energies $E^I$, i.e., the energy of the excited state relative to the ground state, can then be obtained by finding frequencies of the time-dependent perturbation where the density response function has a pole.
By considering only the linear density response, it is then possible to find a non-Hermitian eigenvalue problem whose solutions are the poles of the response function.
This eigenvalue problem can be written as:
\begin{equation}
  \begin{pmatrix}
    \mathbf{A}&\mathbf{B}\\
    \mathbf{B}&\mathbf{A}
  \end{pmatrix}
  \begin{pmatrix}
    \mathbf{X}^I\\
    \mathbf{Y}^I
  \end{pmatrix}
  =E^I
  \begin{pmatrix}
    1&0\\
    0&-1
  \end{pmatrix}
  \begin{pmatrix}
    \mathbf{X}^I\\
    \mathbf{Y}^I
  \end{pmatrix}
  .
  \label{eq:tddft}
\end{equation}
Here, the matrices $\mathbf{A}$ and $\mathbf{B}$ are called orbital rotation Hessians~\cite{Furche2002JCP}, and depend on orbital energy differences and integrals over the two-electron Coulomb and exchange operators.
Once a solution to~\eqref{eq:tddft} has been found, in addition to the excitation energy $E^I$ one also obtains the response vector $(\mathbf{X}^I,\mathbf{Y}^I)$.
This vector is similar to the CI vector from a configuration interaction or multi-reference calculation.
The vector $\mathbf{X}^I$ corresponds to excitations from occupied to virtual orbitals, e.g., vector element $X^I_{ia}$ is the coefficient corresponding to an excitation from an occupied orbital $i$ to a virtual orbital $a$.
The vector $\mathbf{Y}^I$ corresponds to ``de-excitations'' from virtual into occupied orbitals.
These de-excitations, which follow from the derivation of LR TD-DFT equations, do not possess a clear physical meaning~\cite{Dreuw2005CR} and are usually dismissed in the discussions. 
Ignoring the de-excitations in equation \eqref{eq:tddft} is known as the Tamm-Damcoff approximation (TDA)~\cite{Hirata1999CPL}:
\begin{equation}
  \mathbf{A}\mathbf{X}^I
  =
  E^I\mathbf{X}^I.
  \label{eq:tda}
\end{equation}
The TDA is often applied in TD-DFT for computational efficiency reasons, as it replaces a non-Hermitian eigenvalue problem by a Hermitian eigenvalue problem of half the size.
Even more, it has been pointed out that the TDA sometimes leads to improved results by avoiding triplet instabilities for range-separated hybrid functionals \cite{Peach2012JPCA}.
For these reasons, we will restrict ourselves to the TDA in the following.
More information regarding wave function analysis for full TD-DFT can be found in Ref.~\cite{Mewes2015JCP} and regarding mathematical considerations relevant to de-excitations in Ref.~\cite{Etienne2015JCP}.

\begin{figure}[tb]
  \centering
  \includegraphics[scale=1]{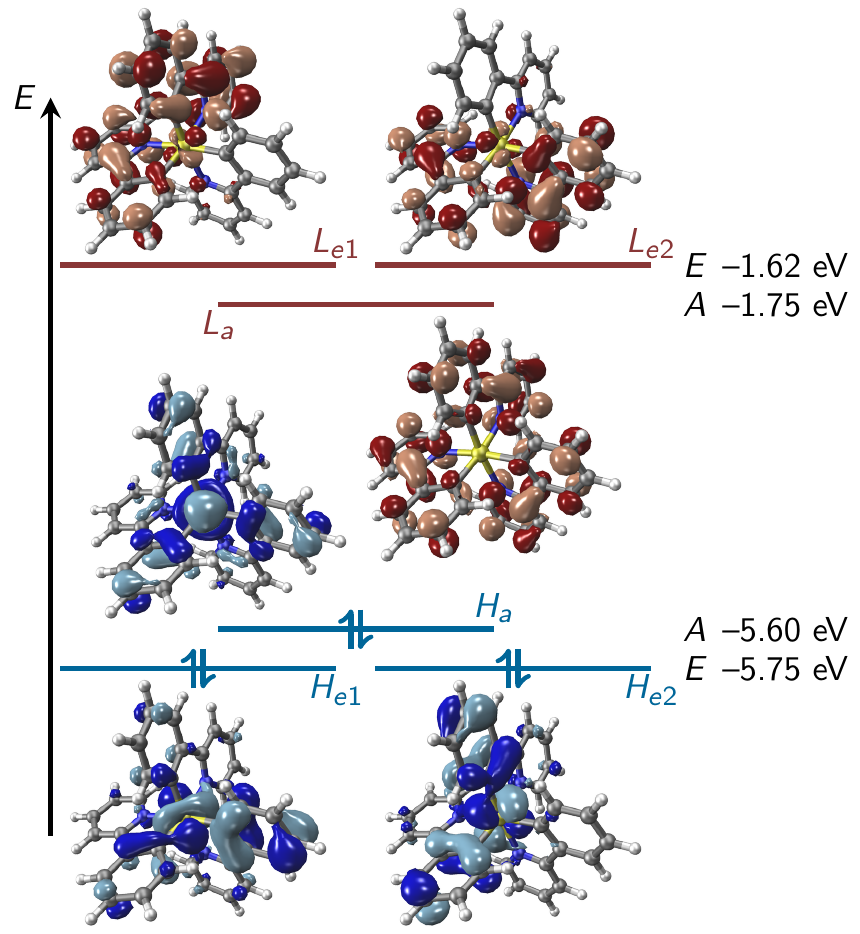}
  \caption{
  Depiction of the frontier canonical orbitals for [Ir(ppy)$_3$].
  The figure also defines the orbital labels $H_a$, $H_{e1}$, $H_{e2}$ for the highest occupied orbitals, and $L_a$, $L_{e1}$, $L_{e2}$ for the lowest unoccupied orbitals.
  (TDA-B3LYP/TZP-DZ(P), COSMO(water); see Appendix A for full computational details)
  }
  \label{fig:orbitals_Irppy3_canonical}
\end{figure}

One of the complexes considered in this tutorial is [Ir(ppy)$_3$] (ppy=2-phenylpyridyl), which is a strongly phosphorescent compound used in OLED design~\cite{Dedeian1991IC} (for a more comprehensive introduction of this complex, see Section~\ref{ssec:IrNuclear}).
Its frontier orbitals, i.e., the ones mainly involved in the low-lying excitations, are depicted in Figure~\ref{fig:orbitals_Irppy3_canonical}.
The three highest occupied orbitals ($H_a$, $H_{e1}$, $H_{e2}$) are combinations of metal $5d$ orbitals and ligand $\pi$ orbitals, whereas the three lowest unoccupied orbitals ($L_a$, $L_{e1}$, $L_{e2}$) are linear combinations of ligand $\pi^*$ orbitals with little metal contributions.
These frontier orbitals are characteristic for many pseudo-octahedral complexes with $D_3$ or $C_3$ symmetry~\cite{Powell2015CCR}.
Note, however, that although our calculations used a symmetric geometry, the calculations were performed without symmetry---as is practiced often.

In order to perform an analysis of the computed excited state, the excitation vector $\mathbf{X}^I$ is needed.
Most commonly, the excited states are analyzed by identifying the largest contributions $X^I_{ia}$ to the excitation vector.
By manual inspection, the orbitals $i$ and $a$ are then assigned to some orbital class---e.g., metal-centered or ligand centered.
Then, if $i$ is a metal-centered orbital and $a$ is a ligand-centered orbital, the electronic state is characterized as an MLCT state.
In Table~\ref{tab:orbitals_Irppy3_canonical}, we compile the excitation energies, symmetries, and leading excitations for the first six triplet states of [Ir(ppy)$_3$].
As can be seen, the lowest six triplet states are very close in energy (within 0.3~eV), and due to the symmetry of the molecule, are grouped into (near-)degenerate sets of states.
The states $T_4$ to $T_6$ are actually so close in energy that it is not possible to identify which is the $A$ state and which are the $E$ states, and are thus labeled as a linear combination of both.
Table~\ref{tab:orbitals_Irppy3_canonical} lists all the contributions with a weight $|X^I_{ia}|^2$ of at least 0.15.
These contributions represent at least 58\% of each of the shown states and it is thus possible to provide a qualitative state assignment based on this table.
Because the occupied orbitals are mixed metal-/ligand-centered and the virtual orbitals are mostly ligand-centered, it can be said that the six triplet states all have mixed MLCT, IL, and LLCT character.

\begin{table}[tb]
  \setlength{\tabcolsep}{3pt}
  \centering
  \caption{Excitation energies ($E$), symmetry (Sym.), and leading excitations with coefficients $X_{ia}$ for the six lowest triplet states of [Ir(ppy)$_3$].
  (TDA-B3LYP/TZP-DZ(P), COSMO(water))
  }
  \label{tab:orbitals_Irppy3_canonical}
  \begin{tabular}{lcclll}
    \hline
                &$E$ (eV)       &Sym.   &\multicolumn{3}{l}{Leading excitations}\\
    \hline
    $T_1$       &2.89           &$A$      &$+0.85H_a    L_a$
                                          &$+0.32H_{e1} L_{e2}$
                                          &$+0.32H_{e2} L_{e1}$\\
    $T_2$       &2.93           &$E$      &$+0.58H_a    L_{e1}$
                                          &$+0.49H_a    L_{e2}$\\
    $T_3$       &2.93           &$E$      &$+0.58H_a    L_{e2}$
                                          &$-0.49H_a    L_{e1}$\\
    $T_4$       &3.17           &$A$+$E$  &$+0.48H_{e1} L_{e1}$
                                          &$+0.44H_{e2} L_{e1}$
                                          &$+0.41H_{e1} L_a$\\
    $T_5$       &3.17           &$A$+$E$  &$+0.56H_{e2} L_a$
                                          &$+0.46H_{e2} L_{e2}$
                                          &$-0.42H_{e2} L_{e1}$\\
    $T_6$       &3.17           &$A$+$E$  &$-0.56H_{e1} L_a$
                                          &$+0.53H_{e1} L_{e2}$
                                          &$+0.45H_a    L_{e2}$\\
    \hline
  \end{tabular}
\end{table}

Beyond this basic qualitative finding, it is hard to gain further knowledge from the coefficients in Table~\ref{tab:orbitals_Irppy3_canonical}.
It is difficult to discern any differences between these six states, as they are complicated linear combinations of different excitations.
Moreover, it is very hard to answer questions like ``Are the ligand-centered contributions combinations of local excitations (IL) or charge transfer (LLCT) excitations?''
It is also not clear from Table~\ref{tab:orbitals_Irppy3_canonical} whether the states shown are actually multi-configurational states, or whether an orbital representation could be found where a single configuration is sufficient to describe one of the states.


One of the basic problems of the above analysis is that, when expressed in the basis of the canonical Kohn-Sham orbitals, the excited states are often linear combinations of many different excitations with similar weights---as exemplified in Table~\ref{tab:orbitals_Irppy3_canonical}.
In such a case, the actual state character is determined by the (non-trivial) interference patterns between the excitations, as visible by the different signs of the coefficients in the table.
In order to simplify the interpretation, it is advantageous to seek an orbital representation which is more suitable to express the excited states in a simple and compact form.
Ideally, this representation should also be method independent, because there are subtle differences in the interpretation of, e.g., canonical Kohn-Sham orbitals versus Hartree-Fock or CASSCF orbitals, and between the response vector from TD-DFT versus the configuration interaction vectors from wave function-based methods~\cite{Meer2014JCTC,Dreuw2005CR}.

To this end, in the following we will heavily use the one-electron transition density matrix (1TDM) $\mathbf{T}^{IJ}$, whose elements in general can be defined as \cite{Plasser2014JCP1}:
\begin{equation}
  T^{IJ}_{rs}
  =
  \left\langle
    \Psi^I
    \middle|
    \hat{a}^\dagger_r\hat{a}_s
    \middle|
    \Psi^J
  \right\rangle.
  \label{1TDM-general}
\end{equation}
Here, $\hat{a}^\dagger_r$ is the creation operator for orbital $\phi_r$, $\hat{a}_s$ the annihilation operator for orbital $\phi_s$, and $I$ and $J$ are two excited state indices.
We will mostly consider transition density matrices between the ground state and an excited state $I$, which we write as $\mathbf{T}^I$.
Such a 1TDM is closely related to physical observables like the transition dipole moment \cite{Plasser2014JCP1}, showing that the 1TDM is a suitable basis for a more physically meaningful analysis of excited-state character.
For methods which allow describing higher excitations (multi-reference, quadratic response, etc) there will also be two-electron and higher transition density matrices, but these contribute only little for states which are predominantly described as single-excitations.

In TD-DFT, which employs a single-determinant ground state, the 1TDM $\mathbf{T}^I$ is directly related to the response vector.
Within the TDA, the following assignment is made:
\begin{equation}
  T^{I}_{ia}=X^I_{ia},
\end{equation}
where $i$ runs over the occupied orbitals and $a$ over the virtual ones.

With the 1TDM defined, we can now introduce the concept of \emph{natural transition orbitals} (NTOs)~\cite{Luzanov1976TEC,Martin2003JCP,Plasser2012JCTC}.
These are obtained by a singular value decomposition of the 1TDM:
\begin{equation}
  \mathbf{T}^I
  =
  \mathbf{V}^I
  \begin{pmatrix}
    \lambda_1^I\\
    &\lambda_2^I\\
    &&\ddots
  \end{pmatrix}
  \mathbf{W}^{I\dagger}.
  \label{eq:ntos}
\end{equation}
Analogously, it can be stated that the unitary matrices $\mathbf{V}^I$ and $\mathbf{W}^I$ contain the eigenvectors of the \emph{hole} and \emph{particle} density matrices ($\mathbf{T}^I\mathbf{T}^{I\dagger}$ and $\mathbf{T}^{I\dagger}\mathbf{T}^I$, respectively)~\cite{Plasser2014JCP1}.
From the matrix $\mathbf{V}^I$ one can construct the NTOs of the \emph{excitation hole} by summing over the appropriate occupied orbitals
\begin{equation}
  \psi^{\mathrm{h},I}_p(\mathbf{r})
  =
  \sum_{i\in \mathrm{occ}}
  V^I_{i p}\phi_i(\mathbf{r})
\end{equation}
and from the matrix $\mathbf{W}^I$ the NTOs of the \emph{excited electron} by summing over the virtual orbitals
\begin{equation}
  \psi^{\mathrm{e},I}_p(\mathbf{r})
  =
  \sum_{a\in \mathrm{virt}}
  W^I_{a p}\phi_a(\mathbf{r}).
\end{equation}
The value $(\lambda_p^I)^2$ then gives the weight of the excitation $\psi^{\mathrm{h},I}_p\rightarrow \psi^{\mathrm{e},I}_p$ in the excited state $I$.
The strength of the NTO transformation is that usually, only a few $(\lambda^I_p)^2$ are significantly larger than zero, and by considering only those few excitations, a very compact representation for the excited state can be obtained.
The number of significant NTOs contributing to a given state can be quantified by the NTO participation ratio (PR) \cite{Plasser2012JCTC}:
\begin{equation}
\label{eq:PRNTO}
  \mathrm{PR}_\mathrm{NTO}^I
  =
  \frac{
  \left(\sum_p (\lambda_p^I)^2\right)^2
  }{
  \sum_p (\lambda^I_p)^4
  },
\end{equation}
which is a measure of how many single-particle excitations are necessary to describe a state and, thus, its multi-configurational character.
This value is always greater or equal to one.
$\mathrm{PR}_\mathrm{NTO}$ has an important practical implication, i.e., it shows how many orbitals should be visualized.
More importantly, the number of non-vanishing NTO amplitudes have a clear physical meaning that can be interpreted in the context of static electron correlation~\cite{Surjan2007CPL}, exciton formation \cite{Mayer2007CPL, Plasser2012JCTC}, and entanglement between the \emph{electron} and \emph{hole} quasiparticles \cite{Plasser2016JCP_Ent}.
The $\mathrm{PR}_\mathrm{NTO}$ value has been applied practically to quantify charge resonance interactions during excimer formation of organic molecules \cite{Plasser2012JCTC, Iyer2017} and to understand the intermediate states involved in two-photon absorption \cite{Nanda2017JPCL}.


        \begin{figure*}
          \centering
          \includegraphics[scale=1]{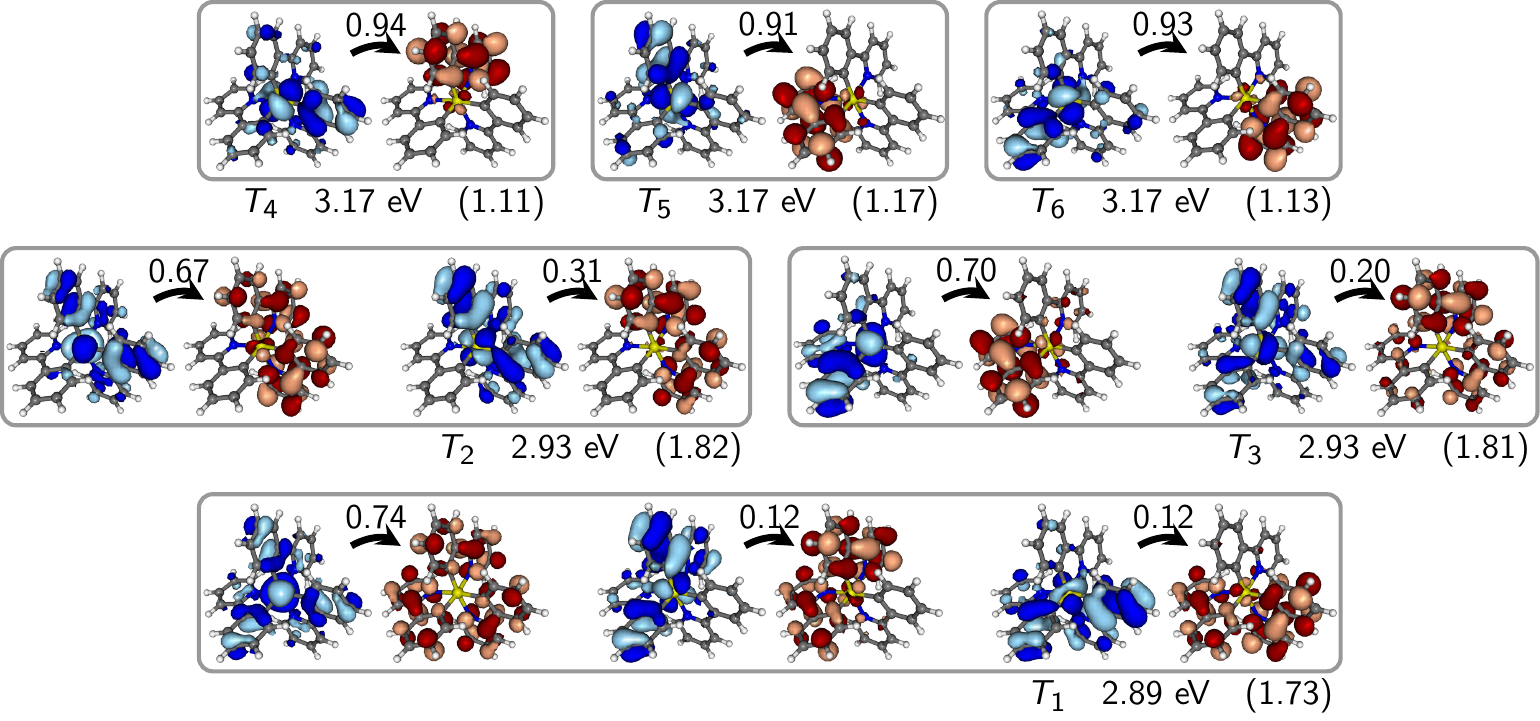}
          \caption{
          Depiction of the leading NTO pairs for the first six triplet states of [Ir(ppy)$_3$], with the weights of the NTO pairs given above the arrows.
          The labels under each set of NTOs indicates the shown state, the excitation energy, and the $\mathrm{PR}_\mathrm{NTO}$ in parenthesis.
          (TDA-B3LYP/TZP-DZ(P), COSMO(water))
          }
          \label{fig:orbitals_Irppy3_nto}
        \end{figure*}

In Figure~\ref{fig:orbitals_Irppy3_nto}, we plot the most important NTOs for the six triplet states discussed above.
The NTOs represent the excitations more compactly than the canonical orbitals did, as can be seen from the sums of weights. 
Indeed, for all states, the shown NTOs represent at least 90\%, which is much better than the 58\% represented in Table~\ref{tab:orbitals_Irppy3_canonical}.
Whereas each individual state can be represented with few NTO pairs, it is now necessary to plot (generally different) NTOs for each state, so that a large number of NTOs will be needed to represent a large set of excited states.

The analysis of the NTOs in Figure~\ref{fig:orbitals_Irppy3_nto} allows obtaining qualitative insight into the state characters beyond the insight gained from Table~\ref{tab:orbitals_Irppy3_canonical}.
First, $\mathrm{PR}_\mathrm{NTO}$ allows judging the multi-configurational character of the states, showing that the first three triplets are indeed multi-configurational---as could be anticipated above.
On the contrary, $T_4$, $T_5$, and $T_6$ are identified as virtually single-configurational states---a fact that could hardly be expected from the canonical orbital representation.
Moreover, in the NTO basis we can qualitatively distinguish between IL and LLCT transitions.
$T_1$, $T_2$, and $T_3$ seem to mostly involve local IL excitations, besides the MLCT character.
This is unexpected, since in the canonical representation, $T_2$ and $T_3$ were mixtures of excitations from the $H_a$ orbital, which is fully delocalized.
The higher triplets constitute a set of LLCT states---again with some MLCT admixture---where the electron is excited counterclockwise (in the used orientation and for the employed enantiomer of [Ir(ppy)$_3$]) from one ligand to another one.
Hence, the six triplet states can be described as a set of three MLCT/IL states and three higher-energy LLCT/MLCT states.

On a side note, $T_4$, $T_5$, and $T_6$ do not reflect the proper symmetry of the molecule because our calculation did not consider explicit symmetry.
This symmetry breaking can be explained by the fact that the three states are nearly degenerate and therefore can mix.
Now, apparently, the computation converged to three states where the excitations are clearly localized.
Oppositely, one can expect that the properly symmetrized states would be more delocalized, forming one $A$ state which is a positive linear combination of the three states, and a pair of $E$ states.
Nevertheless, these symmetrized states would still retain their counterclockwise CT characters, even though this might not be immediately visible in the NTO representation of the symmetrized states.
This discussion shows that in this complex the states have a tendency to easily break symmetry.

Whereas the NTOs contain the full information of the initial TDA computation, there are also ways to reduce this information and represent the excitation in an even more compact way.
A weighted sum of the \emph{electron} and \emph{hole} NTOs leads to the \emph{electron} and \emph{hole} densities \cite{Plasser2014JCP1}
\begin{align}
\rho_h^I(\mathbf{r})&=\sum_p (\lambda_p^I)^2 \left(\psi^{\mathrm{h},I}_p(\mathbf{r})\right)^2\\
\rho_e^I(\mathbf{r})&=\sum_p (\lambda_p^I)^2 \left(\psi^{\mathrm{e},I}_p(\mathbf{r})\right)^2
\end{align}
also termed (unrelaxed) attachment and detachment densities \cite{Head-Gordon1995JCP}.
The difference between these two leads to the unrelaxed difference density
\begin{equation}
\rho_{\mathrm{diff}}^{I}(\mathbf{r})=\rho_h^I(\mathbf{r})-\rho_e^I(\mathbf{r}),
\end{equation}
while the sum
\begin{equation}
\rho_{\mathrm{spin}}^{I}(\mathbf{r})=\rho_h^I(\mathbf{r})+\rho_e^I(\mathbf{r}),
\end{equation}
corresponds to the spin-density in a high-spin triplet computation.

If one goes beyond the TDA, e.g., with full TD-DFT, wave function based methods, or by including orbital relaxations as present in the theory of analytical energy gradients~\cite{Handy1984JCP}, these simple rules connecting the NTOs with the difference and spin densities no longer apply and a wealth of new information can be extracted from the deviations of these quantities, provided suitable analysis methods are chosen \cite{Plasser2014JCP1}.
In particular, a comparison of the NTOs with the eigenvectors of the relaxed difference density matrix---termed \emph{natural difference orbitals}---proved to be very useful for analyzing secondary orbital relaxation effects accompanying the main excitation process \cite{Plasser2014JCP2}.
It was shown that these effects were particularly pronounced for the MLCT states of a small model iridium complex, where the occupied $d$-orbitals contracted after one electron was removed \cite{Plasser2015JPCA}.


\section{Quantification of excited state localization and charge transfer}
\label{sec:CTnumbers}

In the following section, we show how it is possible to identify state characters based on the computation of descriptors, while avoiding entirely the visual inspection of orbitals.
As alluded to in the introduction, a numerical characterization of excited states promises several advantages: it allows avoiding laborious manual work and can be automatized for a large number of excited states; it is objective, reproducible, and more precise; it allows spotting small differences, trends, or continuous changes; and finally, can reveal physics which cannot be represented through orbital pictures.

One such descriptor, the $\mathrm{PR}_\mathrm{NTO}$ value of equation \eqref{eq:PRNTO}, was already introduced above.
In the following, we will introduce another kind of descriptor, the \emph{charge transfer number} \cite{Plasser2012JCTC, Luzanov2010IJQC}, which will be our main tool for the remainder of the text.
The following subsections will first generally outline the underlying theory and show examples for the most common case of non-relativistic (or scalar relativistic) wave functions.
Subsequently, we will present the first application of the same methodology to spin-orbit coupled relativistic states.
Finally, motivated by the fact that in certain situations the different computed charge transfer numbers are strongly correlated, we present a novel clustering ansatz which allows automatically dividing a molecule into chromophoric subunits.


\subsection{Charge transfer numbers}
\label{ssec:CTnumbers}

Excited states are usually discussed in terms of where the excitation originates---i.e., where the \emph{excitation hole} is localized---and where it goes to---i.e., where the \emph{excited electron} is localized.
For example, the excitation in an MLCT state originates at a metal orbital and goes to one or more ligands; an IL state originates at a ligand and goes to a different orbital at the same ligand.
The purpose of the charge transfer numbers is to formalize this concept.
Again, the 1TDM $\mathbf{T}^I$ is considered as the central object, but instead of visualizing it, we partition it among different fragments of the molecular system under study.
Mathematically speaking, the fragments are mutually disjoint subsets of the atoms of the molecule.
Usually, the fragments are defined by chemical intuition, e.g., each ligand of a complex is treated as a separate fragment, but later on we will also discuss strategies for automatic fragmentation. 

The first step is a transformation of the 1TDM from the MO basis to the atomic orbital (AO) basis
\begin{equation}
  \bar{\mathbf{T}}^I
  =
  \mathbf{C}\mathbf{T}^I\mathbf{C}^{\mathrm{T}},
\end{equation}
where $\mathbf{C}$ is the MO-coefficient matrix.
The square of an element of this matrix, $(\bar{T}^I_{\mu\nu})^2$, measures the contribution of an excitation originating on the atomic orbital (AO) $\chi_\mu$ and going to the AO $\chi_\nu$.
The charge transfer number $\Omega_{AB}^{I}$ is now intended to measure the total contributions of excitations that originate at any AO centered on an atom of fragment $A$ and go to any AO on fragment $B$.
In a naive implementation, one could simply sum over all $(\bar{T}^I_{\mu\nu})^2$ elements where $\mu$ lies on $A$ and $\nu$ lies on $B$.
However, for a rigorous computation it is necessary to apply a population analysis scheme that takes into account the non-orthogonality of the AOs.
There is no unique way of dividing the density among atoms and a number of different population analysis schemes have been devised~\cite{Bachrach2007RCC}.
Two such analysis schemes were adapted for analyzing the 1TDM \cite{Plasser2014JCP1,Voityuk2014JCP} and implemented in TheoDORE.
First, it is possible \cite{Plasser2014JCP1} to partition the 1TDM in the spirit of a Mulliken analysis
\begin{equation}
  \Omega_{AB}^I
  =
  \frac{1}{2}
  \sum_{\mu\in A}\sum_{\nu\in B}
  \left[
    (\bar{\mathbf{T}}^I\mathbf{S})_{\mu\nu}
    (\mathbf{S}\bar{\mathbf{T}}^I)_{\mu\nu}
    +
    \bar{T}^I_{\mu\nu}(\mathbf{S}\bar{\mathbf{T}}^I\mathbf{S})_{\mu\nu}
  \right],
\label{eq:OmM}
\end{equation}
where $\mathbf{S}$ is the AO overlap matrix, yielding an equation that is related to Mayer's bond order~\cite{Mayer1986IJQC}.
Alternatively, a L\"owdin orthogonalization can be applied to $\bar{\mathbf{T}}^I$ \cite{Voityuk2014JCP}, leading to the orthogonal matrix
\begin{equation}
\tilde{\mathbf{T}}^I=
\mathbf{S}^{1/2}\bar{\mathbf{T}}^I\mathbf{S}^{1/2},
\end{equation}
which can be directly used for summation
\begin{equation}
  \Omega_{AB}^I
  =
  \sum_{\mu\in A}\sum_{\nu\in B}
  (\tilde{T}^I_{\mu\nu})^2
.
\label{eq:OmL}
\end{equation}
In the implementation used here, the L\"owdin orthogonalization proceeds by using the following identity
\begin{equation}
\tilde{\mathbf{T}}^I=
\mathbf{S}^{1/2}\bar{\mathbf{T}}^I\mathbf{S}^{1/2}=
(\mathbf{P}\mathbf{Q}^{\mathrm{T}})\mathbf{T}^I(\mathbf{P}\mathbf{Q}^{\mathrm{T}})^{\mathrm{T}}
\label{eq:lowdin-id}
\end{equation}
where $\mathbf{P}$ and $\mathbf{Q}$ are the matrices containing the left and right singular vectors of the MO-coefficient matrix $\mathbf{C}$.
It is, thus, possible to evaluate Equation~\eqref{eq:OmL} without knowledge of the AO overlap matrix $\mathbf{S}$, and a similar consideration \cite{Plasser2012JCTC} also applies for Equation~\eqref{eq:OmM}.
Whereas Equation~\eqref{eq:OmM} was initially implemented in TheoDORE, it is advisable to rather apply the L\"owdin style partitioning of Equation~\eqref{eq:OmL} since it is computationally more efficient and in most cases numerically more stable (i.e., the $\Omega_{AB}^I$ values are strictly positive, which is not necessarily the case for the Mulliken partitioning).
It is in principle possible to extend this formalism to other, more involved, population analysis schemes, but this has not been attempted yet.

Finally, the quantity
\begin{equation}
  \Omega^I=\sum_{AB} \Omega_{AB}^I=
  \sum_{rs}(T_{rs}^I)^2=
  \sum_p(\lambda_p^I)^2
  ,
\end{equation}
which is equivalent to the Frobenius norm of $\mathbf{T}$ (or $\tilde{\mathbf{T}}$), can be defined \cite{Plasser2014JCP1, Matsika2014JPCA}.
In the case of a normalized CIS or TDA wave function this value is always equal to 1, whereas it is generally less than 1 for correlated \emph{ab initio} methods \cite{Plasser2014JCP2, Mewes_sub}.
In the latter case, the $\Omega^I$ value can be applied as a method-independent measure for the single-excitation character where $\Omega^I=1$ corresponds to a pure singly excited state and smaller values indicate contributions from double or higher excitations.
The physical meaning of $\Omega^I$ is that it allows making some statements about transition properties of one-electron operators \cite{Matsika2014JPCA}.
Specifically, from $\Omega^I=0$ it follows that $\langle\Psi^0|\hat{O}^{(1)}|\Psi^I\rangle=0$ for any possible one-electron operator $\hat{O}^{(1)}$, for example the dipole moment operator or a spin-orbit operator in a mean field approximation.

For a system of $n_{\mathrm{frag}}$ fragments, the charge transfer number analysis produces an $n_{\mathrm{frag}}\times n_{\mathrm{frag}}$ matrix with all the possible contributions.
Here, the diagonal elements $\Omega^I_{AA}$ correspond to local excitations on fragment $A$, while $\Omega^I_{AB}, A\neq B$ are charge transfer contributions.
The matrix elements $\Omega^I_{AB}$ can be directly plotted in the form of a two-dimensional matrix plot (sometimes called electron-hole pair correlation plot), as has been discussed, for example, in Refs~\cite{Zojer2000JCP, Rissler2001PRB, Tretiak2002CR, Mewes2016PCCP}.

\begin{figure}[tb]
  \centering
  \includegraphics[scale=1]{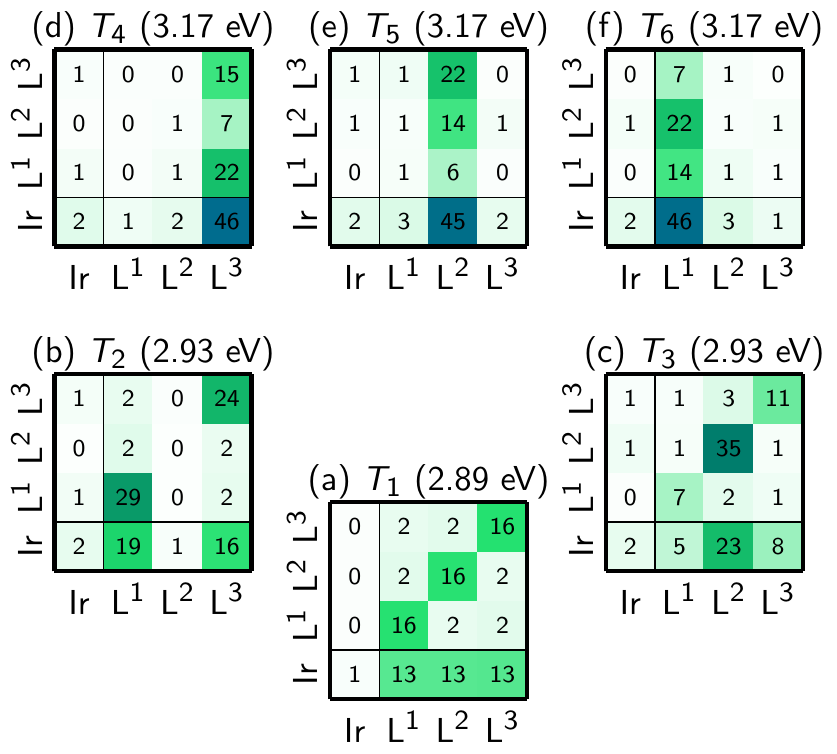}
  \caption{
  Matrix plots of the charge transfer numbers for the first six triplet states of [Ir(ppy)$_3$].
  In the plots, the magnitude of the elements of the $4\times 4$ matrices are given by the color, the numbers give the numerical value (in \%).
  The labels $L^1$, $L^2$, and $L^3$ denote the three ppy ligands.
  The \emph{excitation hole} is plotted on the vertical axis, the \emph{excited electron} on the horizontal axis.
  (TDA-B3LYP/TZP-DZ(P), COSMO(water))
  }
  \label{fig:theodore_Ir_matrices}
\end{figure}

Such matrix plots are shown in Figure~\ref{fig:theodore_Ir_matrices} for the $T_1-T_6$ states of [Ir(ppy)$_3$], denoting the three ppy ligands as $L^1$, $L^2$  and $L^3$. 
Starting with the $T_1$ state (Figure~\ref{fig:theodore_Ir_matrices}a), it is observed that the three strongest contributions (16\% each) are situated on the diagonal, corresponding to local excitations on each of the three ligands.
In addition, three contributions (13\% each) are seen corresponding to Ir$\rightarrow$L excitations.
The same types of transitions are also present for the $T_2$ and $T_3$ (Figure~\ref{fig:theodore_Ir_matrices}b-c) states with the exception that the symmetry is broken and the ligands do not contribute equally to these states.
The states $T_4$ to $T_6$ (Figures~\ref{fig:theodore_Ir_matrices}d-f) show distinctly different plots as opposed to the first three states.
In these cases, only one column of the matrix shows significant contributions, meaning that the \emph{excited electron} is localized on one of the ligands.
The \emph{excitation hole} always has its strongest contribution on Ir (46\%), the second contribution (22\%) is of LLCT type, (L$^i\rightarrow$L$^j$, $i\neq j$), and only the third contribution (14--15\%) is of IL type (L$^i\rightarrow$L$^i$).

\begin{figure}[tb]
  \centering
  \includegraphics[scale=1]{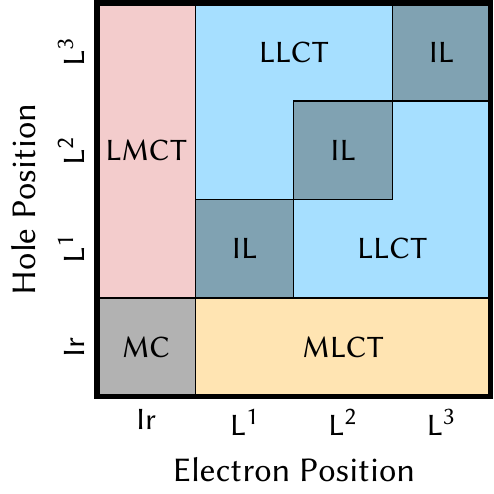}
  \caption{
  Division of the charge transfer matrix of [Ir(ppy)$_3$] into the five text-book classes of TMC excited states.
  }
  \label{fig:theodore_Ir_Om_division}
\end{figure}

While the plots shown in Figure~\ref{fig:theodore_Ir_matrices} provide a compact and rigorous representation of the excitations, it is often convenient to further compress the information.
For this purpose, for each state partial sums are computed over all the contributions that correspond to one of the five classes of TMC excited states: MC, MLCT, LMCT, IL, and LLCT.
Here, the Ir$\rightarrow$Ir matrix element gives the MC contribution to the state, the sum of the Ir$\rightarrow$L$^i$ elements gives the MLCT contribution, and so on.
This generic partitioning of the charge transfer number matrix is shown in Figure~\ref{fig:theodore_Ir_Om_division}.
It can be applied to homoleptic complexes, or more generally to TMCs where all ligands are treated as equivalent.
Instead of plotting the whole matrix of $\Omega^I_{AB}$, it is now possible to identify one or two dominant contributions and represent them in tabular form.
Naturally, the downside of this compression is that some information is lost---in this case the information about excitation localization on the individual ligands. 
Hence, one should carefully decide on the degree of data compression depending on the application.

For the six triplets of [Ir(ppy)$_3$], the results of this analysis are shown in Table~\ref{tab:orbitals_Irppy3_omega}.
It can be readily seen that the six states form two strongly related sets of states: $T_1$, $T_2$, and $T_3$ have almost identical state character contributions, and the three other states likewise.
Furthermore, it can be argued that the obtained numeric values provide more insight than that gained above from the NTOs.
Accordingly, it can be seen that the three lowest-lying triplet states are predominantly ligand-centered, with around 50\% IL character, and only around 35\% MLCT.
In contrast, the higher states are of predominant MLCT (50\%) character and possess LLCT character as a secondary contribution (31\%).

\begin{table}[tb]
  \centering
  \caption{Excitation energies ($E$), symmetry (Sym.), and most important state character contributions for the six lowest triplet states of [Ir(ppy)$_3$].
  (TDA-B3LYP/TZP-DZ(P), COSMO(water))
  }
  \label{tab:orbitals_Irppy3_omega}
  \begin{tabular}{lcclll}
    \hline
                &$E$ (eV)       &Sym.   &\multicolumn{2}{l}{State character}\\
    \hline
    $T_1$       &2.89           &$A$      &49\% IL
                                          &38\% MLCT\\
    $T_2$       &2.93           &$E$      &53\% IL
                                          &35\% MLCT\\
    $T_3$       &2.93           &$E$      &53\% IL
                                          &35\% MLCT\\
    $T_4$       &3.17           &$A$+$E$  &49\% MLCT
                                          &31\% LLCT\\
    $T_5$       &3.17           &$A$+$E$  &50\% MLCT
                                          &31\% LLCT\\
    $T_6$       &3.17           &$A$+$E$  &50\% MLCT
                                          &32\% LLCT\\
    \hline
  \end{tabular}
\end{table}

\begin{figure}[tb]
  \centering
  \includegraphics[scale=1]{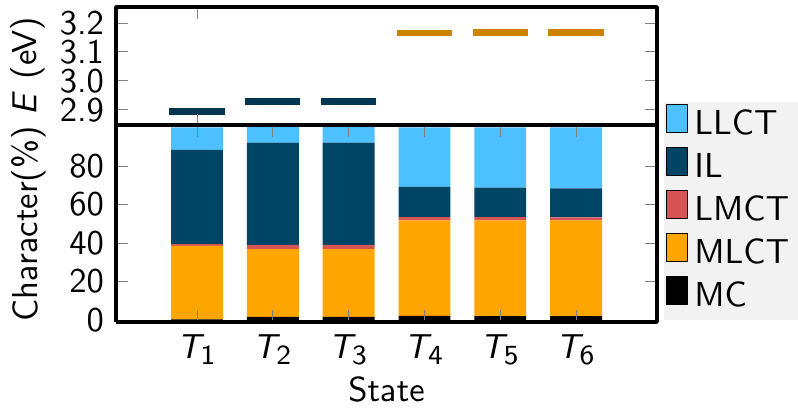}
  \caption{
  Decomposition of the charge transfer number matrix of [Ir(ppy)$_3$] into the five text-book classes of TMC excited states.
  (TDA-B3LYP/TZP-DZ(P), COSMO(water))
  }
  \label{fig:theodore_Ir_analysis}
\end{figure}

The information contained in the table can also be depicted in compact form, as shown in Figure~\ref{fig:theodore_Ir_analysis}, where the energy is depicted with horizontal bars, the main state character is indicated by the color of these horizontal bars, and the contributions to the state character are shown in stacked bar plots.
The advantage of this figure is that it can be easily scaled to a relatively large number of states by simply making the bars narrower. 
For instance, 30--50 states can be depicted in a figure of similar size (as will be shown below), while a table conveying the same information for 50 states would probably take more than half a page.
Moreover, spotting trends or patterns within the table would be much more cumbersome than glancing at the colors of the figure.

In summary, the presented protocol enables a completely automatized, quantitative, and reproducible assignment of state characters in TMCs.
In Sections~\ref{ssec:IrNuclear} to \ref{ssec:ReMethod} we will show that the methodology can be applied beneficially in cases where a larger number of states are to be analyzed, where detailed understanding of ligand effects is needed, or to rationalize the effects of geometric modifications.


\subsection{Charge transfer numbers for spin-mixed states}
\label{ssec:CTnumbersSOC}

In TMCs, SOC is very relevant due to the large nuclear charge of the metal atom.
SOC is often included in the electronic Hamiltonian as a perturbation~\cite{Vallet2000JCP, Neese2007CCR, Marian2012WCMS, Mai2014JCP_reindex, Wang2005JCP_sopert}
\begin{equation}
  \hat{H}^\mathrm{el}
  =
  \hat{H}^\mathrm{MCH}
  +
  \hat{H}^\mathrm{SOC},
\end{equation}
where $\hat{H}^\mathrm{MCH}$ is the \emph{molecular Coulomb Hamiltonian} (MCH) \cite{Mai2015IJQC}, which contains the (scalar relativistic) kinetic energy and Coulomb potential energies of the electrons.
In practice, one usually first computes a number of eigenstates of the MCH ($\Psi^\mathrm{MCH}$; also called spin-free states), e.g., singlets and triplets, and subsequently evaluates the SOC matrix elements within this set of states.
The spin-orbit-coupled states (or \emph{diagonal} states) can then be obtained in a perturbative fashion by a diagonalization of the Hamiltonian matrix, where the diagonal elements contain the energies of the MCH states, and the off-diagonal elements are the SOCs.
The approximation of this step is neglecting the SOC matrix elements with all higher states not computed in the first step.
The diagonalization of $\mathbf{H}^\text{MCH}$---with elements $H^\mathrm{MCH}_{IJ}=\langle \Psi_I^\mathrm{MCH}|\hat{H}^\mathrm{el}| \Psi_J^\mathrm{MCH}\rangle$---can be written as
\begin{equation}
  \mathbf{H}^\text{diag}=\mathbf{U}^\dagger\mathbf{H}^\text{MCH}\mathbf{U},
\end{equation}
where $\mathbf{U}$ is the eigenvector matrix of $\mathbf{H}^\text{MCH}$.
The matrix $\mathbf{U}$ allows reconstructing the wave function of the diagonal state in terms of the MCH states:
\begin{equation}
  |\Psi^\text{diag}_\alpha\rangle
  =
  \sum_I
  |\Psi^\text{MCH}_I\rangle U_{I\alpha}.
  \label{eq:diagMCH}
\end{equation}
Assuming that the ground state $\Psi_0$ is not affected by this transformation, which is often a good approximation, the 1TDM between the ground state and an excited diagonal state is given by
\begin{equation}
  T^{\alpha}_{rs}
  =
  \left\langle
    \Psi_0
    \middle|
    \hat{a}^\dagger_r\hat{a}_s
    \middle|
    \Psi^\text{diag}_\alpha
  \right\rangle,
  \label{1TDM-SOC_1}
\end{equation}
which after insertion of Equation~\eqref{eq:diagMCH} reads
\begin{equation}
  T^{\alpha}_{rs}
  =
  \sum_I
    \left\langle
    \Psi_0
    \middle|
    \hat{a}^\dagger_r\hat{a}_s
    \middle|
    \Psi^\text{MCH}_I
  \right\rangle U_{I\alpha}
  =
  \sum_I
   T^{I}_{rs}
U_{I\alpha}.
  \label{1TDM-SOC_2}
\end{equation}
Insertion into Equation~\eqref{eq:lowdin-id} shows that the same relations also holds for the orthogonalized matrices, i.e.,
\begin{equation}
  \tilde{T}^{\alpha}_{rs}
  =
  \sum_I
   \tilde{T}^{I}_{rs}
U_{I\alpha}
  \label{1TDM-SOC_3}
\end{equation}
and Equation~\eqref{eq:OmL} finally leads to
\begin{equation}
\label{eq:OmSOC}
  \Omega_{AB}^\alpha
  =
  \sum_I
  \sum_J
  U_{I\alpha}U_{J\alpha}
  \sum_{\mu\in A}\sum_{\nu\in B}
  \tilde{T}^I_{\mu\nu}\tilde{T}^J_{\mu\nu}
  =
  \sum_I
  \sum_J
  U_{I\alpha}U_{J\alpha}
  \Omega^{I-J}_{AB}
  .
\end{equation}
Here, the diagonal elements $\Omega^{I-I}_{AB}$ are equivalent to the charge transfer numbers $\Omega^{I}_{AB}$ of the MCH states, as defined above, and the off-diagonal $\Omega^{I-J}_{AB}$ terms are computed in an analogous fashion using the 1TDMs of two different states.
Importantly, the evaluation of Equation~\eqref{eq:OmSOC} does not require any significant additional computational effort when compared to the MCH case, since all required $\tilde{\mathbf{T}}^I$ matrices are already available.
This transformation has been newly implemented into TheoDORE in the course of this work and we will illustrate its potential below.  

\begin{figure}[tb]
  \centering
  \includegraphics[scale=1]{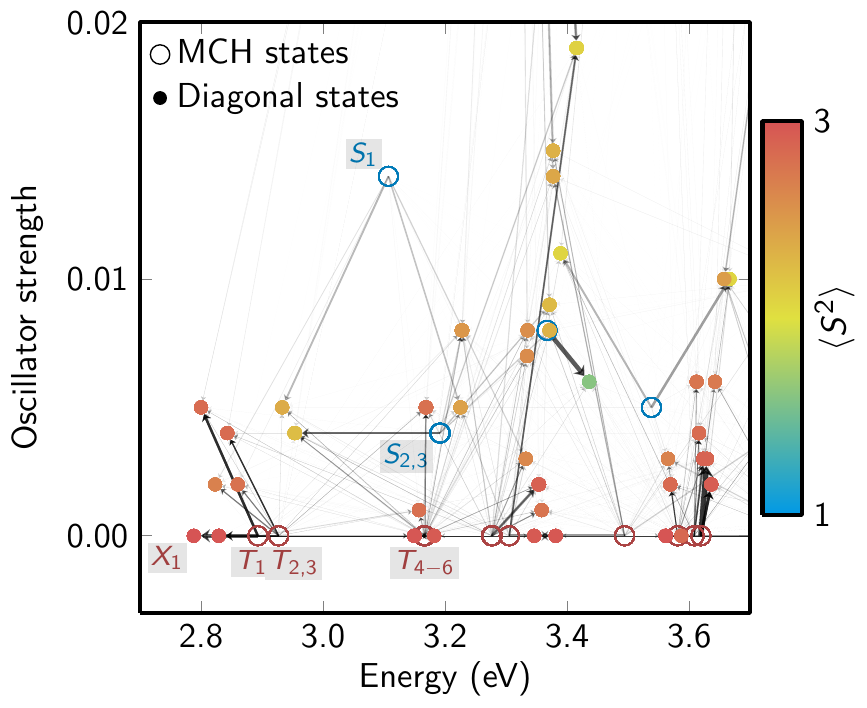}
  \caption{
  Spin-orbit coupling decomposition plot, showing energies and oscillator strengths of the few lowest excited states of [Ir(ppy)$_3$] in the MCH representation (open rings) and in the diagonal representation (filled circles).
  For each state, the color indicates the spin multiplicity---red or blue for the spin-pure MCH states, or any intermediate color for the spin-mixed diagonal states.
  The arrows indicate which MCH states contribute to which diagonal states, with the arrow thickness indicating the weight of the contributions.
  $X_1$ denotes the lowest diagonal state.
  (TDA-B3LYP/TZP-DZ(P), COSMO(water))
  }
  \label{fig:theodore_Ir_SOC_E_f}
\end{figure}

In order to visualize the resulting data, we first consider a plot of the energies versus oscillator strengths. 
Figure~\ref{fig:theodore_Ir_SOC_E_f} shows these two quantities for the lowest excited states of [Ir(ppy)$_3$].
The spin-pure (singlet and triplet) states in the MCH representation are denoted by open rings and the spin-mixed states in the diagonal representation by filled circles.
As the diagonal states do not have a well-defined total spin value, we indicate the total spin expectation value of these states by color.
Additionally, we draw arrows between the MCH and diagonal states, in such a way that it can be discerned which MCH states contribute to which diagonal states.
In that way, the plot can be used to analyze the electronic structure of the diagonal states in a visual way.
At first glance, the complexity of the diagonal states can be appreciated, i.e., in most cases the diagonal states are a linear combination of a large number of different MCH states.

\begin{figure}[tb]
  \centering
  \includegraphics[scale=1]{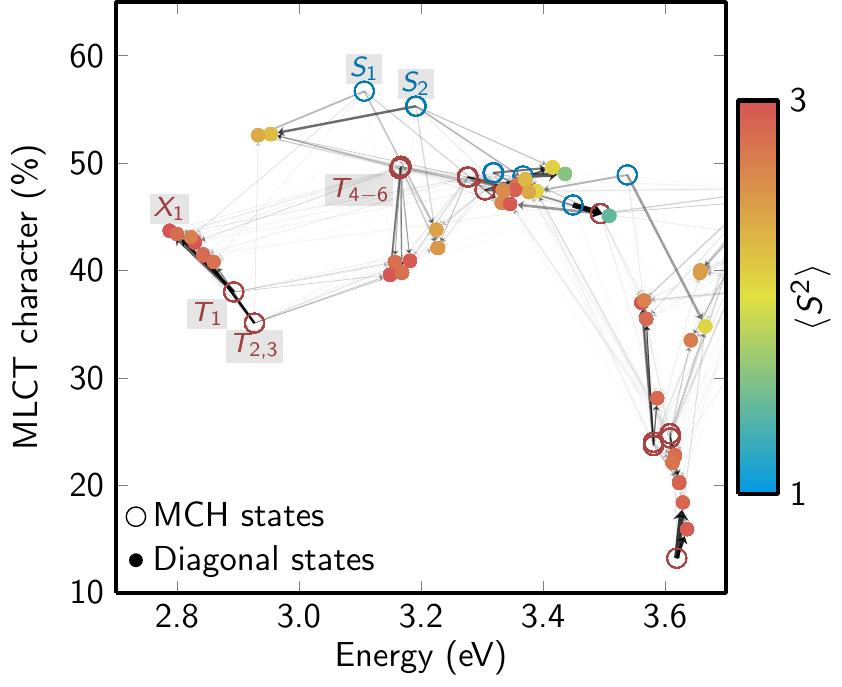}
  \caption{
  Spin-orbit coupling decomposition plot, showing energies and MLCT character of the few lowest excited states of [Ir(ppy)$_3$].
  The meaning of the colors and arrows is the same as in Figure~\ref{fig:theodore_Ir_SOC_E_f}.
  (TDA-B3LYP/TZP-DZ(P), COSMO(water))
  }
  \label{fig:theodore_Ir_SOC_MLCT_E}
\end{figure}

The reason to show energies versus oscillator strengths is that for [Ir(ppy)$_3$] and related complexes, the character of the lowest diagonal triplet states (the sublevels of the $T_1$) is decisive for its phosphorescence properties.
Ideally, the lowest triplet sublevel should have significant oscillator strength in order to facilitate fast and efficient phosphorescent decay.
It has been shown by theoretical arguments~\cite{Powell2015CCR,Powell2015SR} that for pseudo-octahedral complexes of trigonal symmetry, the lowest triplet sublevel is forbidden from decaying radiatively while the higher sublevels are allowed to phosphoresce.
This result is recovered in our calculations, as can be seen in Figure~\ref{fig:theodore_Ir_SOC_E_f}, where the lowest diagonal state (labeled ``$X_1$'') has an oscillator strength of zero.
The plot also shows that some of the other sublevels of $T_1$, $T_2$, and $T_3$ indeed acquire small oscillator strengths.
Figure~\ref{fig:theodore_Ir_SOC_E_f} also reveals that the intensity of these states does not originate from the lowest singlet states $S_1-S_3$, as one might have assumed initially, but it actually derives from the relatively bright $S_4$ and $S_5$ state pair (not shown due to scale).
A possible explanation for this finding is given by the general properties of the spin-orbit operator which govern the magnitude of the SOCs~\cite{Marian2012WCMS}.
Accordingly, in a TMC one can expect very large SOCs between a singlet $I$ and a triplet $J$ if the transition $I\rightarrow J$ is a one-electron excitation localized on the metal atom.
For example, SOCs will be large if $I$ and $J$ are two MLCT states involving different $d$ orbitals and the same $\pi^*$ orbital.
Now, $S_1-S_3$ and $T_1-T_3$ all constitute excitations out of the $H_a$ orbital (see Figure~\ref{fig:orbitals_Irppy3_canonical}), leading to rather small SOCs between these states, and therefore to only minor intensity borrowing.

In Figure~\ref{fig:theodore_Ir_SOC_MLCT_E}, we show the results of a charge transfer number analysis including SOC.
The figure is analogous to the previous one, but it plots the MLCT contribution against the energy of the states.
The most interesting aspect of Figure~\ref{fig:theodore_Ir_SOC_MLCT_E} is to see which states exchange MLCT character and how the energy is affected by that.
For the low-energy group of triplet states ($T_1$, $T_2$, and $T_3$), it can be seen that when they spin-orbit couple to the higher states, they acquire additional MLCT contributions, whereas the higher states (specifically, $T_{4-6}$) lose MLCT contributions.
Another observation is that states with less MLCT character (on the bottom) do not get shifted in energy as much as the states with larger MLCT contributions (on the top).
This is a well-known effect \cite{Yersin2011CCR, Powell2015CCR}, but here it is plotted in a comprehensive way.
Finally, it can be seen that, in general, the MCH states show more extreme positions on the MLCT scale, i.e., the MCH states can be found at very low and very high MLCT percentages, whereas the diagonal states are generally more ``average''.


\subsection{Correlations between charge transfer numbers}
\label{ssec:correlation}

In the case of multiple excited states, the charge transfer numbers constitute a three-dimensional $n_\mathrm{states}\times n_\mathrm{frag}\times n_\mathrm{frag}$ array.
With such a data set, it is possible that some of the charge transfer numbers are linearly correlated with each other, i.e., if one charge transfer number is large for any state, another number will also be large for that state, and vice versa.
Such correlations might be due to different reasons, for example due to the fact that excitations between MOs are delocalized over multiple fragments, or due to simultaneous, coupled excitations involving multiple fragments.


In order to illustrate the presence of such correlations, we start by scrutinizing the excited states of the complex [Re(Cl)(CO)$_3$(bipy)] (bipy=2,2'-bipyridine); this rhenium carbonyl diimine complex will be more thoroughly examined later in section~\ref{ssec:ReMethod}.
For now, it is sufficient to mention that the excited-state characters of this and related complexes are discussed in the literature due to the strong mixing between Re$\rightarrow$L and (CO)$_3$ $\rightarrow$L excitations \cite{Vlcek2009TOC,Bakova2011CCR}, and in the case of halogeno-complexes also X$\rightarrow$L, with X being the halogen and L the diimine ligand (e.g., bipy) \cite{Stor1992IC,Blanco-Rodriguez2005JPCA}.

\begin{figure}[tb]
  \centering
  \includegraphics[scale=1]{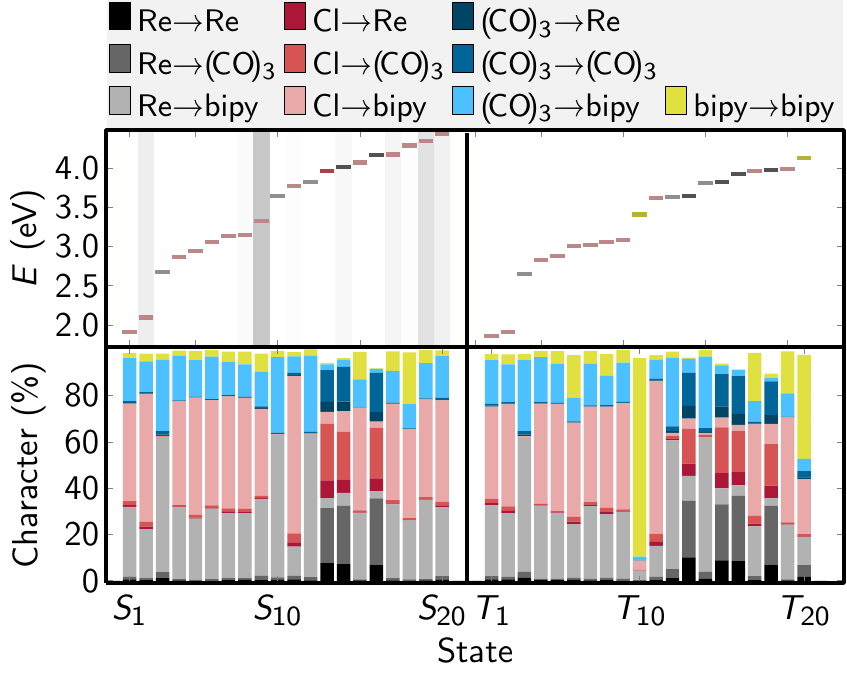}
  \caption{
  Excitation energies (horizontal bars), oscillator strength (shading behind horizontal bars), and charge transfer numbers (stacked bar plots) for the first 20 singlet and 20 triplet states of [Re(Cl)(CO)$_3$(bipy)].
  Note that some charge transfer contributions (e.g., Re$\rightarrow$Cl, Cl$\rightarrow$Cl, (CO)$_3$ $\rightarrow$Cl) are omitted for clarity; therefore, the stacked bars do not add up to exactly one.
  (TDA-B3LYP/TZP-DZ(P), gas phase)
  }
  \label{fig:correlation_analysis}
\end{figure}

This strong mixing is clearly appreciated in Figure~\ref{fig:correlation_analysis}, which displays charge transfer numbers for the first 20~singlets and first 20~triplets of [Re(Cl)(CO)$_3$(bipy)].
For the charge transfer analysis we have divided the molecule into four fragments: (i) Re, (ii) Cl, (iii) (CO)$_3$, and (iv) bipy.
The resulting types of excitations are represented in different colors.
As can be seen, 31 out of the 40 shown states are a mixture of three excitation types: Re$\rightarrow$bipy (in light gray), Cl$\rightarrow$bipy (in light red), and (CO)$_3$ $\rightarrow$bipy (in light blue).
Moreover, it can also be seen that the three types are correlated, in the sense that if one is small, the others tend to be small, too.
This is particularly notable for the Re$\rightarrow$bipy and (CO)$_3$ $\rightarrow$bipy contributions.

\begin{figure}[tb]
  \centering
  \includegraphics[scale=1]{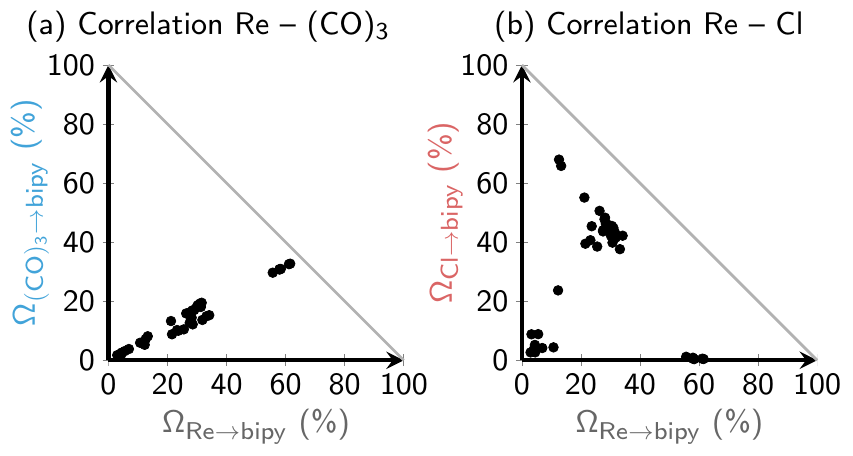}
  \caption{
  Correlations between the (a) Re$\rightarrow$bipy and (CO)$_3$ $\rightarrow$bipy contributions and (b) the Re$\rightarrow$bipy and Cl$\rightarrow$bipy contributions for the first 20 singlet and 20 triplet states of [Re(Cl)(CO)$_3$(bipy)].
  (TDA-B3LYP/TZP-DZ(P), gas phase)
  }
  \label{fig:correlation_scatter}
\end{figure}

These correlations are best identified in a scatter plot like shown in Figure~\ref{fig:correlation_scatter}, which plots pairs of charge transfer numbers (i.e., contributions to the state characters) for all states.
In (a) it can be seen that the Re$\rightarrow$bipy and (CO)$_3$ $\rightarrow$bipy contributions are very well linearly correlated, with a (Pearson) correlation coefficient of $r=0.98$.
As a contrasting example, in (b) one can see that  Re$\rightarrow$bipy and Cl$\rightarrow$bipy are not correlated at all; the correlation coefficient here is $r=-0.14$.
Other pairs of charge transfer numbers are also correlated in a similar manner (not shown), e.g., Re$\rightarrow$Cl with (CO)$_3$ $\rightarrow$Cl, or Re$\rightarrow$Re with (CO)$_3$ $\rightarrow$Re, which suggests that the fragments Re and (CO)$_3$ are themselves correlated.

The correlation between the different charge transfer numbers shows that the choice of the fragments is not necessary straightforward.
Because the fragments Re and (CO)$_3$ are almost perfectly correlated, no significant information is gained by computing their charge transfer numbers separately.
Instead, one could merge the two fragments into a Re(CO)$_3$ unit, keeping all essential information, while removing redundant information by reducing the total number of fragments for this molecule from four to three (i.e., Re(CO)$_3$, Cl, and bipy).
Besides achieving a reduction of data, the example above shows that splitting Re(CO)$_3$ into two fragments is not adequate from a chemical viewpoint.
It is well known that metal carbonyls have strongly covalent metal-carbon bonds and significant backbonding from metal $d$ orbitals to CO $\pi^*$ orbitals.
This bonding situation leads to orbitals which are delocalized over the metal and the carbonyls, such that excitations always involve both parts of the molecule.
What is exceptional in the previous correlation analysis is that it reveals these delocalized excitations from the physically well defined 1TDM alone, without inspecting any single orbital.


Based on the above reasoning, the correlation between different fragments can be used as a measure to decide how to choose the fragments in a charge transfer analysis.
In general, one would like to form fragments whose charge transfer numbers do not significantly correlate, in order to maximize the amount of information contained.
This is especially interesting for situations where the choice of fragments is not immediately obvious, or where the chemically motivated, intuitive fragmentation might not be optimal.

Here, we propose a new scheme which finds correlations between charge transfer numbers from any given fragmentation, and suggests how these fragments could be combined to form a smaller set of fragments for further consideration.
This new scheme consists of three steps: (i) compute the full correlation matrix between all fragments, (ii) perform a hierarchical clustering~\cite{Murtagh2012WDMKD,Murtagh2017WDMKD} based on the correlation matrix, and (iii) cut the cluster hierarchy at a desired level to learn which fragments could be merged.

For this scheme, we first need a way to quantify the correlation \emph{between fragments}.
This has to be distinguished from the correlations above, which were \emph{between pairs of charge transfer numbers}.
The difference is in the fact that each fragment is involved in multiple charge transfer numbers.
To find correlations between fragments, we propose the following.
To compute the correlation between fragments $A$ and $B$, we need to consider all excitations $A\rightarrow X$ and $B\rightarrow X$, for all fragments $X$; in this way, we can find whether the \emph{holes} on $A$ and $B$ are correlated.
This leads to the following equation for the hole covariance matrix $\mathbf{V}^\mathrm{h}$:
\begin{equation}
\small
  V^\mathrm{h}_{AB}
  =
  \frac{1}{N_\mathrm{states}}
  \sum_X
  \sum_I
  \Omega^I_{AX}
  \Omega^I_{BX}
  \\
  -
  \frac{1}{N_\mathrm{states}^2}
  \sum_X
  \left(
    \sum_I
    \Omega^I_{AX}
  \right)
  \left(
    \sum_I
    \Omega^I_{BX}
  \right),
  \label{eq:cov}
\end{equation}
which is more precisely the covariance matrix of the hole, summed over all possible electron positions $X$.
Here, index $I$ runs over the excited states included in the statistical analysis, and $A$, $B$, and $X$ run over the atomic fragments.
The matrix element $V^\mathrm{h}_{AB}$ describes whether fragments $A$ and $B$ tend to simultaneously release electron density to the same acceptor fragment during excitation.

Likewise, for the \emph{electron} correlation between $A$ and $B$, we need to look at all excitations $X\rightarrow A$ and $X\rightarrow B$, for all fragments $X$.
This gives rise to a similar equation for $\mathbf{V}^\mathrm{e}$, just with the following index replacements: $AX\Rightarrow XA$ and $BX\Rightarrow XB$.
The matrix element $V^\mathrm{e}_{AB}$ then describes whether $A$ and $B$ tend to simultaneously receive part of the excited electron from the same donor fragment.

By normalization, one can obtain the hole \emph{correlation} matrix:
\begin{equation}
  C^\mathrm{h}_{AB}
  =
  \frac{V^\mathrm{h}_{AB}}{\sqrt{V^\mathrm{h}_{AA}V^\mathrm{h}_{BB}}},
  \label{eq:cor}
\end{equation}
which contains the Pearson correlation coefficients (called $r$ above) which measure the linear correlations between all pairs of fragments $A$ and $B$.
The electron correlation matrix $\mathbf{C}^\mathrm{e}$ can be computed analogously.

Based on these two matrices, in principle one could find if any of the fragments behave identically in all excited states ($C^\mathrm{h}_{AB}=1$ or $C^\mathrm{e}_{AB}=1$), and one could redefine these two fragments as a single one.
However, in general, it is not trivial to find such pairs, and a manual inspection of the correlation matrices might be subjective and inefficient.
Instead, we propose to employ hierarchical clustering analysis to find the groups of correlated fragments.
In order to do this, we need to define a metric, which allows converting the correlation matrices into distance matrices:
\begin{equation}
  R^\mathrm{h}_{AB}
  =
  \sqrt{
    \frac{1}{2}
    \left(
      1-C^\mathrm{h}_{AB}
    \right)
  },
  \label{eq:dis}
\end{equation}
and analogously for the electron distance matrix $\mathbf{R}^\mathrm{e}$.
If the correlation coefficient $C_{AB}$ is 1, then the distance $R_{AB}$ will be zero, and for smaller correlation coefficients, the distance will increase.
The usage of the square root in the definition ensures that the computed distances follow the triangular inequality~\cite{Dongen2012A}, which is helpful for the clustering step.
Basically, it ensures that if (i) $A$ and $B$ have a small distance, and (ii) $B$ and $C$ have a small distance, then (iii) $A$ and $C$ also have a sufficiently small distance.

With the distance matrix at hand, we can now carry out the hierarchical clustering algorithm.
We employ the common agglomerative algorithm, where initially, all fragments are separate clusters and are merged sequentially, until only one cluster remains.
A description of the algorithm and further details are given in the Computational Details section in appendix A.

\begin{figure}[tb]
  \centering
  \includegraphics[scale=1]{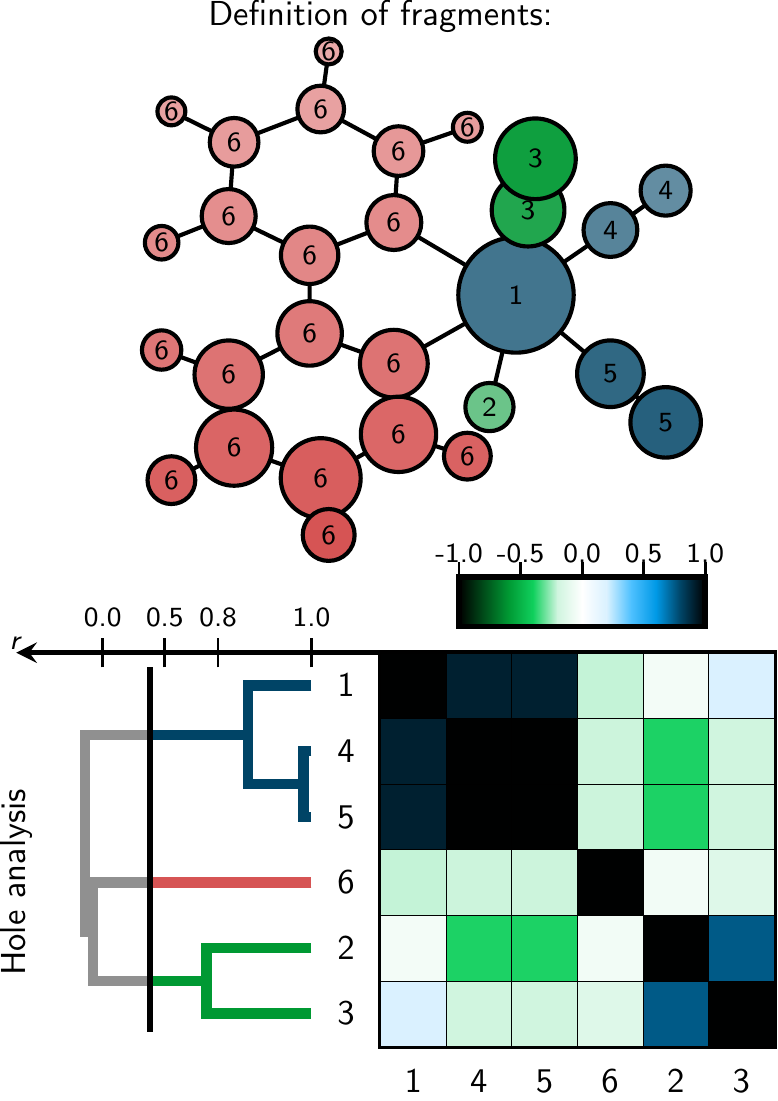}
  \caption{
  Example of the hierarchical clustering of [Re(Cl)(CO)$_3$(bipy)] using the correlations between the excitation hole of all fragments.
  The definition of the six fragments is shown on the top, where fragment 1 is Re, 2 is Cl, 3/4/5 are the three CO molecules,  and 6 is bipy.
  The bottom part shows the correlation matrix $\mathbf{C}^\mathrm{h}$ as a color-coded matrix plot.
  Next to the matrix is the dendrogram showing the clustering of the matrix.
  The black vertical line divides this dendrogram, thus forming the three clusters (blue, green, and gray), which are also indicated in the molecular geometry plot.
  (TDA-B3LYP/TZP-DZ(P), gas phase)
  }
  \label{fig:correlation_matrix}
\end{figure}

      \begin{figure*}[tb]
        \centering
        \includegraphics[width=\textwidth]{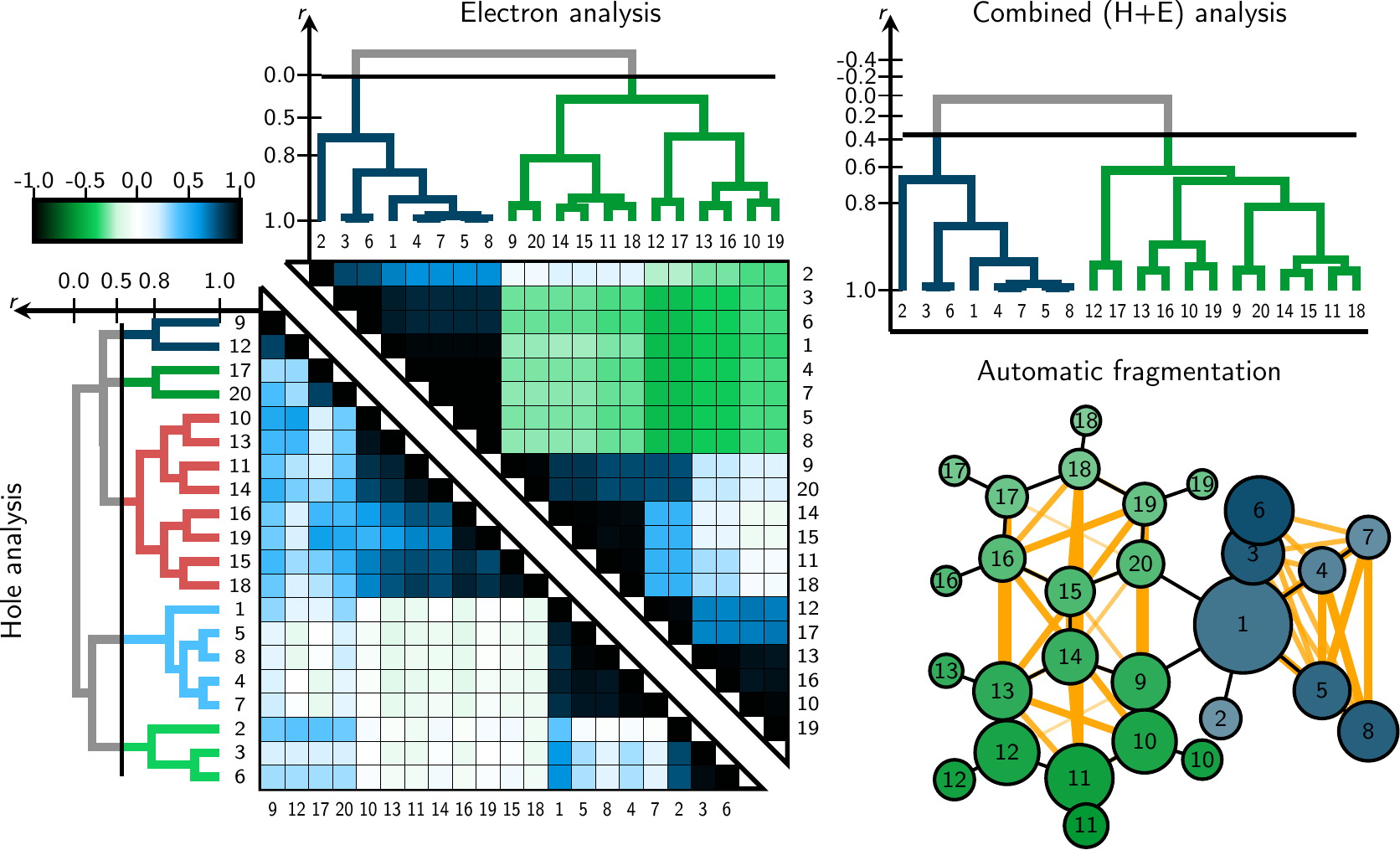}
        \caption{
        Example of the full-atomic hierarchical clustering of [Re(Cl)(CO)$_3$(bipy)] using the combined correlations of excitation hole and excited electron.
        The definition of the 20~fragments is shown on the right, where fragment 1 is Re, 2 is Cl, 3+6/4+7/5+8 are the three CO molecules, and fragments 9 to 20 make up bipy.
        The left part shows the correlation matrices $\mathbf{C}^\mathrm{h}$ (lower left triangular matrix) and $\mathbf{C}^\mathrm{e}$ (upper right) as color-coded matrix plots.
        Next to the matrices are the dendrograms showing the clustering of the matrices, yielding different results for the two analyzes.
        On the top right, the dendrogram of the combined analysis is presented, where the black vertical line forms two clusters (blue and green), which are also indicated in the molecular geometry plot.
        The orange lines in the molecular geometry plot indicate which atoms correlate well with each other.
        (TDA-B3LYP/TZP-DZ(P), gas phase)
        }
        \label{fig:correlation_matrix_full}
      \end{figure*}

Figure~\ref{fig:correlation_matrix} shows an example of such a hierarchical clustering analysis, based on the data presented for [Re(Cl)(CO)$_3$ (bipy)] in Figure~\ref{fig:correlation_analysis}.
The figure contains three parts: (i) a depiction of the molecular geometry, with the fragmentation indicated by the numbering of the atoms, (ii) a matrix plot of the correlation matrix of the excitation hole $\mathbf{C}^\mathrm{h}$, and (iii), the \emph{dendrogram} which presents the result of the hierarchical clustering step.
In the dendrogram, it is shown how the six fragments are subsequently merged to yield intermediate clusters which eventually are all merged to a single system.
Importantly, the horizontal (to the left) axis depicts the $r$ value between the merged clusters.
For example, fragments 4 and 5 (the CO$_\mathrm{eq}$ fragments) have a correlation coefficient of about 0.99 (most likely due to symmetry), whereas the value for fragments 2 and 3 is about 0.75.
Moreover, the correlation coefficient between fragment 1 (Re) and cluster 4+5 (CO$_\mathrm{eq}$)$_2$ is approximately 0.85.
The larger clusters 1+4+5 (Re(CO)$_2$), 2+3 (COCl), and 6 (bipy) are not correlated at all, with values which are slightly negative.
This indicates a negative correlation between the large clusters, which is due to the fact that the sum of all charge transfer numbers is equal to 1 (at least in the present TDA calculations), so that as one number grows larger, all other values tend to decrease.

Based on the dendrogram, the final step of the analysis is to extract a sensible clustering of the fragments.
In order to do so, one defines a threshold value at which the dendrogram is cut.
In the given example, this threshold is indicated by the black vertical line within the dendrogram, at a value of about $r=0.45$.
For this particular dendrogram, of course choosing a value anywhere between 0.0 and 0.75 would have accomplished the same partitioning, so the actual problem is to choose above which merger the cut should be performed.
Here, we employ a heuristic which looks for the largest gap between two subsequent mergers, and defines the threshold accordingly.
In this way, one can separate highly correlated clusters from uncorrelated clusters.

The clustering analysis performed in Figure~\ref{fig:correlation_matrix} is quite interesting, as it proposes yet another fragmentation scheme, different from the one chosen in Figure~\ref{fig:correlation_scatter}.
In the automatic clustering approach, it appears that the Cl atom is mostly correlated with the trans-standing carbonyl, whereas the equatorial carbonyls are correlated with the metal.
The reason for this different finding is that Figure~\ref{fig:correlation_scatter} initially considers all CO molecules as a single fragment, whereas Figure~\ref{fig:correlation_matrix} considers them separately.

Both Figures~\ref{fig:correlation_scatter} and Figure~\ref{fig:correlation_matrix} only consider correlation matrix for the excitation hole, but a similar analysis should be done for the excited electron, so that in the end one obtains two independent clusterings.
Since this might be inconvenient, we suggest to combine the hole and electron distance matrices in
\begin{equation}
  R^\mathrm{combined}_{MN}
  =
  \mathrm{min}\left(R^\mathrm{e}_{MN},R^\mathrm{h}_{MN}\right),
\end{equation}
and perform the clustering with this combined distance measure.
In this way, one obtains only a single clustering, which considers both hole and electron correlation.
Here, the use of the minimum function ensures that two fragments are already considered as correlated if only hole \emph{or} electron are correlated.

In order to obtain an unbiased clustering of the molecule, it is best to provide as little as possible prior knowledge about the fragmentation to the clustering procedure.
This can be achieved by calculating the charge transfer numbers for an atomic fragmentation scheme, and let the clustering procedure figure out all correlations between the atoms; this could be called an ``\emph{ab initio}'' fragmentation scheme.
An example is presented in Figure~\ref{fig:correlation_matrix_full}.
There, the [Re(Cl)(CO)$_3$(bipy)] molecule is initially divided into 20~fragments---each non-hydrogen atom is treated as a separate fragment.
The hydrogens are included in the fragments of the atom to which they are bonded, because hydrogens are not significantly participating in the excitations, hence do not correlate with any other atom, and consequently would be clustered incorrectly.

Admiringly, the automatic ``\emph{ab initio}'' fragmentation procedure fully recovered the results of the above manual fragmentation.
It detected the bipy molecule as one independent subunit of the complex.
The Re(Cl)(CO)$_3$ moiety was identified as a second subunit, where Re and (CO)$_3$ show tighter correlation than Cl with either of the two.
The fact that Re(Cl)(CO)$_3$ is a strongly correlating subunit is consistent with previous experimental findings in the Ref.~\cite{Blanco-Rodriguez2005JPCA}.
There, the authors used the vibrationally active pseudohalide NCS, showing that in [Re(NCS)(CO)$_3$(NN)] (NN=diimine) there is mixing between Re$\rightarrow$NN and NCS$\rightarrow$NN excitations.

The combined analysis also shows that the correlation level of Cl with Re(CO)$_3$ is similar to some of the correlations inside the bipy ligand.
This could indicate that by subdividing the bipy into smaller fragments, one could get additional information from the charge transfer number analysis which might allow distinguishing excited states by the location of the excited electron on the bipy.
This can be understood in the sense that there are multiple $\pi^*$ orbitals on bipy which can participate in the excitation process, and these orbitals show slightly different localization on the different atoms.

We should note that the statistical analysis of the fragment correlations depends on the input data, and in particular on the number and energy range of the states considered in the analysis.
This is especially relevant for systems where one state character appears only at low energies and another character only at high energies.
In such a case, including only the low-energy states in the correlation analysis is likely to lead to different results than would be obtained with the full set of states.
However, when the correlation analysis is done in order to find a fragmentation scheme to be used subsequently, it is best to perform the analysis with the same set of states as will be analyzed later (e.g., in molecular dynamics).
Furthermore, it might be advantageous to include computations at multiple geometries in the correlation analysis, to remove any possible bias coming from the choice of the geometry.


\section{Case studies}
\label{sec:cases}

The quantification of excited state-localization and charge transfer by means of TheoDORE analysis was already exemplified in Sections~\ref{ssec:CTnumbers} and \ref{ssec:CTnumbersSOC} for [Ir(ppy)$_3$].
In the following, the usefulness of this analysis is showcased for five different aspects of excited-state quantum chemistry which affects the state characters:
(i) influence of nuclear relaxation,
(ii) influence of SOCs,
(iii) influence of the ligand sphere,
(iv) influence of the metal center,
and (v) influence of the electronic structure method.

\subsection{Influence of nuclear relaxation: nonradiative decay of [Ir(ppy)$_3$]}
\label{ssec:IrNuclear}

Among phosphorescent emitters, organometallic Ir(III) complexes have been demonstrated to be exceptionally useful~\cite{Yersin2011CCR, Powell2015CCR} due to their relatively short radiative triplet lifetime (about 1.6~$\mu$s in the high-temperature limit \cite{Yersin2011CCR}) and high phosphorescence quantum yields (about 90\% \cite{Yersin2011CCR}).
The emission wavelength of Ir(III) complexes significantly depends on the ligands and their substituents, which can be used to control their properties~\cite{BinMohdYusoff2017TCC, Ho2014JOC, You2012CSR, Chi2010CSR, DeAngelis2007IC, Minaev2009CP}.
Luminescent Ir(III) complexes can be used as well in light-emitting electrochemical cells~\cite{Henwood2016TCC} or as biological probes, imaging reagents, and photocytotoxic agents~\cite{Lo2015ACR}.

In the past decades, a number of theoretical studies, mainly based on TD-DFT with or without inclusion of SOCs, have been devoted to this class of complexes \cite{Hay2002JPCA, Nozaki2006JCCS, DeAngelis2007IC, Zhao2009O, Minaev2009CP, Liu2010JCC, Tian2010EJIC, Li2011EJIC, Smith2011CPC, Su2012IJQC, Brahim2014CTC, Zanoni2014IC, Kleinschmidt2015JCP, Plasser2015JPCA, Smith2016OE, Heil2016MP}.
This huge research activity has provided important insight into the character and emissive properties of the low-lying triplet states, which are responsible for the usefulness of the complexes.

A critical property of [Ir(ppy)$_3$] is its very long nonradiative decay time (about 15--30~$\mu$s in the high-temperature limit \cite{Yersin2011CCR}), which is a prerequisite for a useful phosphorescence yield \cite{Powell2015CCR, Powell2015SR, Yersin2011CCR}.
In this and related TMCs, one of the most important nonradiative decay pathway leads from the $T_1$ minimum of MLCT or IL/LLCT character to states of MC character. 
Computational studies \cite{Yersin2011CCR,Treboux2007CL,Sajoto2009JACS,Yang2008IC} showed that these MC states involve strong elongation or breaking of metal-ligand bonds, accompanied by the formation of a trigonal-bipyramidal metal coordination, and hence lead to easily accessible $T_1/S_0$ crossings.
These nonradiative decay routes are one of the main limitations of blue emitters, because there the emissive states are shifted to higher energies such that they come closer to the MC states.
Hence, in this section, we investigate the contamination of the lowest triplet state by MC states along a possible nonradiative decay route.

\begin{figure}[t]
  \centering
  \includegraphics[scale=1]{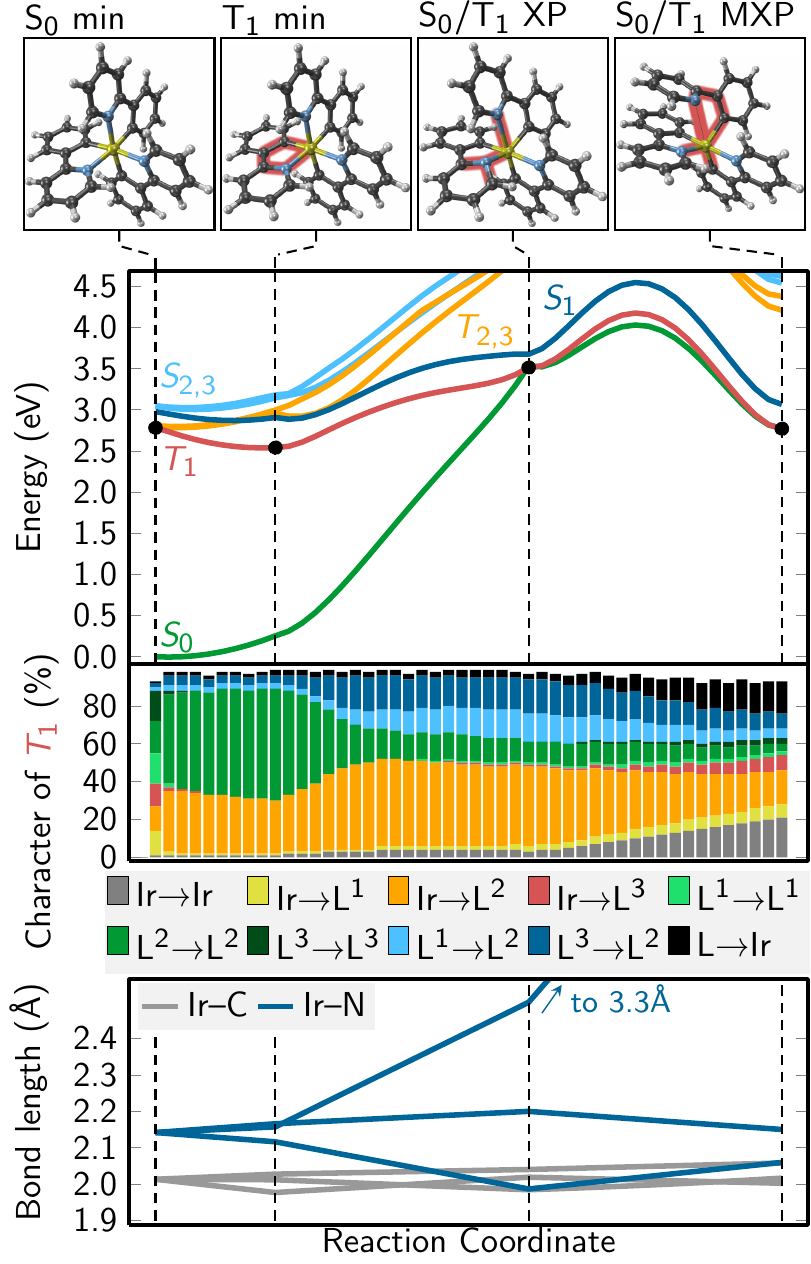}
  \caption{
  Analysis of the nonradiative decay pathway of [Ir(ppy)$_3$].
  On the top, the four optimized geometries of [Ir(ppy)$_3$] ($S_0$ minimum, $T_1$ minimum, $S_0/T_1$ crossing point, and $S_0/T_1$ minimum energy crossing point; see the supporting information for coordinates) are shown.
  Red shade marks the regions where the most significant geometric changes (e.g., bond elongation) occur.
  The upper plot shows the energy profiles of the few lowest states of [Ir(ppy)$_3$] along a linear interpolation in internal coordinates, with the black dots indicating where the minima or crossing points are located.
  The middle bar plot shows the composition of the $T_1$ state along the scan according to a TheoDORE analysis.
  The bottom plot shows the Re--C and Re--N bond lengths along the interpolation scan, where the bond length of the dissociating bond is truncated for the sake of visibility.
  (TDA-B3LYP/TZP-DZ(P), gas phase)
  }
  \label{fig:theodore_Ir_path}
\end{figure}

Figure~\ref{fig:theodore_Ir_path} shows a potential energy profile for the nonradiative decay pathway of [Ir(ppy)$_3$], from the $S_0$ minimum to the $T_1$ minimum and further on to two $T_1/S_0$ crossing points.
These four critical points are indicated in Figure~\ref{fig:theodore_Ir_path} by four black dots.
The $S_0$ minimum has $C_3$ symmetry (left-most black dot), whereas the $T_1$ minimum is slightly non-symmetric due to a localization of the excitation on one of the ligands, denoted here as L$^2$.
While locating the $T_1/S_0$ crossing points, the optimization yielded two qualitatively different crossing geometries.
The first geometry (denoted as $S_0/T_1$ XP; XP=crossing point) is not a true minimum on the crossing seam, but is significant because it is a (relatively low-energy) crossing point where the octahedral coordination of Ir is retained.
Nevertheless, this XP shows a strong elongation of the Ir-N bond to L$^1$, and additionally an out-of-plane puckering of the N atom of L$^2$.
The second crossing geometry corresponds to a minimum on the crossing seam (hence denoted as $S_0/T_1$ MXP; MXP=minimum energy crossing point), and shows a trigonal bipyramidal geometry, where one of the pyridine groups is detached from the Ir, rotated, and stacked on top of one of the other ligands.

The SOCs between $T_1$ and $S_0$ are approximately 400~cm$^{-1}$ at the XP and 1100~cm$^{-1}$ at the MXP, showing that both points might enable ISC if accessed.
The energy profile shows that the first XP is approximately 1~eV above the $T_1$ minimum.
The MXP is only 0.25~eV above the $T_1$ minimum; however, there might be a large barrier (0.7~eV according to the linear interpolation scan, which gives a upper bound) between the XP and MXP.
Hence, both crossing points are too high in energy to be relevant even at room temperature.
The barriers are also significantly larger than the ones previously reported in the literature: Treboux et al.\cite{Treboux2007CL} reported a barrier of 0.28~eV to reach an MC state from the $T_1$ minimum based on a relaxed scan, and Sajoto et al.\cite{Sajoto2009JACS} reported a value of 5000~cm$^{-1}$ (0.6~eV), estimated from computed and experimental data of related complexes.
However, it should be noted that here we actually optimized a $T_1/S_0$ crossing point (for the first time for this complex, to the best of our knowledge), and that the values are therefore not necessarily comparable.
Another possible reason for this discrepancy might be that our linear-interpolation-in-internal coordinate scan leads to a too large barrier, and there is actually a lower-energy pathway from the $T_1$ minimum to the MXP.

In their report on the nonradiative decay pathway of [Ir (ppy)$_3$], Treboux et al.\cite{Treboux2007CL} remarked ``The precise assignation of the respective characters of MLCT and LC is complicated by the presence of a strong metal-ligand mixing in the orbitals ...''
Consequently, the charge transfer analysis of TheoDORE is ideally suited to disentangle this complicated electronic structure situation.
In the bottom panel of Figure~\ref{fig:theodore_Ir_path}, we decompose the transition density (from $S_0$ to $T_1$) into the ten most important contributions, using four fragments (Ir and each ppy separately) which were found to be weakly correlated by a correlation analysis like in Section~\ref{ssec:correlation}.
As already seen in Section~\ref{ssec:CTnumbers}, at the $S_0$ minimum geometry, the $T_1$ state is a mixture of 40\% MLCT character and 50\% IL character, both equally distributed among the three ligands (L$^1$, L$^2$, L$^3$).
It is very interesting to note that, as soon as the trigonal symmetry is broken (moving from the left-most data point to the right), the excitation very quickly localizes on L$^2$, as those ligand's bond lengths are changed.
Hence, the $T_1$ character becomes a mixture of 30\% Ir$\rightarrow$L$^2$ and 60\% L$^2\rightarrow$L$^2$.

When approaching the $T_1/S_0$ XP, the wave function composition changes again, with a shift towards a larger Ir$\rightarrow$L$^2$ contribution (40\%).
The previously large L$^2\rightarrow$L$^2$ contribution of the $T_1$ minimum is changed to an equal mixture of L$^1\rightarrow$L$^2$, L$^2\rightarrow$L$^2$, and L$^3\rightarrow$L$^2$.
Interestingly, at this $T_1/S_0$ geometry there is almost no contribution (3\%) of MC states to the $T_1$.
Instead, the excitation can be regarded as an excitation from a partially metal-localized orbital to an orbital on the strongly puckered pyridine unit of L$^2$.
This kind of puckering of aromatic systems is more commonly known from smaller organic molecules, for example nucleobases~\cite{Merchan2013TCC}.
An interesting ansatz for controlling nonradiative decay in TMCs might thus be to modify the ligands such that puckering is suppressed.

The situation is notably different at the MXP geometry.
Here, the $T_1$ wave function acquires a 20\% contribution of MC excitations, as well as 17\% of LMCT excitations.
Furthermore, the MLCT and IL/LLCT excitations become less localized on the L$^2$ ligand, as can be seen by the reappearance of contributions like Ir$\rightarrow$L$^1$ and Ir$\rightarrow$L$^3$.
Qualitatively, this excitation could be regarded as an MC excitation with significant mixing of the $d$ orbitals with ligand orbitals.


\subsection{Influence of spin-orbit coupling: state character mixing in [Ir(Cl)(CO)(ppy)$_2$]}
\label{ssec:IrSOC}

Relativistic effects play an important role in the photophysics of Ir(III) complexes~\cite{Yersin2011CCR, Brahim2014CTC, Smith2011PRB, Powell2015CCR, Powell2015SR, Li2011EJIC}, because only through these effects can the lowest-energy triplet states acquire any radiative decay rate.
It was shown that the mixing of close-lying low-energy MLCT and IL states is critical for the radiative decay properties~\cite{Powell2015CCR, Powell2015SR, Yersin2011CCR} because the MLCT states lead to enhanced SOCs and the IL states provide a large transition dipole moment.
Hence, the degree of state character mixing is very relevant for this kind of TMCs, and besides the mixed state character observed without SOC, additional mixing is expected when spin-orbit effects are taken into account.
The analysis of the spin-orbit excited-state character becomes particularly challenging due to this additional mixing, as illustrated by the TD-DFT results reported for [Ir(ppy)$_3$] and [Ir(Cl)(CO)(ppy)$_2$] in Ref.~\cite{Brahim2014CTC}.
Here, we report a detailed analysis of the SOC effects on the excited state character of [Ir(Cl)(CO)(ppy)$_2$].
The case of this molecule is particularly instructive, because this complex is characterized by a high density of low-lying singlet and triplet states of mixed XLCT/MLCT character~\cite{Brahim2014CTC} and it is interesting to follow the evolution of the XLCT/MLCT (XLCT = halogen-to-ligand charge transfer) ratio, as well as the MLCT/IL mixing, when applying SOC effects.

\begin{figure}[tbp]
  \centering
  \includegraphics[scale=1]{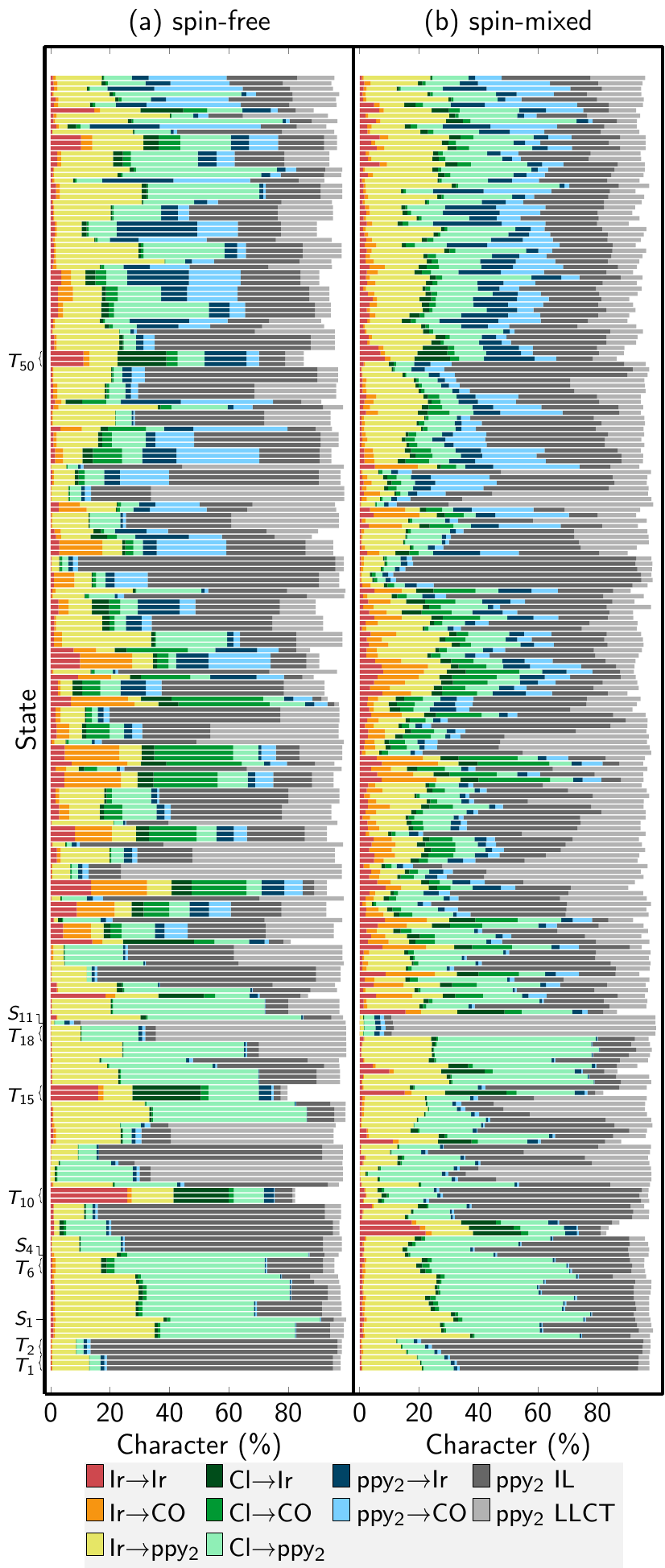}
  \caption{
  Characters of the states of [Ir(Cl)(CO)(ppy)$_2$], without (a) and with (b) SOC included.
  In (a), all triplet components are present (note the triply identical entries), so that both figures have the same number of states.
  States are ordered according to energy, but for reasons of clarity the bars are equally spaced; hence, the energies cannot be read from the plot.
  For a discussion of the labeled states see the text.
  (TDA-B3LYP/TZP-DZ(P), COSMO(water))
  }
  \label{fig:theodore_Ir_SOC_COCl}
\end{figure}

In Figure~\ref{fig:theodore_Ir_SOC_COCl}, we present the state characters for the first 60 singlet and 60 triplet states of [Ir(Cl)(CO)(ppy)$_2$] on the left, and the characters of the resulting 240 spin-mixed states on the right.
Due to the properties of the spin-free--spin-mixed transformation (Equation~\eqref{eq:diagMCH}), the total contribution of each state character to all states (i.e., the total area filled by each color) is the same in both panels.
However, the inclusion of SOC leads to a redistribution of the contributions among the electronic states.

The characters of the lowest states---from $S_1$ to $S_4$ and from $T_1$ to $T_6$, which are all below 3.5~eV (above 354~nm)---are not drastically affected when SOC is included.
The well-separated $T_1$ and $T_2$ are relatively pure IL states, although the minor MLCT and XLCT contributions are slightly enhanced when activating SOC effect.
The subsequent states ($S_1$ to $S_4$ and $T_3$ to $T_6$) are predominantly mixed MLCT/XLCT states, with some significant IL contributions, where MLCT character tends to be more predominant for lower states.
When SOC is taken into account, the state characters become slightly mixed, but the general picture is unaffected.

At higher energies---till states $T_{18}$ and $S_{11}$ at about 4.0~eV (311~nm)---SOC perturbation of the spectrum is more important.
For instance, there are two triplets ($T_{10}$ and $T_{15}$) with dominant MC/XMCT (halogen-to-metal CT) character.
Upon SOC mixing, the lower one of these two triplets changes order with several IL states, and the upper one shows very strong zero field splitting such that the three components become interspersed by other states.
Furthermore, for $T_{18}$, significant de-mixing can be seen, as this state has notable MLCT and XLCT contributions (30\% in total), but after including SOC becomes a very pure LLCT ($>$88\%) state.
Among the remaining states, a clear redistribution of the XLCT and IL contributions and a decrease of the LLCT character can be discerned.

Above $S_{11}$, the analysis of the characters shows the occurrence of ppy$_2\rightarrow$CO and ppy$_2\rightarrow$Ir contributions, as well as excitations towards carbonyl orbitals (Ir$\rightarrow$CO, Cl$\rightarrow$CO).
Additionally, a large number of the states shows major LLCT and IL contributions, whereas XLCT contributions are relatively rare in this range of states and only reappear around $T_{50}$.
The effect of SOC at the top range of the calculated states (i.e., above $T_{50}$) is mostly to make all states more similar---as can be seen, these states are all a mixture of MLCT, XLCT, LLCT, and IL.

It goes without saying that the presented results and discussion are valid only at the Franck-Condon geometry used for the presented excited-state calculation, and for the B3LYP functional employed, within the limitations of TD-DFT.
Multi-reference electronic structure methods---which provide more flexibility, treat charge transfer character differently, and go beyond single excitations---could exhibit a different picture.
Some aspects of the influence of the electronic structure method on the excited-state character will be discussed in Section~\ref{ssec:ReMethod}.

Also worth noting is that a correlation analysis (analogous to the one in Figure~\ref{fig:correlation_matrix}) for [Ir(Cl)(CO)(ppy)$_2$] shows that the metal atom and the carbonyl group are relatively independent units ($r\approx 0.5$).
Actually, the complex shows almost no charge transfer contributions originating at the carbonyl group.
Furthermore, contributions like Cl$\rightarrow$CO or ppy$_2\rightarrow$CO only appear in certain regions of the spectrum.
These findings are in contrast to the strong metal-carbonyl correlation of the Re(CO)$_3$ unit discussed above for [Re(Cl)(CO)$_3$(bipy)].


\subsection{Influence of the ligands: charge transfer analysis of [Ru(L)$_2$(dppz)]$^{2+}$ (L=bipy, phen, tap)}
\label{ssec:RuLigand}

A very interesting class of TMCs is based on the complex [Ru(bipy)$_2$(dppz)]$^{2+}$ (bipy = 2,2'-bipyridine, dppz = dipyrido-[3,2-a:2',3'-c]-phenazine), which has been reported as a molecular light switch for DNA in 1990~\cite{Friedman1990JACS}.
Since then, a huge research activity has developed in this field, both experimentally~\cite{Olson1997JACS, Chen1997JPCB, Coates2001JPCB, Brennaman2002JACS, Brennaman2004JPCA, Sun2010IC, Sun2010ICa,  Sun2010JACS, Sun2010JPCB, Jakubikova2009IC} and theoretically~\cite{Pourtois2004JACS, Fantacci2004CPL, Fantacci2005JACS, Batista2005JPCA, Atsumi2007JPPA, Chantzis2013CPL, Ambrosek2010JIB, Very2014CEJ, Jakubikova2009IC}.
The main goal of these studies was to understand the impact of the ligand sphere on the photophysics of [Ru(L)$_2$(dppz-like)]$^{2+}$ molecules and to assess the character of the low-lying excited states responsible for the luminescence properties.
[Ru(bipy)$_2$(dppz)]$^{2+}$ does not luminesce in water, however, it is slightly luminescent in acetonitrile and highly luminescent in the presence of DNA.
Similarly, [Ru(phen)$_2$(dppz)]$^{2+}$ (phen = 1,10-phenanthroline) is characterized by a rapid, nearly nonradiative decay (time constant of 250 ps) in water~\cite{Olson1997JACS}, and by a moderate luminescence in acetonitrile that increases drastically in DNA~\cite{Olson1997JACS, Coates2001JPCBa, Coates2001JPCB, McKinley2011CCR}.
In contrast to the bipy- and phen-substituted complexes, [Ru(tap)$_2$(dppz)]$^{2+}$ (tap=1,4,5,8-tetraazaphenanthrene)~\cite{Ortmans2004DT, Elias2008CEJ} is luminescent in water and in organic solvents, but interestingly has its luminescence quenched in DNA by electron transfer.

\begin{figure}[tb]
  \centering
  \includegraphics[scale=1]{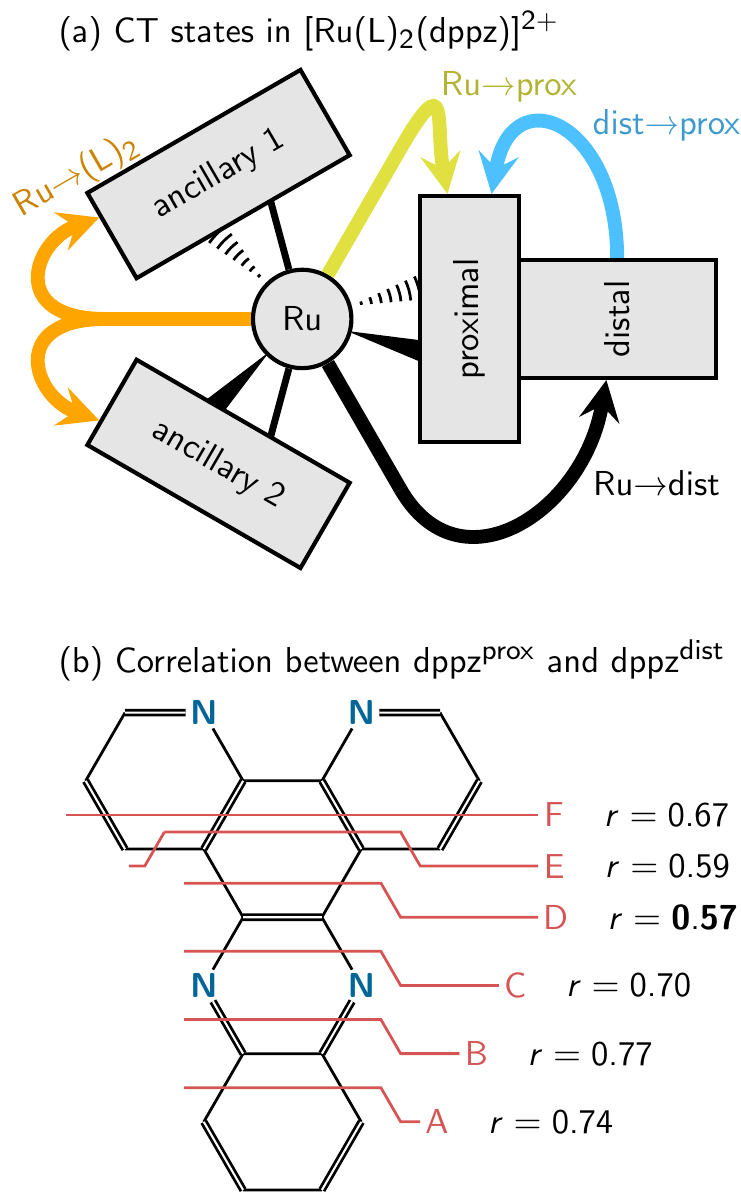}
  \caption{
  In (a), a general scheme of [Re(L)$_2$(dppz)]$^{2+}$ and its most relevant charge transfer states.
  In (b), different fragmentation schemes for the dppz ligand, where the red lines depict where proximal (top part of molecule) and distal (bottom) parts are separated.
  The labels next to the lines give the correlation coefficients between the proximal and distal parts.
  (TDA-B3LYP/TZP-DZP, COSMO(water))
  }
  \label{fig:dppz_scheme}
\end{figure}

It has been shown, both experimentally~\cite{Sun2010IC, Olofsson2004JPCA} and theoretically~\cite{Very2014CEJ, Ambrosek2010JIB}, that the relative energies of the low-lying triplet states of different nature---IL$_\mathrm{dppz}$, MLCT$_\mathrm{anc}$ (anc: ancillary ligands), MLCT$_\mathrm{prox}$, and MLCT$_\mathrm{dis}$ (localized on the dppz ligand, see Figure~\ref{fig:dppz_scheme}a)---govern the luminescence properties of this class of complexes.
This effect can be traced to the different sensitivity of the different states to the environment and to substituents effects.

\begin{figure*}[t]
  \centering
  \includegraphics[width=\textwidth]{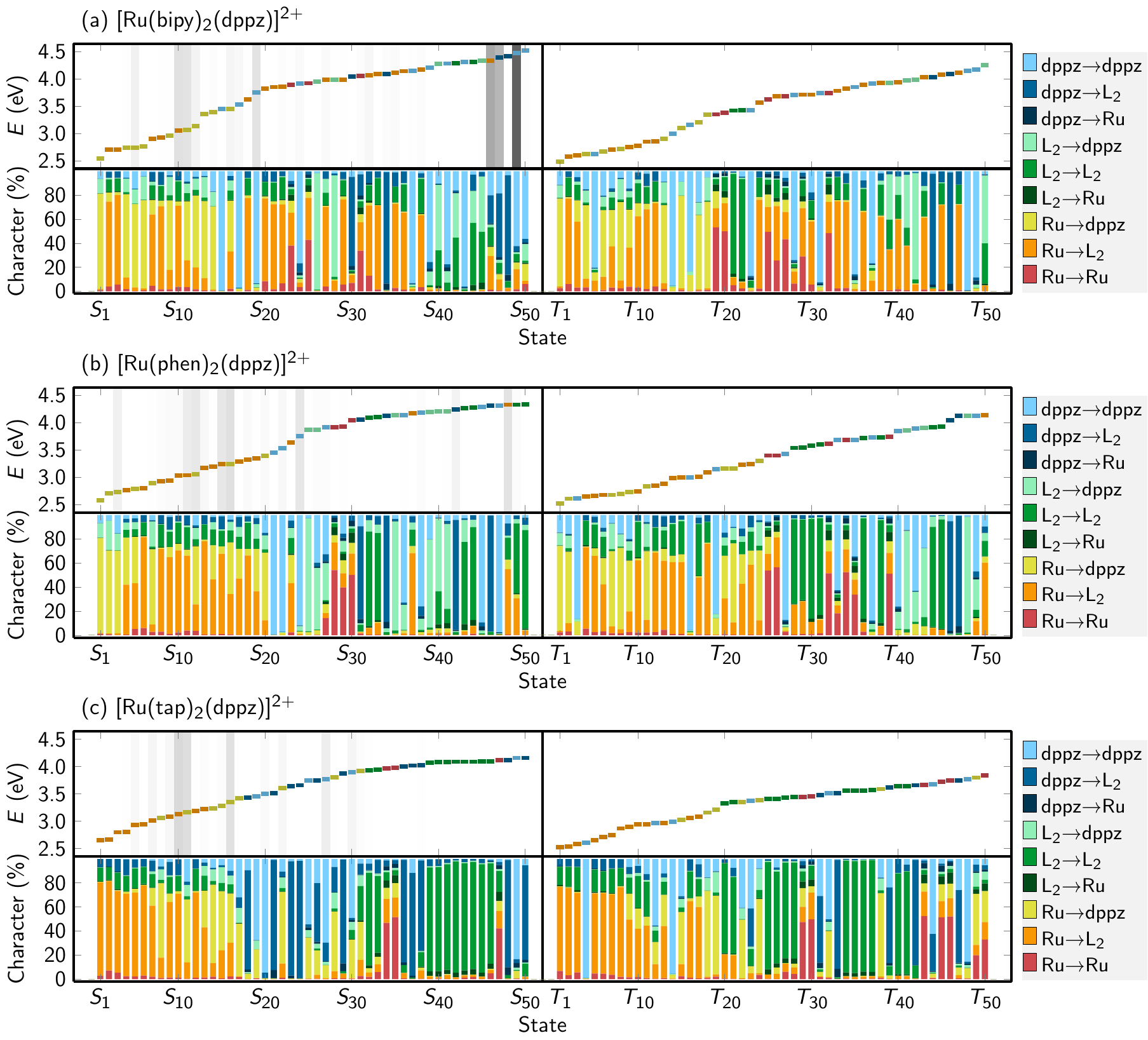}
  \caption{
  Compact depiction of the results of vertical excitation calculations for [Ru(L)$_2$(dppz)]$^{2+}$ (L=bipy, phen, tap), each with 50~singlets and 50~triplets.
  For each molecule, the top portion of the respective panel shows the energies (horizontal bars; energy range equivalent to 275--500~nm), oscillator strength (gray shading behind bars), and state character (color of the bars).
  The bottom part of each panel shows the contributions to each state, as computed with TheoDORE.
  (TDA-B3LYP/TZP-DZP, COSMO(water))
  } 
  \label{fig:theodore_example}
\end{figure*}

In order to quantify the different types of transitions underlying the photophysics of [Ru(bipy)$_2$(dppz)]$^{2+}$, [Ru(phen)$_2$ (dppz)]$^{2+}$, and [Ru(tap)$_2$(dppz)]$^{2+}$, we have performed a charge transfer analysis of the 50 lowest singlet and triplet states of each molecule.
Figure~\ref{fig:theodore_example} shows the excitation energies, oscillator strengths, and charge transfer character for the three complexes, where they have been divided into three fragments: (i) Ru, (ii) L$_2$, and (iii) dppz, as suggested by automatic fragmentation and consistency with previous theoretical studies~\cite{Very2014CEJ, Ambrosek2010JIB}.

For all three complexes, it can be seen that the low-energy states are dominated by MLCT states, involving either the ancillary ligands L$_2$ (orange) or the dppz ligand (yellow).
Which ligands are involved in the lowest singlet ($S_1$) and lowest triplet ($T_1$) states depends on the kind of ancillary ligands present, as bipy, phen, and tap differ considerably in their electron acceptor strength.
Hence, for L=bipy or phen, the $S_1$ and $T_1$ are MLCT states towards the dppz ligand, whereas for L=tap, these states involve the ancillary ligands.
More generally, for L=bipy, the MLCT states involving bipy tend to be higher in energy than the MLCT states involving dppz.
For L=phen, both types of MLCT states are at similar energies (nicely visible on the left of panel (b)), whereas for L=tap, the MLCT states involving dppz are shifted upwards relative to the MLCT states involving tap.

The lowest non-MLCT singlet is only located at around 3.4~eV, making it the $S_{15}$ to $S_{20}$ depending on L.
For L=bipy/phen, this lowest non-MLCT state has IL$_\text{dppz}$ character, whereas for L=tap it has dppz$\rightarrow$L$_2$  character.
For the triplet states, the situation is different, as the lowest non-MLCT state has an energy of only 2.6~eV, which is only slightly higher than the lowest triplet state (at 2.5~eV).
For all three molecules, this low-lying non-MLCT state is of IL$_\text{dppz}$ character.
The low energy of this triplet is due to the fact that such $\pi\pi^*$ transitions typically have a very large singlet-triplet splitting due to large exchange interactions.


          \begin{figure*}[t]
            \centering
            \includegraphics[width=\textwidth]{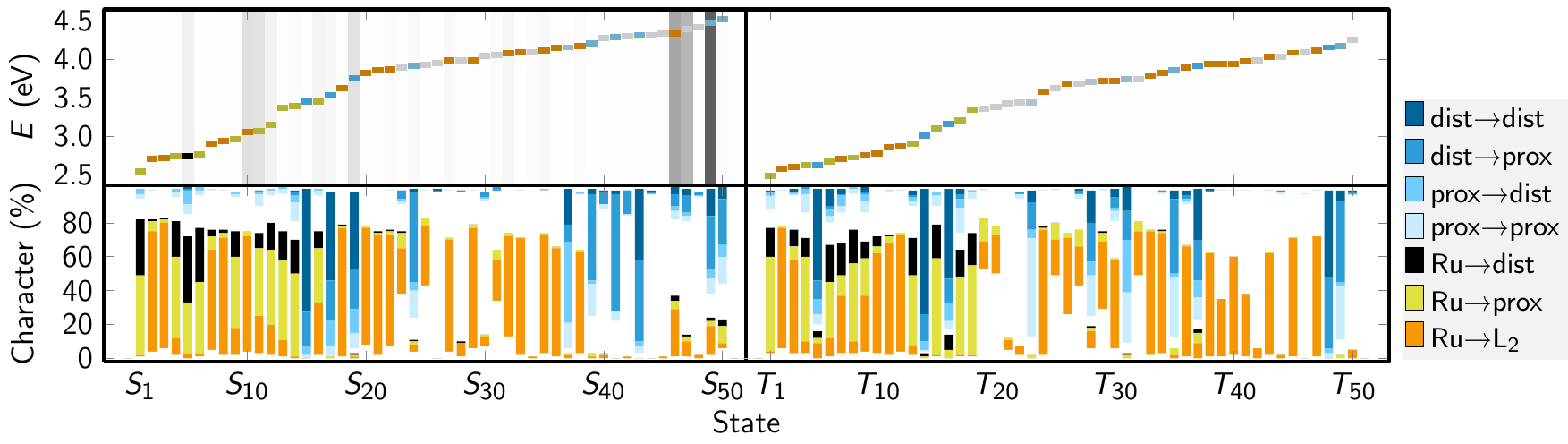}
            \caption{
            Compact depiction of the results of vertical excitation calculations for [Ru(L)$_2$(dppz)]$^{2+}$ (L=bipy), each with 50~singlets and 50~triplets.
            In this plot, we only show the MLCT contributions---distinguishing between the proximal and distal parts of dppz---and the different IL$_\text{dppz}$ contributions.
            In Figure~\ref{fig:dppz_scheme}a, the different charge transfer types are depicted in a similar color code.
            (TDA-B3LYP/TZP-DZP, COSMO(water))
            }
            \label{fig:theodore_example_2}
          \end{figure*}

At higher energies, a large number of IL and LLCT states appear in all three molecules.
However, there are important differences, which can be conveniently extracted from the figure.
For example, states of L$_2$ $\rightarrow$L$_2$ character (medium green) are very rare for L=bipy, and gain importance when going to L=phen or L=tap.
For L=tap, there is actually a set of 8 nearly degenerate L$_2$ $\rightarrow$L$_2$ (6 IL and 2 LLCT states) states at 4.0--4.1~eV.
Similarly, for L=bipy/phen, there are some L$_2$ $\rightarrow$dppz states (light green), whereas L=tap does not exhibit such states, only a large number of dppz$\rightarrow$L$_2$ states.
Of course, this different behavior of the L=tap complex is related to the well-established strong $\pi^*$ acceptor character of tap as compared to bipy or phen.

One very important aspect of the MLCT and IL states of the [Ru(L)$_2$(dppz)]$^{2+}$ complexes is their localization on the dppz ligand.
In particular, it was found that these complexes exhibit states which mostly involve the part of dppz close to the metal atom (called ``proximal'' subunit~\cite{Sun2010IC}) and states which involve the other part (``distal''~\cite{Sun2010IC}).
Using the charge transfer numbers, one can also investigate with more detail the relationship between these two subunits of dppz.
We first employed the correlation analysis to quantify how much the two subunits are actually decoupled.
In order to do so, in Figure~\ref{fig:dppz_scheme}b we show the correlation coefficients between the proximal and distal parts of dppz, depending on where the two units are divided.
A low correlation coefficient indicates that the two units are more independent, i.e., that they participate in different electronic states.
In that sense, the optimal division of dppz results in a bipyridine unit (proximal) and a phenazine unit (distal), which is consistent with chemical intuition.
This finding is the same irrespective of L (bipy, phen, or tap) and solvent (acetonitrile or water), and the same result was also obtained for a completely different dppz-containing complex, namely [Re(CO)$_3$(py)(dppz)]$^+$ (py=pyridine).
Furthermore, \emph{ab initio} fragmentation analyzes (i.e., starting from atomic fragments) of the dppz in the Ru and Re complexes also suggest a division in bipyridine and phenazine units.

With the optimal division of dppz established, it is possible to analyze the excited-state characters of the complexes, shown in Figure~\ref{fig:theodore_example_2} for the lowest 50 singlet and 50 triplet states of [Ru(bipy)$_2$(dppz)]$^{2+}$.
Most importantly, the figure shows how the MLCT$_\mathrm{dppz}$ contributions are split into Ru$\rightarrow$prox (yellow) and Ru$\rightarrow$dist (black) contributions, and how the IL$_\mathrm{dppz}$ contribution is composed.
It can be seen that for MLCT states, the proximal part of dppz is the better electron acceptor.
For IL$_\mathrm{dppz}$ states, it is found that the lower-energy states are mostly local dist$\rightarrow$dist excitations, whereas the higher states involve dist$\rightarrow$prox contributions.

The presented data can also be used to generally quantify the electron donor and acceptor properties of the proximal and distal parts, by calculating the average contribution of, e.g., the sum over all prox$\rightarrow$any excitations.
It is found that the proximal unit participates more as acceptor, with an average contribution of 28\% to all states, compared to 17\% for the distal unit.
On the contrary, the distal part is a better donor (18\%) than the proximal part (12\%).
Using either the detailed analysis of the states or these donor/acceptor descriptors, it should be possible to design substituted dppz ligands, where the proximal/distal contributions can be controlled to tailor the charge transfer character of the low-lying excited states for different applications, as mentioned above.


\subsection{Influence of the metal center: metal-carbonyl interaction in [M(CO)$_3$(im)(phen)]$^+$ (M=Mn, Tc, Re)}
\label{ssec:ReMetal}

Metal carbonyl diimine complexes [M(CO)$_3$(L)(NN)]$^{n+}$ (NN=diimine, L=axial ligand) of group 7 (M=Mn, Tc, Re) have received increased attention due to their electronic flexibility and the possibility to be incorporated in different environments---like polymers, proteins, or DNA.
They are employed in solar cells, photocatalysis, luminescent materials, conformational probes, radiopharmacy or diagnostic and therapeutic tools at the interfaces between chemistry, physics, and biology~\cite{Kumar2010TOC, Warren2012CCR,Twala2015DT}.

For rhenium derivatives [Re(CO)$_3$(L)(NN)]$^{n+}$, recent ultrafast spectroscopic experiments~\cite{ElNahhas2013JPCA, Zalis2011ICA, Nahhas2011IC, Cannizzo2008JACSa} and theoretical investigations \cite{Bakova2011CCR, Heydova2012JPCA, Mai2017PCCP, Fumanal2016JCC, Harabuchi2016JCTC, Eng2015ACR} have provided a comprehensive picture of their optical and photophysical properties.
The imidazole (im) derivative [Re(CO)$_3$(im)(phen)]$^+$ is employed in the study of electron transfer in proteins~\cite{Winkler1992CR,Gray2003QRB,Shih2008S,Warren2012CCR}.
In this particular complex, it has been often noted that the optical orbitals are delocalized over the metal and the carbonyls, thus giving rise to mixed MLCT/LLCT states~\cite{Vlcek2009TOC,Bakova2011CCR}. 
As this situation is very similar to that encountered in [Re(Cl)(CO)$_3$(bipy)] in section~\ref{ssec:correlation}, this molecule seems very suitable to challenge wave function analysis.
In particular, we would like to investigate whether this is an effect specific to the rhenium, or whether it can be encountered in other metals of the same group.
Accordingly, we shall investigate the correlation between the metal center and the carbonyl ligands in the series [M(CO)$_3$(im)(phen)]$^+$ (M=Mn, Tc, Re).
To this aim, we divided the molecule into four fragments: M, (CO)$_3$, im, and phen, and in each case we computed the first 50~singlet and 50~triplet states for a set of 20 randomly generated geometries (sampled from the ground state nuclear vibrational density distribution, see Computational Details), i.e., forming a set of 2000 excited states for each molecule.

\begin{figure}[t]
  \centering
  \includegraphics[scale=1]{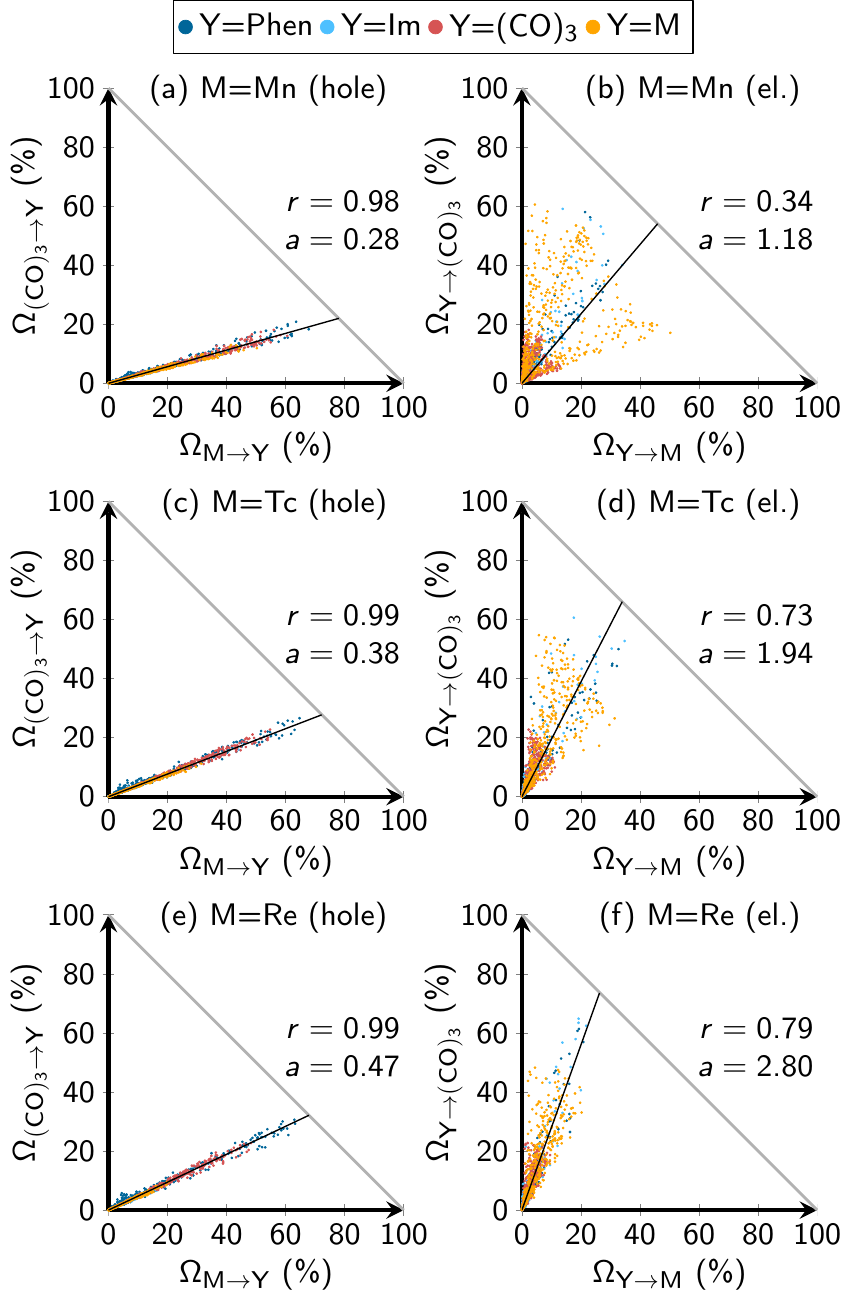}
  \caption{
  Charge transfer numbers for the first 50 singlets and 50 triplets of [M(CO)$_3$(im)(phen)]$^+$ (M=Mn, Tc, Re), based on 20 geometries sampled from the ground state nuclear vibrational density distribution.
  In the left column (panels (a), (c), (e)), the correlation between the charge transfer numbers $\Omega_{\text{M}\rightarrow\text{Y}}$ and $\Omega_{\text{(CO)}_3\rightarrow\text{Y}}$ is shown, thus presenting a correlation analysis for the hole.
  In the right panels ((b), (d), (f)), the correlations are shown for the electron.
  The $r$ and $a$ values in the panels are the Pearson correlation coefficient and slope from a fit with $f(x)=ax$.
  (TDA-B3LYP/TZP-DZP, COSMO(water))
  } 
  \label{fig:correlation_MnTcRe}
\end{figure}

The correlation analyses are shown in Figure~\ref{fig:correlation_MnTcRe}.
The left panels show the correlation between the excitations \emph{out of} the orbitals of M or (CO)$_3$.
Very strong correlations are observed in all three cases, with Pearson correlation coefficients of 0.98 or above.
This strong correlation is certainly not accidental, as it is found for all 2000~states at 20 different geometries, with energies ranging from below 3~eV up to 6~eV, and with widely differing state characters.
Basically, the correlation tells that independent of the state character, the charge transfer contribution of (CO)$_3$ is always a fixed portion of the charge transfer contribution of the metal.

The figure correlates four principally independent pairs of charge transfer numbers:
$\Omega_{\text{M}\rightarrow\text{M}}$      with $\Omega_{\text{(CO)}_3\rightarrow\text{M}}$ (orange);
$\Omega_{\text{M}\rightarrow\text{(CO)}_3}$ with $\Omega_{\text{(CO)}_3\rightarrow\text{(CO)}_3}$ (red);
$\Omega_{\text{M}\rightarrow\text{Im}}$     with $\Omega_{\text{(CO)}_3\rightarrow\text{Im}}$ (light blue); and
$\Omega_{\text{M}\rightarrow\text{Phen}}$   with $\Omega_{\text{(CO)}_3\rightarrow\text{Phen}}$ (dark blue).
Accordingly, we see that not only is MLCT correlated with LLCT , but likewise is MC correlated with LMCT.
Moreover, it is interesting to see that the strong correlation is found regardless of the metal atom, although the slopes of the correlations are different.
For Mn, the shown slope is very shallow (0.28), which indicates that the charge transfer states originate mostly from the metal center, and the participation of the carbonyls is relatively weak.
For the intermediate metal Tc, the slope is already 0.38, whereas for Re it is 0.47.
Hence, with increasing size of the metal center, the participation of the carbonyls in the electronic excitations increase.
In the Re complex, in all charge transfer states, two thirds of the excited electron density comes from the metal and one third comes from the three carbonyls.

The situation is different if the correlations of excitations \emph{into} the orbitals of M and (CO)$_3$ are investigated (right panels).
In this case, the correlations are much weaker, in particular for M=Mn.
This finding indicates that the virtual orbitals are more localized on either M or (CO)$_3$, unlike the delocalized occupied orbitals, and that for M=Re the virtual orbitals are mostly localized on (CO)$_3$.


\subsection{Influence of the method: exchange- and correlation effects in [Re(Cl)(CO)$_3$(bipy)]}
\label{ssec:ReMethod}

As a final application of the wavefunction analysis, we investigate here the influence of the exchange- and correlation functionals on the energies and state characters of the example [Re(Cl)(CO)$_3$(bipy)].
More than often, it is necessary to assess the quality of the results of some excited-state calculation against  another excited-state calculation using a different---possibly more reliable---electronic structure method.
A simple comparison of the excitation energies is in most cases not sufficient, because the order of the states (in terms of state character) might change~\cite{Atkins2017JCTC}. 
Hence, wave function analysis techniques can be very useful, especially in TMCs with their very high density of states. 

\begin{figure*}[t]
          \centering
          \includegraphics[width=\textwidth]{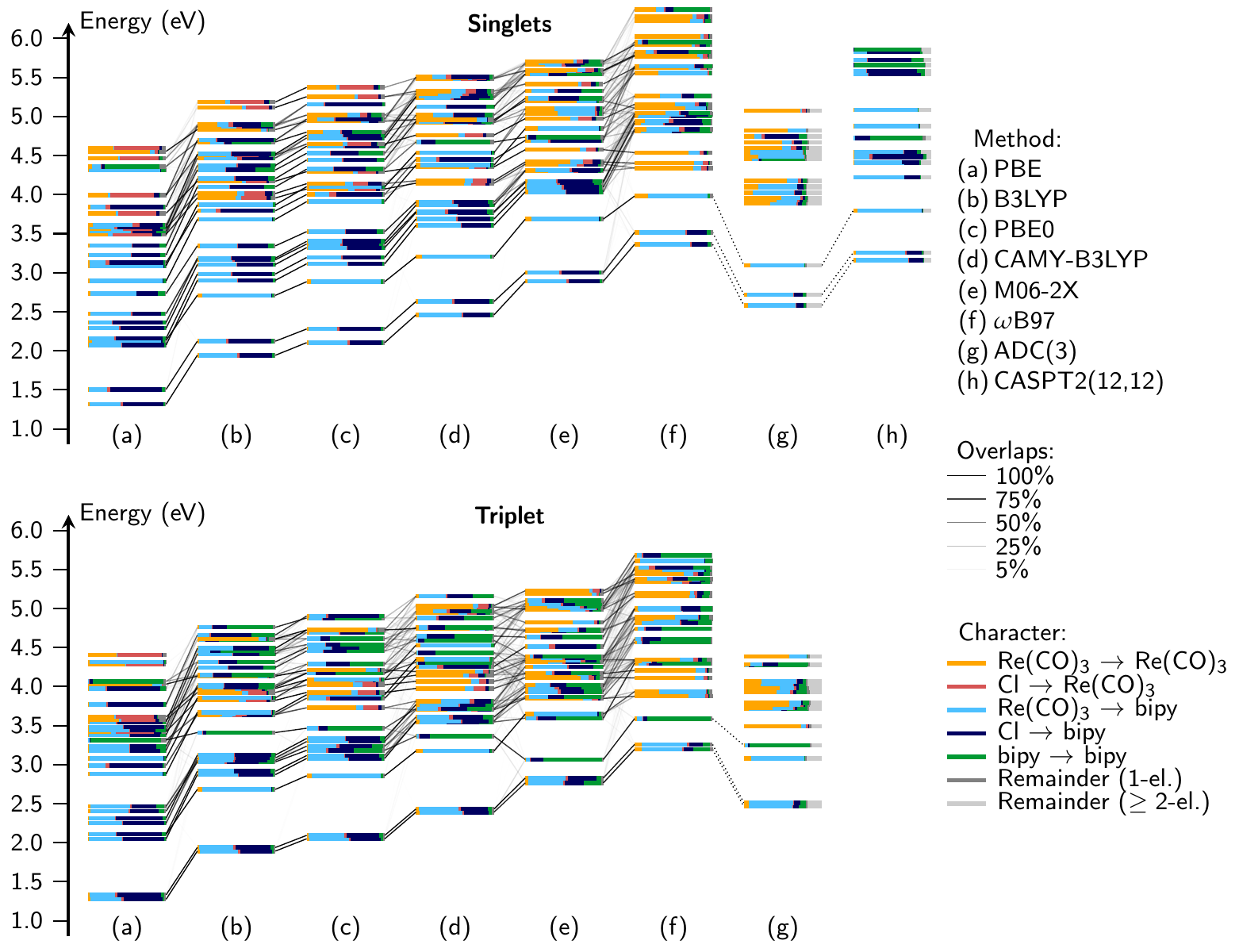}
          \caption{
          Comparison of energies ($y$ axis) and state character (color code of bars) of the low-lying states of [Re(Cl)(CO)$_3$(bipy)] for different electronic structure methods.
          The diagonal black lines correlate states of the same character---obtained through overlap calculations---where the thickness of the line is proportional to the overlap.
          The dotted lines serve the same purpose, but were added manually based on an inspection of the state characters, because overlaps of the TD-DFT results with ADC(3) or CASPT2 cannot be calculated currently.
          (TDA-DFT/TZP-DZ(P), ADC(3)/def2-TZVP, CASPT2/ANO-RCC, all gas phase; see appendix A for full computational details)
          }
          \label{fig:methods_ReCl}
        \end{figure*}

Excited-state calculations for [Re(Cl)(CO)$_3$(bipy)] have been performed with TD-DFT using the functionals PBE~\cite{Perdew1996PRL}, B3LYP \cite{Becke1993JCP}, PBE0~\cite{Adamo1999JCPa}, CAMY-B3LYP~\cite{Seth2012JCTC}, M06-2X~\cite{Zhao2008TCA}, and $\omega$B97~\cite{Chai2008JCP}).
The results of these functionals will be compared to the ones from the \emph{ab initio} ADC(3)~\cite{Schirmer1982PRA, Schirmer2004JCP, Harbach2014JCP} and CASPT2(12,12) \cite{Roos1987ACP,Pulay2011IJQC} methods.
Details on these calculations can be found in the Computational Details section.
Figure~\ref{fig:methods_ReCl} compactly shows the resulting energies of the excited states on the vertical axis, with colored horizontal bars indicating the state character.
The different TD-DFT results are correlated (gray lines) using overlap calculations~\cite{Plasser2016JCTC,Plasser2016JCP}.

Generally speaking, there are three classes of excited states visible in the plot: mixed MLCT/XLCT states (indicated in shades of blue), MC states (orange; might contain admixtures of XMCT states in red), and IL states (green).
Here, we consider Re and (CO)$_3$ as one fragment because they are correlated (see Section~\ref{ssec:correlation}).
The lowest excited states with all methods are several MLCT/XLCT states, even though the energy of this class of states depends heavily on the employed method and generally increases when more Hartree-Fock exchange is considered in the TD-DFT calculations (from left to right).
Moreover, the ratio between MLCT and XLCT depends significantly on the method, where more Hartree-Fock exchange tends to decrease the XLCT contribution; for the two \emph{ab initio} methods, the XLCT contribution is also relatively small.
The MC states only appear at higher energies---around 4~eV---but their energies are far less sensitive to the method.
Only few IL states were computed, although among the triplets this class of states is more common due to the strong triplet stabilization of the IL states.

For this particular complex, it can also be followed from Figure~\ref{fig:methods_ReCl}, that for TD-DFT calculations a relatively large amount of Hartree-Fock exchange is necessary in order to obtain results consistent with the correlated \emph{ab initio} methods.
Compared to the ADC(3) results, the TD-DFT computations using CAMY-B3LYP and M06-2X provided the best results.
On the contrary, the CASPT2 results are qualitatively reproduced by $\omega$B97, although the CASPT2 calculations did not provide any MC states due to active space restrictions.

Without attempting to discuss which result is better (that goes beyond the intention and scope of this review) the previous example shows the easiness in comparing results obtained from different methods when using quantitative wave function analysis.

\section{Concluding remarks}

We have reviewed a quantitative wave function analysis technique based on the concept of charge transfer numbers.
Charge transfer numbers are a mathematical tool based on the transition density matrix for the analysis of excited states in terms of electronic transitions between groups of atoms.
This tool---which is implemented in the TheoDORE software package and interfaced to several quantum chemistry programs---allows quantifying the localization and charge transfer character in an automatized way, without looking at any orbital.
Charge transfer numbers provide significant advantages compared to traditional state characterization based on orbital visualization.
These advantages include efficient analysis of large numbers of states, reproducibility, objectivity, and quantification which allows spotting chemical trends.
In the case of transition metal photochemistry, where a high density of states and the many different state characters are encountered, such a quantitative excited-state analysis is particularly helpful.

In this contribution we also report the extension of charge transfer numbers to the case of spin-orbit-coupled states and introduce a novel clustering approach which allows automatically partitioning the molecule into sensible chromophoric units.
This machinery was subsequently applied to various prototypical transition metal complexes, in a number of case studies devoted to Ir(III) phosphorescent emitters, Ru(II) dipyrido-phenazine photo-switches, and Re(I) $\alpha$-diimine carbonyl complexes.
The case studies demonstrated the potential of the charge transfer number concept for investigating different aspects of transition metal quantum chemistry: (i) the effect of modification of the nuclear geometry along a reaction coordinate, (ii) the effect of spin-orbit-induced mixing, (iii) the effect of the ligand sphere, (iv) the effect of the metal center, and (v) the effect of the electronic structure method.
The presented case studies are primarily intended as exploratory and tutorial-like, as the results depend on the chosen underlying electronic structure method and on the considered nuclear geometry.
Nevertheless, the case studies show beautifully the complexity of the photophysics and photochemistry of the investigated transition metal complexes.

More elaborate applications of the presented analysis tools can be envisaged.
In a very general sense, automatic wave function analysis can aid in the design of metal complexes for specific photonic applications, for example by serving as an optimization target.
Regarding excited-state dynamics simulations, automatic analysis can tremendously help in tracking the electronic states of interest along the nuclear trajectory.
Moreover, it might even be possible to employ the charge transfer numbers in automatized diabatization schemes, by transforming the states such that the descriptors change as little as possible.
Applications are also imaginable in theoretical chemistry method development.
Charge transfer numbers could be used to aid in the design of sensible active spaces for multi-configurational calculations, based on results from less costly TD-DFT computations. 
They also might help to understand the failure of some electronic structure methods at describing charge transfer states.
Finally, charge transfer numbers could be employed to parametrize point charges for excited-state force fields for molecular dynamics or hybrid quantum mechanics/molecular mechanics calculations.
Hence, we expect that the presented analysis techniques will be very useful for future theoretical investigations of transition metal complexes.


\section*{Acknowledgments}

S.M., F.P., and L.G.\ gratefully acknowledge funding from the Austrian Science Fund (FWF) within project I2883 and the University of Vienna. 
The Vienna Scientific Cluster (VSC3) is acknowledged for computational time.
M.F.\ and C.D.\ acknowledge funding from the Agence nationale de la recherche (ANR) within project ANR-15-CE29-0027 and the FRC and Labex CSC (ANR-10-LABX-0026\_CSC).
The authors also want to thank A. Vl\v{c}ek and S. Z\'ali\v{s} for fruitful discussions, and the inspiring COST action CM1305 (ECOSTBio).


\section*{Appendix A. Computational details}

\subsection*{Electronic structure packages}

The quantum chemical calculations were performed with ADF 2017 \cite{ADF2017} for DFT and TD-DFT, with QChem 5.0 \cite{QChem4} for ADC(3) \cite{Harbach2014JCP}, and with a development version of MOLCAS 8.1 \cite{Aquilante2015JCC} for CASSCF/CASPT2  \cite{Roos1987ACP,Pulay2011IJQC}.
The charge transfer number analyzes were done using a development version of TheoDORE 1.5 \cite{Plasser2017TheoDORE}.
For ADF, a newly developed interface between TheoDORE and ADF was used (see below).
Integrated analysis utilities in Q-Chem \cite{Plasser2014JCP1} and Molcas \cite{Plasser2017JCTC} were used to preprocess the output from these programs.

\subsection*{Interface between TheoDORE and ADF}

For the purpose of this work, a new interface between the TheoDORE program package and ADF was implemented.
This interface proceeds by reading all required information (geometries, response vectors, MO-coefficients, locations of the basis functions) from the binary ``TAPE21'' files produced by ADF.
In the case of full TD-DFT computations (rather than TDA), generally the right eigenvectors, i.e., $\mathbf{X}+\mathbf{Y}$ (Equation~\eqref{eq:tddft}), are analyzed.
For the purpose of analyzing spin-orbit coupled charge transfer numbers, the spin-orbit Hamiltonian was extracted from the TAPE21 file and subsequently diagonalized to yield the coefficients $U_{I\alpha}$, which were inserted into Equation~\eqref{eq:OmSOC}.
The described TheoDORE--ADF interface will be made available within the next TheoDORE release along with utility scripts that help in generating the plots shown in this work.

\subsection*{Hierarchical clustering}

The agglomerative hierarchical clustering algorithm~\cite{Murtagh2012WDMKD,Murtagh2017WDMKD} is employed to automatically analyze the correlations between the fragments, which are quantified with Equations~\eqref{eq:cov}, \eqref{eq:cor}, and \eqref{eq:dis}.
The clustering algorithm takes the distance matrix $\mathbf{R}$ as main input.
In each iteration of the algorithm, the smallest value $R_{AB}$ ($A\neq B$) in the distance matrix is looked up, and the two clusters $A$ and $B$ are subsequently merged.
To do this, the columns and rows $A$ and $B$ are deleted from the distance matrix, and instead a new column $C$ is added, holding the effective distances from the just-formed cluster $C=A\cup B$ to all other clusters in the matrix.
The effective distances are computed as a function of the original distances, so in order to compute the distance $R_{CD}$, one computes $R_{CD}=f(\{R_{xy}, x\in C, y\in D\})$.
Here, different choices for $f$ are possible, for example using the minimum of the distances (the algorithm is then called \emph{single-linkage clustering}), the maximum of the distances (\emph{complete-linkage clustering}), or the average of the distances (\emph{average-linkage clustering}).
Among these choices, complete-linkage clustering is generally reported to be very sensitive to outliers, whereas single-linkage clustering has a tendency to yield chain-like clusters.
Hence, we employ average-linkage clustering for our purposes.

\subsection*{Level of theory}

The single point calculation for the analysis presented in Sections \ref{sec:orbs}, \ref{ssec:CTnumbers}, and \ref{ssec:CTnumbersSOC} for [Ir(ppy)$_3$] was performed using the geometry reported in Ref.~\cite{Plasser2015JPCA}, which has $C_3$ symmetry.
For the TD-DFT calculation, the following basis sets were used: TZP for Ir, DZP for C and N, and DZ for H (denoted as ``TZP-DZ(P)'' in the main text).
We used the B3LYP functional and COSMO (for water, with non-equilibrium solvation with $\varepsilon_\infty=1.77$) to compute 30 excited singlet states and 30 triplet states, although in Sections \ref{sec:orbs} and \ref{ssec:CTnumbers} only the lowest six triplets are discussed.

The calculation for the correlation analysis of the complex [Re(Cl)(CO)$_3$(bipy)] in Section \ref{ssec:correlation} was carried out with the geometry reported in~\cite{Harabuchi2016JCTC}, which is of $C_s$ symmetry.
TZP for Re, DZP for C, N and Cl, and DZ for H were used together with the B3LYP functional.
The computation was done in the gas phase, and 20 singlets plus 20 triplets were computed.

The potential energy scan of [Ir(ppy)$_3$] in Section \ref{ssec:IrNuclear} was carried out using the geometry reported in Ref.~\cite{Plasser2015JPCA} as the $S_0$ minimum.
The $T_1$ minimum was optimized with UKS, whereas the $T_1/S_0$ minimum energy path (MXP) was optimized with TD-DFT; both optimizations started from the $S_0$ geometry.
The crossing points were located with the algorithm described in Ref.~\cite{Bearpark1994CPL}, employing the gradients provided by ADF and the ORCA~\cite{Neese2012WCMS} optimizer, as ADF does not allow optimizing crossing points.
The non-minimum crossing point (XP) in Figure~\ref{fig:theodore_Ir_path} was taken from the optimization iterations, such that the $T_1-S_0$ gap was $<$0.02~eV, the octahedral coordination was retained, and the energy was as low as possible.
Subsequently, a linear interpolation in internal coordinates was carried out between $S_0$ minimum, $T_1$ minimum, the XP, and the MXP.
Both optimizations and the single point calculations for the interpolated geometries were done in the gas phase, with TZP for Re, DZP for CN, DZ for H, and the B3LYP functional.

The single point calculation for [Ir(Cl)(CO)(ppy)$_2$] in Section \ref{ssec:IrSOC} employed the geometry reported by Brahim et al. \cite{Brahim2014CTC}, using the geometry data denoted as ``cis-[Ir(ppy)2(CO)(Cl)] 2a'' in the supporting information; (note that in the main text of \cite{Brahim2014CTC}, structure 2a is erroneously labeled as the trans conformer).
The calculation was performed with B3LYP, TZP for Re, DZP for C, N, O, Cl, and DZ for H in the gas phase and requesting 60 singlets and 60 triplets.

The three computations for [Ru(L)$_2$(dppz)]$^{2+}$, L=bipy, phen, or tap, in Section \ref{ssec:RuLigand} are based on geometries optimized in $C_2$ symmetry with B3LYP/TZP and COSMO (water), the one for [Re(CO)$_3$(py)(dppz)]$^{+}$ on a geometry in $C_s$ symmetry.
For the vertical excitation calculations, we employed the B3LYP functional, TZP for Ru, and DZP for C, N, and H (denoted as ``TZP-DZP'' in the main text), again using COSMO (water, $\varepsilon_\infty=1.77$).
50 singlets and 50 triplets were computed for each of the complexes.

The ground state geometries of [M(CO)$_3$(im)(phen)]$^+$, M = Mn, Tc, or Re, in Section \ref{ssec:ReMetal} were optimized with B3LYP/TZP, with COSMO (water).
Out of the possible conformers of these molecules, only one conformer---where the imidazole plane is oriented parallel to the N$_\mathrm{phen}$--N$_\mathrm{phen}$ connection line \cite{Mai2017PCCP}---was considered, as this is the most stable conformer (by 1--4~kJ/mol).
For each optimized geometry, we performed a frequency calculation to obtain the vibrational frequencies and normal modes, which together fully specify the harmonic ground state potential energy surface around the minimum.
From that, we computed the density distribution of the nuclear vibrational ground state wave function \cite{Schinke1995,Dahl1988JCP}, from which for each complex 20 geometries were randomly sampled.
For each of these 60 geometries, we then computed 50 singlet and 50 triplet states using B3LYP, TZP for Re, and DZP for C, N, O, and H.

Finally, the computations in Section \ref{ssec:ReMethod} contrasting quantum chemical methods in [Re(Cl)(CO)$_3$(bipy)] also employed the $C_s$ symmetric geometry of Ref.~\cite{Harabuchi2016JCTC}.
For the TD-DFT calculations, TZP for Re, DZP for C, N, O, Cl, and DZ for H were used.
Excited states were computed with the following functionals: PBE~\cite{Perdew1996PRL}, B3LYP~\cite{Becke1993JCP}, PBE0~\cite{Adamo1999JCPa}, CAMY-B3LYP~\cite{Seth2012JCTC}, M06-2X~\cite{Zhao2008TCA}, and $\omega$B97~\cite{Chai2008JCP} using 30 singlets and 30 triplets.
The ADC(3) computations used Ahlrichs' def2-TZVP basis set with the corresponding ECP for Re, and def2-SV(P) for the remaining atoms~\cite{Andrae1990TCA, Weigend2005PCCP}.
In this case, 40 states were computed in total, each 10 for $^1A^\prime$, $^1A^{\prime\prime}$, $^3A^\prime$, and $^3A^{\prime\prime}$.
The CASSCF/CASPT2 computations \cite{Roos1987ACP,Pulay2011IJQC} used the ANO-RCC basis set~\cite{Roos2004JPCA} with a [24s21p15d11f4g|8s7p5d2f1g] contraction scheme for Re, [17s12p5d|4s3p1d] for Cl, [14s9p4d|3s2p1d] for C, N, O, and [8s|2s] for H.
In the CAS(12,12) calculation the active space contains $2\times p_{\mathrm{Cl}}, 1\times \pi_{\mathrm{bipy}}, 3\times d_{\mathrm{Re}}, 3\times \pi_{\mathrm{bipy}}^* abd 3\times \pi_{\mathrm{CO}}^*$ orbitals.
The CASPT2 uses the default IPEA shift~\cite{Ghigo2004CPL} of 0.25~a.u., an imaginary level shift of 0.2~a.u., and 54 frozen occupied orbitals.
State-averaging in the CASSCF computations was performed over 19 singlet states and state-specific CASPT2 corrections were computed for each state.

All the geometries employed in this contribution, both taken from the literature or optimized within the present work, are reported in the supporting information. 
This also contains the active space orbitals of the CASPT2 calculation on [Re(Cl)(CO)$_3$(bipy)].


\section*{References}


\clearpage

\section*{\Large Supporting Information}

\section*{CASSCF active space orbitals}

\begin{figure}[h]
  \centering
  \includegraphics[scale=0.8]{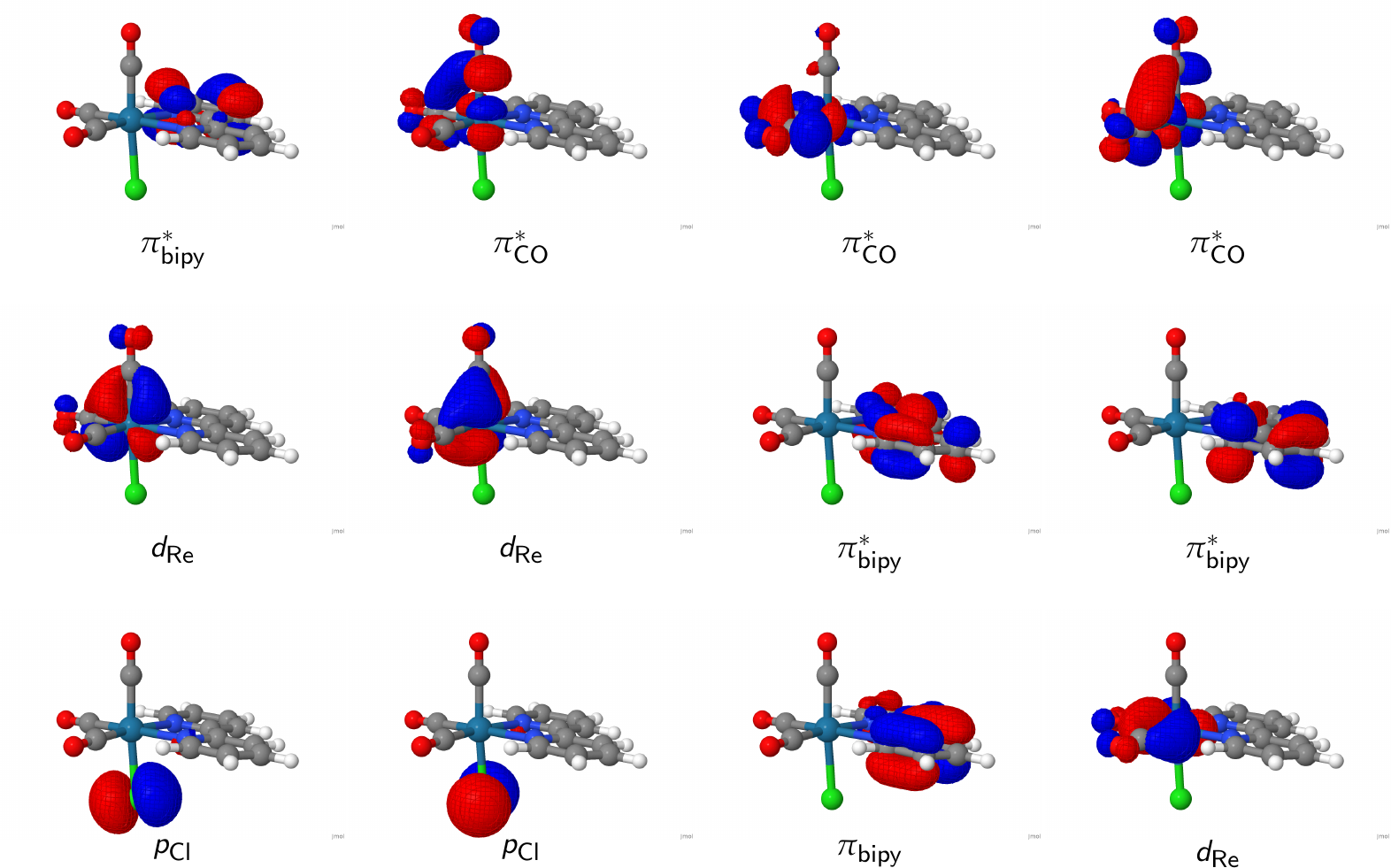}
  \caption{Active space orbitals employed for the CASPT2(12,12) calculations on Re(CO)$_3$(Cl)(bipy).}
  \label{fig:orb}
\end{figure}

Figure~\ref{fig:orb} shows the active space orbitals employed for the CASPT2(12,12) calculations on Re(CO)$_3$(Cl)(bipy) in Sections~\ref{ssec:correlation} and \ref{ssec:ReMethod}.

\section*{Molecular geometries}

On the following pages, all geometries used for calculations in this work are given in XYZ file format.

\scriptsize
\newpage

\begin{multicols}{2}

\bigskip
\begin{minipage}{0.45\textwidth}
\begin{verbatim}
61
Ir(ppy)3: S0 min (Plasser2015JPCA)
Ir +0.000000 +0.000000 -0.004130
N +0.872090 -1.626490 -1.091240
N -1.844620 +0.057990 -1.091240
C -0.595800 +1.616650 +1.037790
C +1.697960 -0.292350 +1.037790
C +2.058050 -2.056070 -0.602390
C +2.539790 -1.326530 +0.561610
N +0.972530 +1.568500 -1.091240
C -1.102160 -1.324300 +1.037790
C +0.751580 +2.810360 -0.602390
C -0.121090 +2.862790 +0.561610
C -2.809630 -0.754290 -0.602390
C -2.418710 -1.536260 +0.561610
C +2.146600 +0.405050 +2.166320
C +3.346570 +0.098370 +2.786510
C +4.158150 -0.923160 +2.301150
C +3.750830 -1.632120 +1.189840
C +2.713190 -3.127730 -1.213780
C +2.153700 -3.748950 -2.309560
C +0.932340 -3.297200 -2.792440
C +0.331210 -2.237130 -2.147150
C +1.771800 +1.405400 -2.147150
C +2.389290 +2.456030 -2.792440
C +2.169830 +3.739640 -2.309560
C +1.352100 +3.913560 -1.213780
C -0.461960 +4.064380 +1.189840
C -1.279590 +4.062650 +2.301150
C -1.758480 +2.849030 +2.786510
C -1.424090 +1.656490 +2.166320
C -3.288880 -2.432250 +1.189840
C -2.878560 -3.139490 +2.301150
C -1.588090 -2.947400 +2.786510
C -0.722520 -2.061540 +2.166320
C -2.103010 +0.831730 -2.147150
C -4.065290 -0.785830 -1.213780
C -4.323540 +0.009310 -2.309560
C -3.321630 +0.841170 -2.792440
H +1.534880 +1.202570 +2.572330
H +3.656510 +0.660800 +3.661810
H +5.096180 -1.161210 +2.788940
H +4.382550 -2.429350 +0.812910
H +3.659240 -3.471620 -0.818000
H +2.660480 -4.582420 -2.782450
H +0.449130 -3.757680 -3.644210
H -0.626110 -1.850990 -2.476620
H +1.916060 +0.383270 -2.476620
H +3.029680 +2.267800 -3.644210
H +2.638260 +4.595250 -2.782450
H +1.176890 +4.904810 -0.818000
H -0.087390 +5.010080 +0.812910
H -1.542450 +4.994020 +2.788940
H -2.400520 +2.836230 +3.661810
H -1.808890 +0.727960 +2.572330
H -4.295160 -2.580720 +0.812910
H -3.553720 -3.832810 +2.788940
H -1.255990 -3.497030 +3.661810
H +0.274020 -1.930530 +2.572330
H -1.289950 +1.467720 -2.476620
H -4.836130 -1.433190 -0.818000
H -5.298730 -0.012830 -2.782450
H -3.478810 +1.489880 -3.644210
\end{verbatim}
\end{minipage}

\bigskip
\begin{minipage}{0.45\textwidth}
\begin{verbatim}
61
Ir(ppy)3: T1 min (U-B3LYP/Re-TZP,CN-DZP,H-DZ)
Ir -0.001360 -0.002890 +0.016651
N +0.864530 -1.621537 -1.116951
N -1.788054 +0.063817 -1.115843
C -0.604282 +1.628727 +1.028030
C +1.722602 -0.301155 +1.041560
C +2.053296 -2.059729 -0.628370
C +2.540486 -1.349606 +0.552649
N +1.001527 +1.563058 -1.094178
C -1.058610 -1.303569 +1.065250
C +0.770279 +2.810317 -0.613998
C -0.115685 +2.867829 +0.544268
C -2.786454 -0.797736 -0.607181
C -2.428702 -1.506881 +0.545899
C +2.180215 +0.413185 +2.152090
C +3.380598 +0.092059 +2.770284
C +4.167664 -0.955174 +2.290481
C +3.749540 -1.670477 +1.182289
C +2.707307 -3.111466 -1.277981
C +2.139546 -3.696411 -2.393704
C +0.911229 -3.236000 -2.862690
C +0.306496 -2.194426 -2.187444
C +1.824195 +1.372887 -2.129337
C +2.450659 +2.418708 -2.777455
C +2.214890 +3.711528 -2.316879
C +1.378613 +3.905768 -1.234366
C -0.472851 +4.069790 +1.166855
C -1.310829 +4.063452 +2.267073
C -1.803751 +2.850110 +2.748547
C -1.457765 +1.654933 +2.136044
C -3.285956 -2.414430 +1.234828
C -2.828995 -3.094713 +2.331601
C -1.505589 -2.923609 +2.820027
C -0.650743 -2.040859 +2.179207
C -2.068859 +0.861572 -2.141517
C -4.040328 -0.853404 -1.283169
C -4.290309 -0.055066 -2.358932
C -3.283636 +0.851846 -2.802383
H +1.577268 +1.225071 +2.545431
H +3.709675 +0.656950 +3.638097
H +5.102354 -1.207908 +2.780032
H +4.370203 -2.481610 +0.816493
H +3.657459 -3.470244 -0.904562
H +2.646995 -4.514070 -2.895635
H +0.428813 -3.680309 -3.725364
H -0.653838 -1.792135 -2.488922
H +1.972502 +0.344116 -2.435775
H +3.110105 +2.224830 -3.615283
H +2.688338 +4.561082 -2.798111
H +1.199796 +4.905837 -0.862062
H -0.099436 +5.018424 +0.796468
H -1.583963 +4.996278 +2.748917
H -2.464190 +2.841866 +3.611120
H -1.844221 +0.720552 +2.528293
H -4.303434 -2.571460 +0.895999
H -3.494721 -3.783946 +2.843450
H -1.178132 -3.479155 +3.692057
H +0.355743 -1.901962 +2.558538
H -1.274098 +1.534463 -2.447892
H -4.799806 -1.539416 -0.927834
H -5.246378 -0.099232 -2.869057
H -3.455339 +1.526279 -3.632144
\end{verbatim}
\end{minipage}

\bigskip
\begin{minipage}{0.45\textwidth}
\begin{verbatim}
61
Ir(ppy)3: XP S0/T1 (TD-B3LYP/Re-TZP,CN-DZP,H-DZ)
Ir -0.055534 +0.191782 +0.116633
N +0.757587 -1.850103 -1.074310
N -1.331813 -0.121677 -1.373523
C -0.680124 +1.761613 +1.158030
C +1.868610 -0.242906 +0.638232
C +1.850271 -2.383882 -0.521883
C +2.490286 -1.483497 +0.442675
N +0.924180 +1.838634 -0.964279
C -1.078281 -1.257257 +1.081808
C +0.660889 +3.053130 -0.448961
C -0.212558 +3.030587 +0.731805
C -2.499214 -0.869105 -0.864995
C -2.223105 -1.630843 +0.362633
C +2.477685 +0.716065 +1.448379
C +3.645685 +0.401144 +2.135883
C +4.233541 -0.846405 +1.984215
C +3.661638 -1.779208 +1.129972
C +2.297515 -3.649618 -0.890029
C +1.572716 -4.360790 -1.836101
C +0.434810 -3.791854 -2.391673
C +0.057874 -2.520951 -1.978999
C +1.734963 +1.696203 -2.010662
C +2.321739 +2.785381 -2.628285
C +2.046609 +4.052320 -2.131241
C +1.213168 +4.190202 -1.032826
C -0.573172 +4.169825 +1.437806
C -1.399713 +4.067009 +2.553790
C -1.874214 +2.829042 +2.966213
C -1.518487 +1.678403 +2.271791
C -2.975984 -2.700064 +0.838914
C -2.618552 -3.349899 +2.020017
C -1.494714 -2.957657 +2.734551
C -0.718635 -1.902122 +2.250936
C -1.686554 +0.962814 -2.254509
C -3.672204 -0.822385 -1.532790
C -3.878069 -0.001995 -2.679801
C -2.868473 +0.987275 -2.899244
H +2.048486 +1.700781 +1.578946
H +4.101702 +1.147610 +2.777926
H +5.146178 -1.088832 +2.515163
H +4.135302 -2.745494 +1.004356
H +3.190012 -4.076425 -0.450254
H +1.896132 -5.353066 -2.135524
H -0.159960 -4.320710 -3.127912
H -0.813593 -2.003569 -2.363151
H +1.891689 +0.680844 -2.353260
H +2.972608 +2.642326 -3.483177
H +2.479550 +4.930790 -2.597969
H +0.993312 +5.173291 -0.638575
H -0.223294 +5.149283 +1.133063
H -1.679373 +4.964219 +3.095887
H -2.528780 +2.758304 +3.828897
H -1.900767 +0.715702 +2.583563
H -3.840919 -3.050386 +0.284322
H -3.222006 -4.181470 +2.372377
H -1.213690 -3.474098 +3.646728
H +0.177858 -1.604821 +2.786918
H -0.936426 +1.707703 -2.415169
H -4.492743 -1.419508 -1.140604
H -4.863302 +0.111122 -3.108628
H -3.045736 +1.791947 -3.600586
\end{verbatim}
\end{minipage}

\bigskip
\begin{minipage}{0.45\textwidth}
\begin{verbatim}
61
Ir(ppy)3: MXP S0/T1 (TD-B3LYP/Re-TZP,CN-DZP,H-DZ)
Ir -0.145947 +0.564575 +0.165466
N +0.928223 -1.840875 -1.821194
N -1.379993 -0.054515 -1.363300
C -0.703964 +2.033977 +1.429061
C +1.775002 -0.104439 +0.479767
C +1.357097 -2.303459 -0.637632
C +2.216679 -1.397153 +0.156122
N +0.758100 +2.281079 -0.761781
C -0.958832 -1.053760 +1.020019
C +0.584478 +3.450487 -0.098610
C -0.242497 +3.338122 +1.099763
C -2.189938 -1.131937 -1.061889
C -1.924712 -1.727309 +0.231418
C +2.648999 +0.744692 +1.168211
C +3.906609 +0.308157 +1.569764
C +4.323747 -0.983869 +1.273856
C +3.478731 -1.827510 +0.565295
C +0.986315 -3.552453 -0.142460
C +0.127943 -4.342345 -0.888515
C -0.324830 -3.865330 -2.111112
C +0.107356 -2.615429 -2.530859
C +1.573466 +2.211135 -1.819179
C +2.228002 +3.322587 -2.313692
C +2.038077 +4.542809 -1.670801
C +1.219808 +4.605862 -0.557134
C -0.554920 +4.421447 +1.923132
C -1.330098 +4.235709 +3.057051
C -1.796539 +2.961454 +3.383933
C -1.487250 +1.874535 +2.580847
C -2.504756 -2.919165 +0.679016
C -2.131648 -3.458127 +1.896548
C -1.173980 -2.807693 +2.679905
C -0.598580 -1.623622 +2.245993
C -1.498001 +0.547419 -2.569832
C -3.134796 -1.558749 -1.988994
C -3.250790 -0.940548 -3.219790
C -2.402824 +0.134867 -3.514567
H +2.340341 +1.753850 +1.421571
H +4.561763 +0.983233 +2.112732
H +5.307214 -1.327963 +1.577883
H +3.808898 -2.827474 +0.297933
H +1.323599 -3.864617 +0.838491
H -0.201676 -5.304160 -0.509956
H -1.008943 -4.442307 -2.724203
H -0.236166 -2.202634 -3.476319
H +1.703021 +1.222293 -2.246304
H +2.884976 +3.229584 -3.170623
H +2.540339 +5.435972 -2.027730
H +1.085503 +5.544685 -0.035041
H -0.199424 +5.418738 +1.686248
H -1.574939 +5.083377 +3.688714
H -2.408310 +2.822756 +4.270731
H -1.858962 +0.889700 +2.839514
H -3.235434 -3.441733 +0.070518
H -2.579276 -4.385069 +2.240170
H -0.877233 -3.237556 +3.632479
H +0.161027 -1.140609 +2.853195
H -0.835291 +1.382830 -2.747956
H -3.780420 -2.390490 -1.736408
H -3.987579 -1.278432 -3.939285
H -2.454155 +0.650149 -4.466669
\end{verbatim}
\end{minipage}

\bigskip
\begin{minipage}{0.45\textwidth}
\begin{verbatim}
28
Re(CO)3(Cl)(bipy): S0 min (Harabuchi2016JCTC)
C -3.275103 -0.044247 +1.502075
C -3.196555 -0.052099 +2.892632
C -2.100549 -0.066125 +0.740179
H -4.106718 -0.034369 +3.493527
C -1.939620 -0.083561 +3.496676
N -0.878233 -0.094186 +1.337309
H -1.823889 -0.091815 +4.580356
C -0.810560 -0.103192 +2.683419
H +0.188925 -0.125931 +3.115734
H -4.248407 -0.020601 +1.016325
C -2.100549 -0.066125 -0.740179
N -0.878233 -0.094186 -1.337309
C -3.275103 -0.044247 -1.502075
C -0.810560 -0.103192 -2.683419
C -3.196555 -0.052099 -2.892632
C -1.939620 -0.083561 -3.496676
H -4.248407 -0.020601 -1.016325
H +0.188925 -0.125931 -3.115734 
H -4.106718 -0.034369 -3.493527 
H -1.823889 -0.091815 -4.580356 
Re +0.894823 -0.052419 +0.000000
C +2.265517 +0.103478 +1.361644 
C +1.070050 -1.967897 +0.000000 
C +2.265517 +0.103478 -1.361644 
Cl +0.509433 +2.478712 +0.000000
O +1.176732 -3.127265 +0.000000 
O +3.066717 +0.208112 +2.195548 
O +3.066717 +0.208112 -2.195548 
\end{verbatim}
\end{minipage}

\bigskip
\begin{minipage}{0.45\textwidth}
\begin{verbatim}
44  
Ir(CO)(Cl)(ppy)2: S0 min (Brahim2014CTC)
C +0.263266 -1.632625 +2.665885 
C +0.723630 -2.043724 +3.917764 
C +1.779326 -1.374338 +4.550821 
C +2.374717 -0.288470 +3.922698 
C +1.913866 +0.131515 +2.660906 
C +0.844071 -0.537695 +2.013079 
C +2.477168 +1.266785 +1.937072 
N +1.897397 +1.520647 +0.725439 
C +2.326835 +2.544498 -0.034846 
C +3.364088 +3.376550 +0.356442 
C +3.976629 +3.137455 +1.590675 
C +3.531594 +2.084759 +2.375781 
Ir +0.320355 +0.259705 +0.206373
C +1.421556 -1.058029 -0.708865
O +2.032462 -1.852827 -1.274433
C -0.975454 +1.651673 +1.045719
C -1.597864 +2.532878 +0.123098
C -2.480957 +3.535637 +0.563179
C -2.767834 +3.674956 +1.915756
C -2.174117 +2.805477 +2.834992
C -1.292794 +1.812508 +2.401452
C -1.293905 +2.332234 -1.299595
C -1.882354 +3.043758 -2.359608
C -1.537281 +2.756067 -3.672401
C -0.604082 +1.748249 -3.928887
C -0.056978 +1.075581 -2.845330
N -0.377571 +1.361933 -1.571410
Cl -1.593377 -1.196652 -0.190960
H +1.808999 +2.679522 -0.976566
H +3.677314 +4.189216 -0.289639
H +3.992756 +1.885902 +3.336007
H +4.791542 +3.768090 +1.934161
H +3.193454 +0.228844 +4.416327
H +0.255975 -2.894572 +4.407875
H +2.132433 -1.701353 +5.524807
H -0.856061 +1.143678 +3.136255
H -2.951059 +4.211943 -0.145515
H -2.401900 +2.899950 +3.894334
H -3.451157 +4.450630 +2.250512
H +0.658281 +0.273137 -2.987369
H -0.313243 +1.479897 -4.938652
H -2.618753 +3.809839 -2.148707
H -1.997318 +3.304541 -4.489593
H -0.558505 -2.157732 +2.190307
\end{verbatim}
\end{minipage}

\bigskip
\begin{minipage}{0.45\textwidth}
\begin{verbatim}
73
[Ru(bipy)2(dppz)]2+: S0 min (B3LYP/TZP)
Ru +0.000000 +0.000000 +1.449479
N -0.992742 -1.055937 +2.959913
C -0.445075 -2.255534 +3.287288
C +0.727450 -2.656787 +2.490587
C +1.386578 -3.871365 +2.649228
C +2.481127 -4.167513 +1.852931
C +2.897074 -3.239347 +0.908338
C +2.203650 -2.046997 +0.797049
N +1.145261 -1.751497 +1.567751
C -0.980351 -3.029886 +4.311202
C -2.085802 -2.573366 +5.010534
C -2.637782 -1.346085 +4.670335
C -2.062678 -0.619292 +3.642772
N +1.006891 +0.871571 -0.174848
C +1.994402 +1.768860 -0.129160
C +2.566095 +2.304708 -1.279235
C +2.106682 +1.903083 -2.516324
C +1.069124 +0.967669 -2.588381
C +0.527817 +0.483309 -3.853931
N +1.033088 +0.948199 -4.981383
C +0.526491 +0.484537 -6.138533
C -0.526491 -0.484537 -6.138533
N -1.033088 -0.948199 -4.981383
C -0.527817 -0.483309 -3.853931
C -1.069124 -0.967669 -2.588381
C -2.106682 -1.903083 -2.516324
C -2.566095 -2.304708 -1.279235
C -1.994402 -1.768860 -0.129160
N -1.006891 -0.871571 -0.174848
C -0.539800 -0.480275 -1.390374
C +0.539800 +0.480275 -1.390374
C -1.036969 -0.955264 -7.374197
C -0.521594 -0.481145 -8.549357
C +0.521594 +0.481145 -8.549357
C +1.036969 +0.955264 -7.374197
N +0.992742 +1.055937 +2.959913
C +2.062678 +0.619292 +3.642772
C +2.637782 +1.346085 +4.670335
C +2.085802 +2.573366 +5.010534
H +2.506211 +3.167587 +5.808383
C +0.980351 +3.029886 +4.311202
C +0.445075 +2.255534 +3.287288
C -0.727450 +2.656787 +2.490587
N -1.145261 +1.751497 +1.567751
C -2.203650 +2.046997 +0.797049
C -2.897074 +3.239347 +0.908338
C -2.481127 +4.167513 +1.852931
H -2.998103 +5.108434 +1.969656
C -1.386578 +3.871365 +2.649228
H +2.998103 -5.108434 +1.969656
H -2.506211 -3.167587 +5.808383
H -2.461633 +0.339491 +3.347998
H -3.497574 -0.949065 +5.186817
H +3.742212 -3.425827 +0.264261
H +2.497190 -1.298289 +0.076926
H +2.461633 -0.339491 +3.347998
H +3.497574 +0.949065 +5.186817
H -3.742212 +3.425827 +0.264261
H -2.497190 +1.298289 +0.076926
H +2.337257 +2.061239 +0.852028
H +3.362890 +3.025461 -1.184942
H +2.533060 +2.296448 -3.425459
H +1.831717 +1.687397 -7.357098
H +0.908961 +0.839228 -9.492212
H -0.908961 -0.839228 -9.492212
H -1.831717 -1.687397 -7.357098
H -2.533060 -2.296448 -3.425459
H -3.362890 -3.025461 -1.184942
H -2.337257 -2.061239 +0.852028
H +1.048298 -4.581822 +3.384287
H -0.537432 -3.978234 +4.564316
H -1.048298 +4.581822 +3.384287
H +0.537432 +3.978234 +4.564316
\end{verbatim}
\end{minipage}

{
\scriptsize
\bigskip
\begin{minipage}{0.45\textwidth}
\begin{verbatim}
77
[Ru(phen)2(dppz)]2+: S0 min (B3LYP/TZP)
Ru +0.000000 +0.000000 +1.153207
N -1.031338 -1.047808 +2.649192
C -0.459450 -2.248875 +2.956082
C +0.707296 -2.620931 +2.224543
C +1.341410 -3.850821 +2.484940
C +2.485701 -4.165133 +1.727986
C +2.929791 -3.273130 +0.778125
C +2.244192 -2.068550 +0.582564
N +1.162331 -1.740342 +1.285729
C +0.796574 -4.708061 +3.495243
C -0.313184 -4.352633 +4.193480
C -0.978265 -3.108682 +3.942920
C -2.133506 -2.688488 +4.628767
C -2.699923 -1.474391 +4.312043
C -2.121881 -0.677318 +3.316952
N +1.041191 +0.825410 -0.462362
C +2.083002 +1.661157 -0.412379
C +2.699562 +2.148879 -1.561623
C +2.227863 +1.766695 -2.801952
C +1.133060 +0.897451 -2.877589
C +0.563217 +0.444794 -4.145018
N +1.104715 +0.869872 -5.273634
C +0.563594 +0.442520 -6.432045
C -0.563594 -0.442520 -6.432045
N -1.104715 -0.869872 -5.273634
C -0.563217 -0.444794 -4.145018
C -1.133060 -0.897451 -2.877589
C -2.227863 -1.766695 -2.801952
C -2.699562 -2.148879 -1.561623
C -2.083002 -1.661157 -0.412379
N -1.041191 -0.825410 -0.462362
C -0.565478 -0.449272 -1.680022
C +0.565478 +0.449272 -1.680022
C -1.111763 -0.869723 -7.668505
C -0.559539 -0.437208 -8.844227
C +0.559539 +0.437208 -8.844227
C +1.111763 +0.869723 -7.668505
N +1.031338 +1.047808 +2.649192
C +2.121881 +0.677318 +3.316952
C +2.699923 +1.474391 +4.312043
C +2.133506 +2.688488 +4.628767
H +2.563877 +3.320886 +5.394200
C +0.978265 +3.108682 +3.942920
C +0.459450 +2.248875 +2.956082
C -0.707296 +2.620931 +2.224543
N -1.162331 +1.740342 +1.285729
C -2.244192 +2.068550 +0.582564
C -2.929791 +3.273130 +0.778125
C -2.485701 +4.165133 +1.727986
H -3.001873 +5.100898 +1.898163
C -1.341410 +3.850821 +2.484940
C -0.796574 +4.708061 +3.495243
C +0.313184 +4.352633 +4.193480
H +3.001873 -5.100898 +1.898163
H -2.563877 -3.320886 +5.394200
H -2.550518 +0.278490 +3.052723
H -3.587197 -1.119833 +4.816894
H -0.717956 -5.007062 +4.954685
H +1.290989 -5.650894 +3.690280
H +3.803419 -3.481883 +0.177221
H +2.578921 -1.353473 -0.154691
H +2.550518 -0.278490 +3.052723
H +3.587197 +1.119833 +4.816894
H +0.717956 +5.007062 +4.954685
H -1.290989 +5.650894 +3.690280
H -3.803419 +3.481883 +0.177221
H -2.578921 +1.353473 -0.154691
H +2.430679 +1.941183 +0.570879
H +3.540925 +2.819938 -1.465123
H +2.686992 +2.126740 -3.710875
H +1.964057 +1.536925 -7.651679
H +0.975769 +0.762037 -9.789185
H -0.975769 -0.762037 -9.789185
H -1.964057 -1.536925 -7.651679
H -2.686992 -2.126740 -3.710875
H -3.540925 -2.819938 -1.465123
H -2.430679 -1.941183 +0.570879
\end{verbatim}
\end{minipage}
}

\bigskip
\begin{minipage}{0.45\textwidth}
\begin{verbatim}
73
[Ru(tap)2(dppz)]2+: S0 min (B3LYP/TZP)
Ru +0.000000 +0.000000 -1.158577
N +1.162983 -1.734480 -1.265064
C +0.689890 -2.629635 -2.176146
C -0.465272 -2.269047 -2.912355
N -1.031573 -1.058556 -2.647624
C -2.109435 -0.737499 -3.350222
C -2.630726 -1.622276 -4.315096
N -2.100512 -2.801593 -4.579794
C -0.997323 -3.145608 -3.873239
C -0.351786 -4.407422 -4.081553
C +0.753130 -4.756144 -3.369636
C +1.310594 -3.871908 -2.390238
N +2.412045 -4.219967 -1.683019
C +2.853229 -3.341100 -0.802498
C +2.239581 -2.092991 -0.578877
H -2.576239 +0.216879 -3.158425
H -3.512946 -1.329181 -4.869450
H -0.772548 -5.073014 -4.823123
H +1.240821 -5.708943 -3.525241
H +3.732157 -3.601276 -0.226859
H +2.631836 -1.400843 +0.150716
N -1.162983 +1.734480 -1.265064
C -0.689890 +2.629635 -2.176146
C +0.465272 +2.269047 -2.912355
N +1.031573 +1.058556 -2.647624
C +2.109435 +0.737499 -3.350222
C +2.630726 +1.622276 -4.315096
N +2.100512 +2.801593 -4.579794
C +0.997323 +3.145608 -3.873239
C +0.351786 +4.407422 -4.081553
C -0.753130 +4.756144 -3.369636
C -1.310594 +3.871908 -2.390238
N -2.412045 +4.219967 -1.683019
C -2.853229 +3.341100 -0.802498
C -2.239581 +2.092991 -0.578877
H +2.576239 -0.216879 -3.158425
H +3.512946 +1.329181 -4.869450
H +0.772548 +5.073014 -4.823123
H -1.240821 +5.708943 -3.525241
H -3.732157 +3.601276 -0.226859
H -2.631836 +1.400843 +0.150716
N -1.055470 -0.810407 +0.459583
C -0.572365 -0.440016 +1.675577
C +0.572365 +0.440016 +1.675577
N +1.055470 +0.810407 +0.459583
C +2.112156 +1.625885 +0.407501
C +2.737420 +2.101291 +1.557305
C +2.257332 +1.727751 +2.796205
C +1.147258 +0.877836 +2.872673
C +0.570228 +0.435012 +4.140040
C -0.570228 -0.435012 +4.140040
C -1.147258 -0.877836 +2.872673
C -2.257332 -1.727751 +2.796205
C -2.737420 -2.101291 +1.557305
C -2.112156 -1.625885 +0.407501
N +1.118484 +0.852421 +5.268068
C +0.570894 +0.433580 +6.426040
C +1.124571 +0.853645 +7.662534
C +0.565830 +0.429186 +8.837852
C -0.565830 -0.429186 +8.837852
C -1.124571 -0.853645 +7.662534
C -0.570894 -0.433580 +6.426040
N -1.118484 -0.852421 +5.268068
H +2.467692 +1.899436 -0.574400
H +3.592219 +2.754809 +1.460913
H +2.722567 +2.079141 +3.705410
H +1.986375 +1.508481 +7.645916
H +0.986545 +0.747821 +9.782939
H -0.986545 -0.747821 +9.782939
H -1.986375 -1.508481 +7.645916
H -2.722567 -2.079141 +3.705410
H -3.592219 -2.754809 +1.460913
H -2.467692 -1.899436 -0.574400
\end{verbatim}
\end{minipage}

\bigskip
\begin{minipage}{0.45\textwidth}
\begin{verbatim}
49
[Re(CO)3(py)(dppz)]+: S0 min (B3LYP/TZP)
Re -0.037760 -0.172581 +0.000000
N -0.314508 +1.557127 +1.333112
N -0.314508 +1.557127 -1.333112
C -1.907856 -0.659416 +0.000000
C +0.341701 -1.495220 -1.352453
C +0.341701 -1.495220 +1.352453
C -0.396602 +1.517080 +2.663900
C -0.358827 +2.666020 +3.446023
C -0.220325 +3.894692 +2.835086
C -0.155383 +3.960603 +1.438957
C -0.224326 +2.763420 +0.722029
C -0.224326 +2.763420 -0.722029
C -0.155383 +3.960603 -1.438957
C -0.220325 +3.894692 -2.835086
C -0.358827 +2.666020 -3.446023
C -0.396602 +1.517080 -2.663900
C -0.010350 +5.218975 +0.714943
C -0.010350 +5.218975 -0.714943
O -3.029255 -0.935997 +0.000000
O +0.578972 -2.265930 -2.181036
O +0.578972 -2.265930 +2.181036
H -0.479351 +0.539285 +3.109806
H -0.424617 +2.577492 +4.518405
H -0.424617 +2.577492 -4.518405
H -0.479351 +0.539285 -3.109806
H -0.166316 +4.803703 +3.412838
H -0.166316 +4.803703 -3.412838
N +0.128994 +6.337138 +1.403250
N +0.128994 +6.337138 -1.403250
C +0.278319 +7.483599 -0.716852
C +0.278319 +7.483599 +0.716852
C +0.445140 +8.707766 -1.411275
H +0.444805 +8.690836 -2.491751
C +0.445140 +8.707766 +1.411275
H +0.444805 +8.690836 +2.491751
C +0.604950 +9.871211 -0.709972
H +0.734907 +10.805655 -1.236594
C +0.604950 +9.871211 +0.709972
H +0.734907 +10.805655 +1.236594
C +2.696074 +0.904680 +1.152995
C +3.935047 +1.518525 +1.193811
C +4.567569 +1.841454 +0.000000
H +5.531840 +2.328977 +0.000000
C +3.935047 +1.518525 -1.193811
H +4.386297 +1.738648 -2.149061
N +2.067935 +0.608063 +0.000000
C +2.696074 +0.904680 -1.152995
H +2.180709 +0.646260 -2.064653
H +4.386297 +1.738648 +2.149061
H +2.180709 +0.646260 +2.064653
\end{verbatim}
\end{minipage}

\bigskip
\begin{minipage}{0.45\textwidth}
\begin{verbatim}
38
[Mn(CO)3(im)(phen)]+: S0 min (B3LYP/TZP)
Mn -0.462641 +0.871149 +0.224406
N +1.032095 +0.168187 +1.521057
N +0.802574 +0.210318 -1.296362
N -1.002827 -1.137193 +0.115222
N +2.993137 -0.287769 +2.384035
C +2.337150 +0.366533 +1.408314
C -1.738897 +1.346945 -0.984395
C -1.561503 +1.266749 +1.620727
C +0.215787 +2.559161 +0.274183
C +1.672748 +0.912303 -2.015349
C +2.532013 +0.321489 -2.951058
C +2.484719 -1.038934 -3.151352
C +1.564967 -1.809811 -2.416050
C +0.739136 -1.132201 -1.498199
C -0.226393 -1.855415 -0.738085
C -0.349226 -3.249123 -0.898282
C -1.327260 -3.905777 -0.127931
C -2.116028 -3.168942 +0.725202
C -1.926241 -1.784057 +0.818234
C +1.428653 -3.229054 -2.554389
C +0.512925 -3.918908 -1.826026
C +2.070487 -0.940823 +3.166428
C +0.853881 -0.649365 +2.623097
O -2.542254 +1.637373 -1.754394
O -2.256798 +1.518994 +2.502283
O +0.652345 +3.623094 +0.311633
H +1.701827 +1.978439 -1.842996
H +3.222327 +0.947471 -3.498149
H -2.879838 -3.635584 +1.330589
H -2.536316 -1.192121 +1.484891
H +2.073350 -3.741322 -3.256858
H +0.414471 -4.990953 -1.937411
H +3.141268 -1.520786 -3.864136
H -1.450487 -4.977328 -0.215314
H +3.994011 -0.296089 +2.509353
H -0.120792 -0.973916 +2.939776
H +2.825164 +0.954526 +0.652517
H +2.353195 -1.539690 +4.013220
\end{verbatim}
\end{minipage}

\bigskip
\begin{minipage}{0.45\textwidth}
\begin{verbatim}
38
[Tc(CO)3(im)(phen)]+: S0 min (B3LYP/TZP)
Tc -0.504419 +0.943061 +0.234166
N +1.083560 +0.154720 +1.579801
N +0.826069 +0.173721 -1.353140
N -1.015243 -1.200783 +0.087748
N +3.049417 -0.390967 +2.375187
C +2.391837 +0.322219 +1.443840
C -1.873475 +1.480409 -1.035591
C -1.680503 +1.363905 +1.720420
C +0.214436 +2.745530 +0.280723
C +1.706272 +0.862469 -2.074220
C +2.574361 +0.255279 -2.988929
C +2.526356 -1.108291 -3.160208
C +1.599940 -1.865849 -2.419421
C +0.758442 -1.176405 -1.522106
C -0.210013 -1.902512 -0.757518
C -0.306620 -3.301951 -0.900819
C -1.274801 -3.972735 -0.130341
C -2.085723 -3.250323 +0.713069
C -1.927354 -1.862165 +0.793779
C +1.482435 -3.287446 -2.540211
C +0.569818 -3.976667 -1.809860
C +2.127013 -1.052594 +3.150384
C +0.906737 -0.706002 +2.650246
O -2.683800 +1.787640 -1.795683
O -2.385084 +1.588355 +2.605706
O +0.670184 +3.805015 +0.297671
H +1.731334 +1.931761 -1.922021
H +3.269949 +0.869587 -3.542277
H -2.841430 -3.730098 +1.318181
H -2.553912 -1.276930 +1.451152
H +2.140017 -3.800678 -3.229811
H +0.484408 -5.051428 -1.903805
H +3.188740 -1.606081 -3.856424
H -1.371662 -5.047534 -0.211074
H +4.052106 -0.430539 +2.478352
H -0.069062 -1.016679 +2.977964
H +2.880599 +0.929112 +0.703190
H +2.412735 -1.692893 +3.965231
\end{verbatim}
\end{minipage}

\bigskip
\begin{minipage}{0.45\textwidth}
\begin{verbatim}
38
[Re(CO)3(im)(phen)]+: S0 min (B3LYP/TZP)
Re -0.506387 +0.945561 +0.234076
N +1.079472 +0.145452 +1.563853
N +0.827178 +0.166913 -1.343083
N -1.004040 -1.199665 +0.089193
N +3.043709 -0.394781 +2.360887
C +2.389012 +0.316952 +1.429217
C -1.870502 +1.487578 -1.021487
C -1.673956 +1.368340 +1.711148
C +0.202020 +2.739241 +0.279311
C +1.705973 +0.861594 -2.063586
C +2.573862 +0.257139 -2.978911
C +2.531259 -1.106830 -3.153076
C +1.607459 -1.869331 -2.414337
C +0.765507 -1.183923 -1.516787
C -0.201687 -1.908345 -0.754692
C -0.302877 -3.305933 -0.897471
C -1.275852 -3.970960 -0.128166
C -2.085771 -3.243021 +0.712346
C -1.922137 -1.856466 +0.795163
C +1.489077 -3.291074 -2.535055
C +0.574681 -3.980251 -1.806031
C +2.121667 -1.057774 +3.135889
C +0.901452 -0.716211 +2.634818
O -2.687028 +1.801263 -1.780745
O -2.380684 +1.594721 +2.601541
O +0.657251 +3.805167 +0.293658
H +1.724983 +1.929791 -1.906323
H +3.267444 +0.874551 -3.531043
H -2.845025 -3.718920 +1.315836
H -2.544361 -1.265087 +1.450382
H +2.147055 -3.804762 -3.223814
H +0.488990 -5.054833 -1.900913
H +3.196140 -1.600125 -3.849971
H -1.377901 -5.045326 -0.207013
H +4.046379 -0.432359 +2.465782
H -0.075648 -1.025659 +2.958531
H +2.873471 +0.926337 +0.688377
H +2.407868 -1.696657 +3.951475
\end{verbatim}
\end{minipage}

\end{multicols}

\end{document}